\documentclass[apj,numberedappendix]{emulateapj}
\begin{document}
\slugcomment{Accepted for publication in ApJ}
\title{The Stellar Populations of Lyman Break Galaxies at $z\sim5$}
\author{Kiyoto Yabe\altaffilmark{1}, Kouji Ohta\altaffilmark{1}, Ikuru Iwata\altaffilmark{2}, Marcin Sawicki\altaffilmark{3,1}\\
Naoyuki Tamura\altaffilmark{4}, Masayuki Akiyama\altaffilmark{4,5}, Kentaro Aoki\altaffilmark{4}}
\email{kiyoyabe@kusastro.kyoto-u.ac.jp}
\altaffiltext{1}{Department of Astronomy, Kyoto University, Kyoto, 606-8502, Japan}
\altaffiltext{2}{Okayama Astrophysical Observatory, National Astronomical Observatory of Japan, Okayama 719-0232, Japan}
\altaffiltext{3}{Department of Astronomy and Physics, St. Mary's University, 923 Robie St., Halifax, Nova Scotia, B3H 3C3, Canada}
\altaffiltext{4}{Subaru Telescope, National Astronomical Observatory of Japan, 650 North A'ohoku Place, Hilo, HI 96720, USA}
\altaffiltext{5}{Astronomical Institute, Tohoku University, Sendai 980-8578, Japan}
\begin{abstract}
We present the results of Spectral Energy Distribution (SED) fitting analysis for Lyman Break Galaxies (LBGs) at $z\sim5$ in the GOODS-N and its flanking fields (the GOODS-FF). With the publicly available IRAC images in the GOODS-N and IRAC data in the GOODS-FF, we constructed the rest-frame UV to optical SEDs for a large sample ($\sim100$) of UV-selected galaxies at $z\sim5$. Comparing the observed SEDs with model SEDs generated with a population synthesis code, we derived a best-fit set of parameters (stellar mass, age, color excess, and star formation rate) for each of sample LBGs. The derived stellar masses range from $10^{8}$ to $10^{11}M_{\odot}$ with a median value of $4.1\times10^{9}M_{\odot}$. The comparison with $z=2-3$ LBGs shows that the stellar masses of $z\sim5$ LBGs are systematically smaller by a factor of $3-4$ than those of $z=2-3$ LBGs in a similar rest-frame UV luminosity range. The star formation ages are relatively younger than those of the $z=2-3$ LBGs. We also compared the results for our sample with other studies for the $z=5-6$ galaxies. Although there seem to be similarities and differences in the properties, we could not conclude its significance. We also derived a stellar mass function of our sample by correcting for incompletenesses. Although the number densities in the massive end are comparable to the theoretical predictions from semi-analytic models involving AGN feedback, the number densities in the low-mass part are smaller than the model predictions. By integrating the stellar mass function down to $10^{8}M_{\sun}$, the stellar mass density at $z\sim5$ is calculated to be $(0.7-2.4)\times10^{7}M_{\sun}\textrm{Mpc}^{-3}$. The stellar mass density at $z\sim5$ is dominated by massive part of the stellar mass function. Compared with other observational studies and the model predictions, the mass density of our sample is consistent with general trend of the increase of the stellar mass density with time.
\end{abstract}


%
%
\section{Introduction}
\label{sec:introduction}
A large number of Lyman Break Galaxies (LBGs) at $z=2-4$ have been
found and studies of their properties have been made extensively \citep[e.g.,][]{steidel96a,steidel96b,giavalisco96,lowenthal97,giavalisco98,pettini98,steidel98,shapley01,shapley03,steidel04,papovich04,reddy05,reddy06,reddy08,sawicki05,sawicki06}. Studies of LBGs at $z\sim5$ \citep{iwata03,iwata07,ouchi04a,ouchi04b,lehnert03,ando04,ando06,ando07,yoshida06,beckwith06,kashikawa06,verma07,stark07,bouwens07,wiklind08} and even at $z\ga6$ have been progressing \citep{stanway03,shimasaku05,bouwens06,yan06,eyles07,bouwens07}. With the gathering of growing samples of galaxies at high redshifts, broad descriptions of galaxy evolution at early cosmic time have been revealed. Particularly, the evolution of the UV-luminosity function and the UV-luminosity density of galaxies from $z\sim6$ to present are extensively studied. Since the rest-frame UV luminosity density traces star-formation activity in galaxies, the recent studies are revealing the cosmic star-formation history. A compilation of the results by  \citet{hopkins06} shows that the cosmic star formation rate (SFR) density increases from $z=6-10$  to $z=2-3$ and decreases to present. It is also important to investigate the stellar mass assembly in the high-redshift universe. It is thought that galaxies are assembled from smaller systems through mergers in the framework of the cold dark matter (CDM) hierarchical structure formation scenario. Thus a stellar mass of an individual galaxy and a stellar mass density of galaxies are expected to be increasing with redshift. The stellar mass together with its age, color excess, and star formation rate are estimated by fitting of spectral energy distribution (SED) of an LBG \citep{sawicki98}. This method is now widely used to constrain  stellar population of galaxies. For instance, \citet{papovich01} and \citet{shapley01} studied stellar population of LBGs at $z\sim3$ and found that the  LBGs have stellar masses of $\sim10^{10} M_{\sun}$ and they are dominated by relatively young (several tens to several hundreds of Myr) stellar populations. Similar study was made at $z\sim2$ ($\sim 1$ Gyr after $z\sim3$) for BX galaxies \citep{shapley05}; the massive side of stellar mass distribution at $z\sim2$ seems to increase slightly, that might indicate the stellar mass evolution of an LBG. Stellar masses for other kinds of galaxy populations at $z=2-4$ as well as for galaxy samples with photometric redshifts have also been studied; these observations  show the gradual growth of the stellar mass density at $z<5$ \citep[e.g.][]{shapley05, daddi04, franx03, brammer07, brinchmann00, dickinson03b, drory05, fontana06, rudnick06, perez-gonzalez08,elsner08}.

Most of the stellar masses of $z<5$ galaxies  have been derived based on observations at wavelengths from optical to near-infrared (NIR); NIR data is necessary  to cover  the  rest-frame wavelength region larger than 4000 \AA, which is sensitive to the stellar mass of a galaxy. In order to push the redshift higher, the observations in mid-infrared are required to constrain stellar masses. With the advent of the Spitzer Space Telescope (hereafter the Spitzer) we can now  access the longer wavelengths. The Infrared Array Camera \citep[hereafter IRAC]{fazio04} on the Spitzer allows us to reach rest-frame optical wavelength for $z\ga5$ galaxies. Only recently, there have been studies of the stellar populations of galaxies at $z\ga5$ in the southern field of the Great Observatories Origins Deep Survey \citep[hereafter the GOODS-S]{dickinson03a}, where deep optical to MIR images obtained with IRAC are available including NIR data \citep{yan06,labbe06,stark07, verma07, eyles07}. However, the results are only obtained in the GOODS-S. Considering relatively small sizes of the samples and a possible cosmic variance, we need to study with a larger sample and in another field to reach robust consensus.

In this paper, we reveal stellar populations, especially stellar mass, of LBGs at $z\sim5$ in a field containing the Great Observatories Origins Deep Survey North \citep[hereafter the GOODS-N]{dickinson03a}, where our group constructed  a large sample of LBGs at $z\sim5$,  studied the evolution of  UV-luminosity function \citep{iwata03, iwata07}, and made spectroscopic follow-up observations \citep{ando04,ando07}. In addition to public IRAC data in the GOODS-N, we made Spitzer/IRAC observations in the flanking fields (hereafter the GOODS-FF) of the GOODS-N. Although the depth of the GOODS-FF is $1 \sim 1.5$ mag shallower than the GOODS-N, it gives us more than twice as large area as the GOODS-S and consequently making the number of our sample galaxies large. This would especially be important to increase the sample size of massive  side of galaxy stellar mass function. We perform the SED fitting analysis with this large sample and derive the stellar populations of galaxies and aim at investigating the evolution of stellar populations from $z\sim5$ to $z=2-3$. We also derive  the stellar mass function and density at $z\sim5$. This paper is organized as follows. In \S 2 we summarize data and sample selection. We outline our population synthesis modeling in \S 3 and present results in \S 4. In \S 5, comparison with LBGs at lower and higher redshifts and  estimation of stellar mass function and stellar mass density are presented. Summary is given in \S 6. Throughout this paper, we adopt the concordance cosmology, $(\Omega_{M}, \Omega_{\Lambda}, h) = (0.3, 0.7, 0.7)$. All magnitudes are on AB system \citep{oke83}.

%
%
\section{Data and Lyman Break Galaxy Sample}
\label{sec:data}
%
%

\subsection{Optical Data}
\label{sec:data_optical}

The optical imaging data used in this work are taken from \citet{iwata07}. They carried out deep and wide imaging observations of two independent blank fields, namely, the field including the GOODS-N and the J0053+1234 field, with the Suprime-Cam \citep{miyazaki02} attached to the Subaru Telescope. Here we use a sub-sample of the GOODS-N field with an effective survey area of 508.5 arcmin$^2$. The images were taken through the $V$, $I_c$, and $z'$ filters. The FWHMs of the reduced data are $\sim1.1\arcsec$. The detailed descriptions of the image properties and the data reduction are presented by \citet{iwata07}. Object detection and photometry were made by using SExtractor \citep{bertin96}. They measured MAG\_AUTO and $1.6\arcsec$ diameter aperture magnitudes for total magnitudes in $z'$-band and $I_c-z'$ colors, respectively. For total magnitudes in $I_c$-band, we calculated them from total magnitudes in $z'$-band and $I_c-z'$ colors. $V$-band magnitudes are obtained by $1.6\arcsec$ diameter aperture photometry. The limiting magnitudes for $V$, $I_c$, and $z'$-band images are 28.1, 26.8, and 26.6 mag, respectively (3$\sigma$, $1.6\arcsec$ diameter aperture). We do not use the $V$-band magnitudes in the SED fitting analysis.

%
%
\subsection{Mid-Infrared Data}
\label{sec:data_midir}
A part of the Subaru imaging area was observed with the IRAC. We use the publicly available GOODS-N IRAC data in this work. We use the First Data Release (DR1) and the Second Data Release (DR2) of IRAC data products from the GOODS Spitzer Legacy Science program, which consist of imaging data in 3.6, 4.5, 5.8, and 8.0 $\mu$m passbands with the total effective area of $\sim$150 arcmin$^2$. The 3$\sigma$ limiting magnitudes in $2.4\arcsec$ diameter apertures are 25.9, 25.6, 23.7, and 23.6 mag in the 3.6, 4.5, 5.8, and 8.0 $\mu$m bands, respectively.

In addition to the GOODS-N data, we obtained IRAC data for the GOODS-FF to cover the most part of the Subaru imaging area. The IRAC data in the the GOODS-FF were obtained in 2005 December and 2006 June as General Observer (GO) program 20218, and are $1-1.5$ mag shallower but $\sim$100 arcmin$^2$ wider than the GOODS-N IRAC imaging data.

We used the Basic Calibrated Data processed by the IRAC data reduction pipelines (version S14.0.0) at the Spitzer Science Center (SSC), and the MOPEX package (version 030106) was used for further reduction. After removing artifacts (mux bleed and column pulldown) on the images, background counts of individual images are estimated and subtracted. Then the pointing refinement was made to improve the consistency of positions of individual images. Finally, the individual frames were drizzled and mosaiced to create a single image of the GOODS-FF for each IRAC channel. The pixel scale of the mosaiced images was set to be $0.606\arcsec$, which is approximately half of the native IRAC pixel scale and almost comparable to the pixel scale of the public images in the GOODS-N. The 3$\sigma$ limiting magnitudes in $2.4\arcsec$ diameter apertures are 24.8, 24.1, 22.2, and 22.3 mag in the 3.6, 4.5, 5.8, and 8.0 $\mu$m bands, respectively. Combining the GOODS-N IRAC data and the GOODS-FF data, we covered a total effective area of $\sim$400 arcmin$^2$, which covers $\sim$80\% of the area taken with Subaru.

Source detection and photometry were made by using SExtractor. As we discuss below, because the IRAC images are crowded and neighboring objects may affect photometry, we performed aperture photometry in all channels with a diameter of $2.4\arcsec$ and applied aperture corrections to obtain total magnitudes. We examined the best value of the aperture size and chose the value of $2.4\arcsec$, which maximizes the signal-to-noise ratio(S/N). The correction factors from aperture magnitudes to total magnitudes were derived from Monte Carlo simulations in which artificial objects with IRAC point spread functions (PSFs) were put into the images, then detected and measured with the same SExtractor parameters as we adopted. In these simulations, our targets were assumed to be point sources because their apparent size is small enough as compared to the PSF. The PSF of the IRAC images was made by stacking the IRAC images of objects showing SExtractor "Stellarity" indices (1 for point sources and 0 for extended sources) larger than $0.98$ in the Subaru data. For the sample of the GOODS-N, the factors are $-0.69$, $-0.72$, $-0.99$, and $-1.06$ mag in 3.6, 4.5, 5.8, and 8.0 $\mu$m bands, respectively. The uncertainties of the correction are $\sim2\%$, $\sim3\%$, $\sim10\%$, and $\sim10\%$ in 3.6, 4.5, 5.8, and 8.0 $\mu$m bands, respectively. For the sample of the GOODS-FF, the factors are $-0.70$, $-0.73$, $-0.95$, and $-1.01$ mag in 3.6, 4.5, 5.8, and 8.0 $\mu$m bands, respectively. The uncertainties of the correction are $\sim3\%$, $\sim5\%$, $\sim30\%$, and $\sim30\%$ in 3.6, 4.5, 5.8, and 8.0 $\mu$m bands, respectively. The errors of the resulting magnitudes are taken to be 1$\sigma$ standard deviation of sky background. Because of the too low S/N to provide useful upper limits in the SED fitting and large uncertainty of the correction factors in the $5.8\mu$m and $8.0\mu$m bands, we do not use data of these bands for the SED fitting.

%
%
\subsection{Sample}
\label{sec:data_sample}

\begin{figure}
\epsscale{1.0}
\plotone{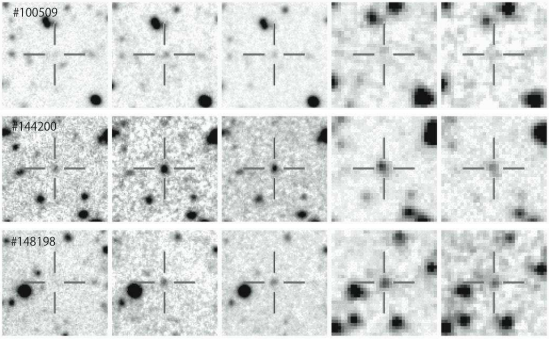}
\caption{Postage stamps ($20\arcsec$$\times$$20\arcsec$) of three representative objects in 5 passbands. From left to right, $V$, $I_{c}$, $z'$, $3.6\mu$m, and $4.5\mu$m-band images are shown and the LBG candidate is indicated by a cross in each panel. North is up and east is to the left.\label{fig:montage}}
\end{figure}

The LBG sample we use in this work is selected by the following color selection criteria \citep{iwata07}:
\begin{equation}
\begin{array}{c}
V - I_{c} > 1.55 \ \ \& \ \ V - I_{c} > 7.0(I_{c} - z') + 0.15\ .
\end{array}
\label{eq:colorselectioncriteria}
\end{equation}
The number of objects which satisfy these selection criteria is 617 in $z' < 26.5$ mag. These criteria are designed to select LBG candidates at $z\sim5$ (hereafter, we refer them as LBGs at $z\sim5$, though they are candidates in the strict sense.) efficiently without heavy contamination by interlopers such as objects at lower redshift and stars in the Galaxy. Follow-up spectroscopy of the candidates confirms that the selection criteria effectively extract star-forming galaxies at $z\sim5$ \citep{ando04,ando07}, though, the number of spectroscopically confirmed objects is still limited to $\sim10$. \citet{iwata03, iwata07} estimated a fraction of the interlopers employing a resampling method, and found  the estimated fractions of interlopers are $\sim50$\%, $\sim20$\%, and $\sim 10$\% in $z' = 23.0-24.0$ mag, $24.0-26.0$ mag, and $26.0-26.5$ mag, respectively \citep[see Table 5. of][]{iwata07}. We correct for the factors, when we derive the stellar mass function in \S \ref{sec:discussions_mass_function}.

\begin{figure*}
\epsscale{0.90}
\plotone{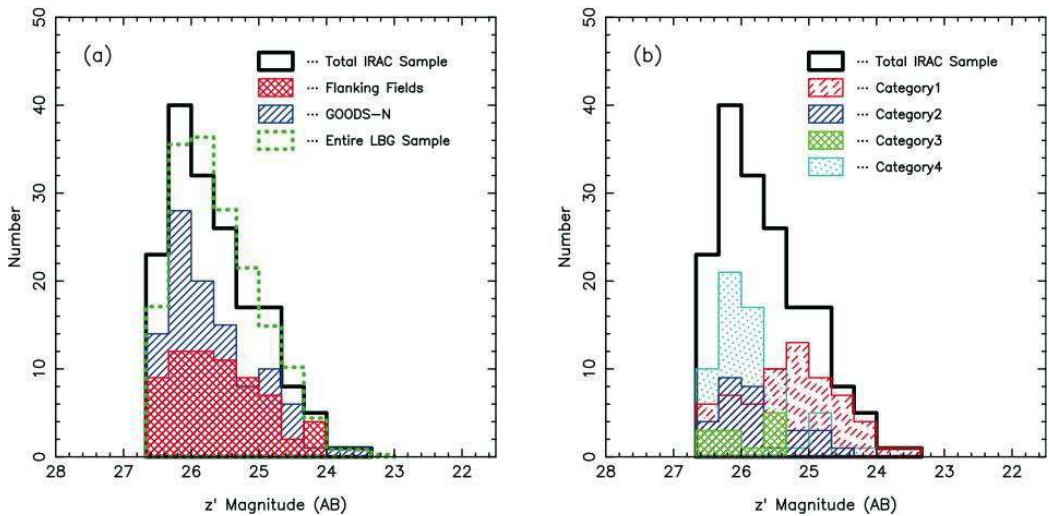}
\caption{(a) Distributions of the $z'$-band magnitudes of the sample. The distribution of the IRAC sample used in this work is indicated by thick solid line. The distribution of the entire LBG sample (617 objects) by \citet{iwata07} is scaled to the total number of the IRAC sample (170 objects) and is indicated by dotted line. The distributions of the GOODS-N sample and the GOODS-FF are indicated by hatched region and cross-hatched region, respectively. (b) Distributions of $z'$-band magnitudes for the total IRAC sample and the samples categorized by the IRAC detection (from Category 1 to 4; see text for the categorization).\label{fig:zmag_distribution}}
\end{figure*}

There are deep X-ray observations with \textit{Chandra} in the GOODS-N \citep{alexander03}. About 60\% of our sample LBGs lie in the region covered by \textit{Chandra}. We cross-matched the LBGs with X-ray sources and found one object is an active galactic nucleus (AGN) at $z=5.186$ with $L_{X}=6.8\times10^{43}$ ergs s$^{-1}$ (2-10 keV) \citep{barger02,ando04}. We do not use this object in the SED fitting analysis. All others are not detected at the $3\sigma$ flux limit of  $2.5\times10^{-17}$ erg s$^{-1}$ cm$^{-2}$ (0.5-2.0 keV), which corresponds to $L_{X} \sim 6 \times10^{42}$ erg s$^{-1}$ (2-10 keV). This luminosity level is that for Seyfert class AGNs, and hence the LBGs do not harbor X-ray luminous AGNs or they may be obscured AGNs. \citet{hasinger08} suggested that the fraction of type-2 AGN increases with redshift, and at $z=3-5$ the type-2 fraction in luminosity range of $L_{X}=10^{42}-10^{43}$ erg s$^{-1}$ is $\sim0.9$. Hence, in any cases, AGN components presumably do not affect the SEDs of the host galaxies and the results of the SED fitting.

Because the mean FWHM of the IRAC PSFs is $\sim$$1.8\arcsec$ as contrasted with that of Subaru optical PSF of $\sim$$1.1\arcsec$, some objects in the IRAC images are seriously contaminated by surrounding objects. For this reason, we checked 617 objects by eye whether they have neighbors in their close vicinity on the high-resolution $z'$-band image. Furthermore, we also examined the IRAC images by eye whether the neighboring objects around the LBG position affect the photometry to  make a sample of the  LBGs secure for the SED fitting. After these inspections, we selected 170 objects for subsequent analyses. In Figure \ref{fig:montage}, representative objects of the sample LBGs that are  detected both in 3.6$\mu$m and 4.5$\mu$m bands are shown.  

The distribution of the $z'$-band magnitudes for the sample is shown in Figure \ref{fig:zmag_distribution}. The $z'$-band magnitudes for the total sample range from 26.5 to 23.5 mag (M$_{1500\textrm{\AA}}$=$-19.8$ to $-22.8$ mag) with a median value of 25.8 mag (M$_{1500\textrm{\AA}}$=$-20.5$ mag), where the absolute magnitudes are calculated with $z=4.8$ and the spectral indices obtained from $I_C-z'$ colors. In Figure \ref{fig:zmag_distribution}a, the magnitude distribution of the sample we use in this work (170 objects) with that of the entire LBG sample (617 objects) is plotted. There seems not to be much difference between them. We made the Kolmogorov-Smirnov test for the distribution of $z'$-band magnitude; the hypothesis that the IRAC LBG sample is chosen from the the same sample as the original LBG sample cannot be rejected at the 5\% confidence level. We also made the test for $I_C - z'$ color distribution. Again we cannot reject the hypothesis. A fraction of the uncontaminated objects in IRAC images in each $z'$-band is almost constant within $1\sigma$ error and is independent of the $z'$-band magnitude. These tests support the idea that the sub-sample is not biased by our selection of objects based on their lack of neighbors, i.e. it is randomly selected out of the total sample. According to the expected fraction of the interlopers in each $z'$-band magnitude bin presented by \citet{iwata07}, the expected number of interlopers is 25 among the IRAC sample of 170 objects. Among the IRAC sample, four objects meet IRAC-selected extremely red object (IERO) criteria \citep{yan04}. However, even if these objects are truly $z\la3$ IEROs, they do not affect main results in this paper.

\begin{figure*}
\begin{center}
\epsscale{0.90}
\plotone{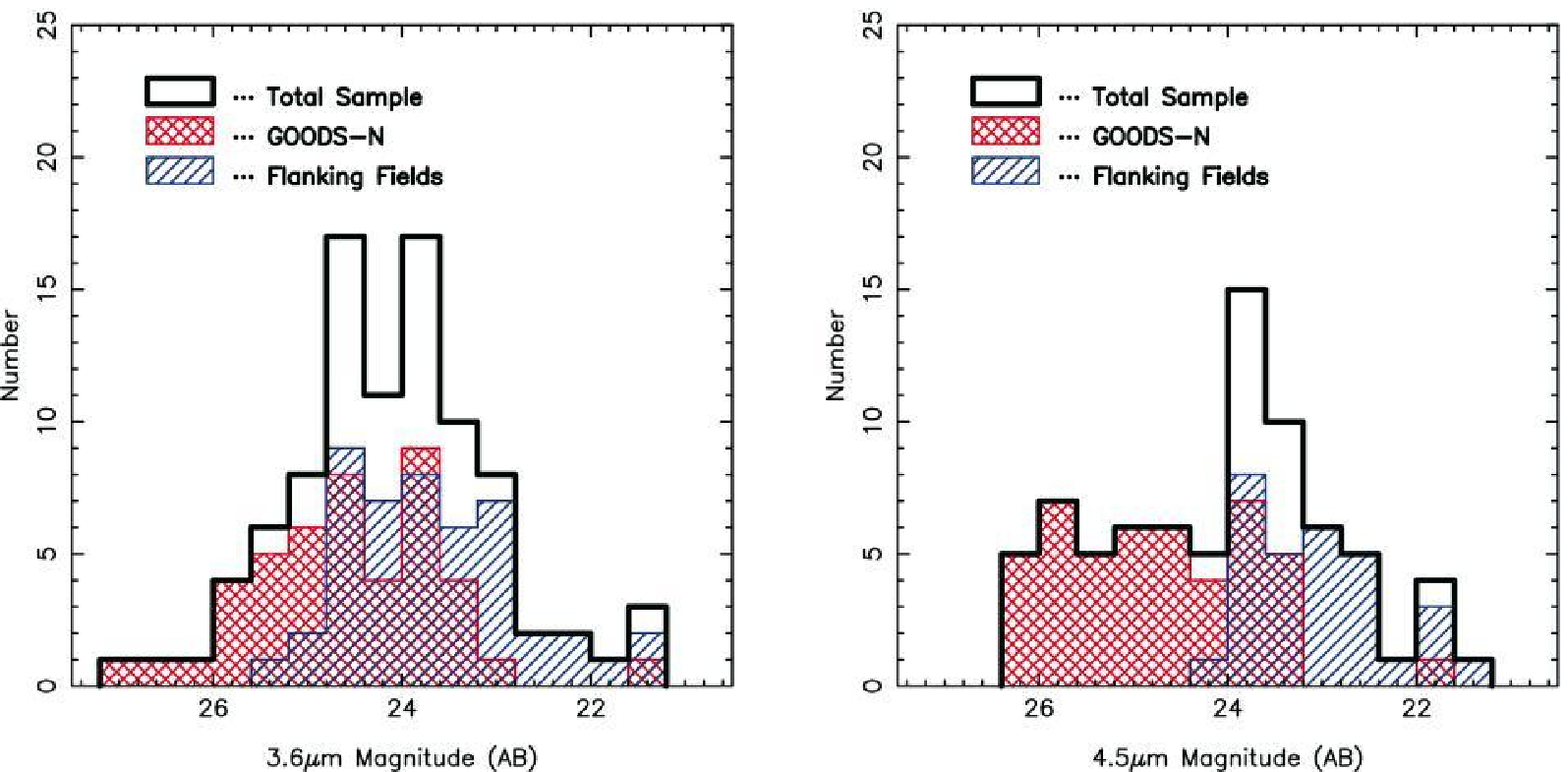}
\caption{Distributions of magnitudes in $3.6\mu$m (left panel) and $4.5\mu$m (right panel) band of the sample, and those for the GOODS-N and the GOODS-FF region.\label{fig:ch1_ch2_distribution_region}}
\plotone{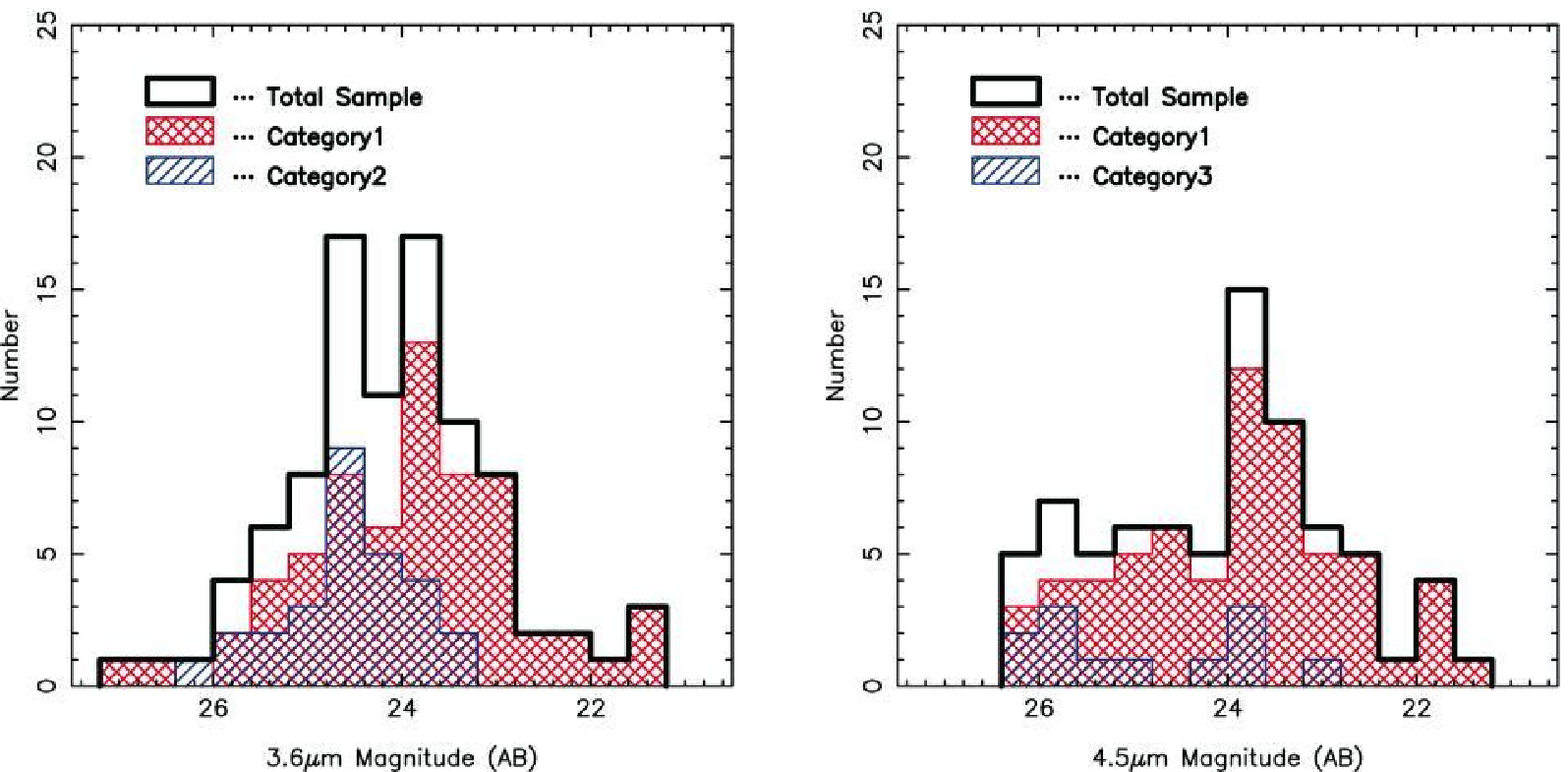}
\caption{Distributions of magnitudes in $3.6\mu$m (left panel) and $4.5\mu$m (right panel) band of the sample, and those for the Category 1, 2, and 3.\label{fig:ch1_ch2_distribution_category}}
\end{center}
\end{figure*}

The cross-matching between sources in the Subaru images and IRAC images was made with a $1.2\arcsec$ search radius. Most of the IRAC counterparts are found within $0.2\arcsec-0.4\arcsec$, which is almost comparable to the accuracy of our astrometry. With this search radius, 105 objects are detected in IRAC 3.6$\mu$m and/or 4.5$\mu$m. Among these objects 64 are detected both in 3.6$\mu$m and 4.5$\mu$m bands, and 29 and 12 are detected only in 3.6$\mu$m and 4.5$\mu$m band, respectively. The other 65 objects are detected neither in 3.6$\mu$m nor 4.5$\mu$m bands. These objects are  grouped into four categories: objects detected both in 3.6$\mu$m and 4.5$\mu$m (Category 1), objects detected only in 3.6$\mu$m (Category 2), objects detected only in 4.5$\mu$m (Category 3), and objects detected neither in 3.6$\mu$m nor 4.5$\mu$m (Category 4). Note that the Category 2-4 objects are not undetected due to the blending with the neighboring sources but are intrinsically faint in IRAC bands because we selected objects that are isolated and free from contaminations from nearby sources in IRAC bands. Figure \ref{fig:zmag_distribution}b shows the distributions of $z'$-band magnitudes of the samples categorized from 1 to 4. It is notable that objects in the Category 1 sample are generally brighter in $z'$-band than those of the other categories. The magnitudes of Category $1-3$ objects are presented in Table \ref{table1}, which will used for SED fittings.

The distributions of the magnitudes in 3.6$\mu$m and 4.5$\mu$m band are shown in Figure \ref{fig:ch1_ch2_distribution_region} and \ref{fig:ch1_ch2_distribution_category}. In 3.6$\mu$m band, the magnitudes of our sample range from $m_{3.6\mu m}$=21.5 to 26.9 mag, which corresponds to a range of $M_{r'}$=$-24.8$ to $-19.4$ mag at $z=4.8$. In the 4.5$\mu$m band, the magnitudes of our sample range from $m_{4.5\mu m}$=21.3 to 26.3 mag, which corresponds to a range of $M_{i'}$=$-25.0$ to $-20.0$ mag at $z=4.8$. Figure \ref{fig:ch1_ch2_distribution_region} shows the absence of the faint objects in the GOODS-FF due to the shallower limiting magnitudes both in 3.6$\mu$m and 4.5$\mu$m bands. The figure also shows that the absence of the bright objects in the GOODS-N compared to the GOODS-FF. This may be due to a fact that the LBGs brighter in $z'$-band tend to reside in the GOODS-FF rather than in the GOODS-N, probably by chance. Figure \ref{fig:ch1_ch2_distribution_category} shows that the Category 2 and 3 objects are generally fainter than the Category 1 objects.

\begin{figure*}
\epsscale{1.00}
\plotone{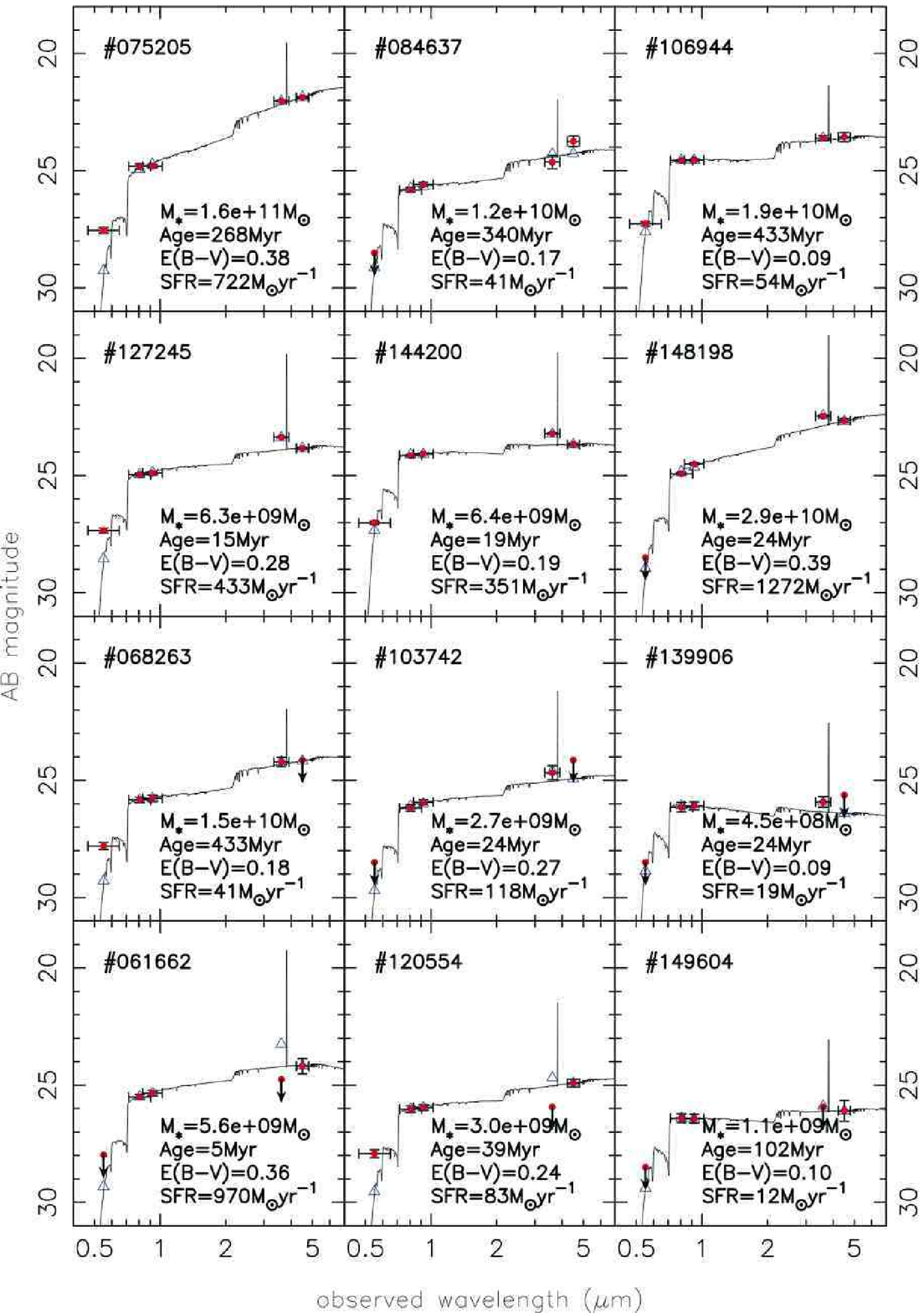}
\caption{Best-fit model SEDs (\textit{open triangles}) and observed SEDs (\textit{filled circles}). Best-fit model spectrum is also shown with solid line. Although $V$-band magnitudes are plotted, the data were not used in the fitting procedure. In each panel, object ID and its best-fitted parameters are shown.\label{fig:fit_results}}
\end{figure*}

%
%
\section{Population Synthesis Models and SED fitting}
\label{sec:models}
We then perform SED fitting by using the \textit{SEDfit} software package\footnote{This code uses a standard $\chi^2$ minimization algorithm. The main differences compared to \citet{sawicki98} are that this code uses the \citet{calzetti00} law instead of the \citet{calzetti97} and the population synthesis model by \citet{bruzual03} instead of that by \citet{bruzual93}. This package will soon be made publicly available and in the meantime can be obtained by contacting M. Sawicki at sawicki@ap.smu.ca.} (Sawicki, in prep.), which employs essentially the same algorithm as that by \citet{sawicki98}.

We generated a set of model SEDs with a population synthesis model as follows: We used the \citet{bruzual03} (hereafter BC03) population synthesis code with Padova 1994 evolutionary tracks. The Salpeter IMF \citep{salpeter55} with the mass range of $0.1M_{\sun}-100M_{\sun}$ was used. Although this combination of the modeling might not be modern now, we intend to compare our results with previous studies of LBGs at the lower redshifts to see the evolution. Metal abundance was adopted to be 0.2$Z_{\sun}$, by considering that metallicity of galaxies at $z=2-3$ are still sub-solar \citep{pettini01,erb06,halliday08} and the metallicity of $z\sim5$ LBGs  is suggested to be at least $Z\sim0.1Z_{\sun}$ \citep{ando07}. We adopted the constant star formation history (CSF). In Appendix \ref{appendix:effects}, we investigate effects on the  results by adopting other metallicities (1.0$Z_{\odot}$ and 0.02$Z_{\odot}$) and star formation histories (instantaneous burst,  exponentially declining models with $\tau$=1 Myr, 10 Myr, 100 Myr, 1 Gyr, and 10 Gyr, and two-component star formation history models). The universe at $z\sim5$ is $\sim$1.2 Gyr old and no object can be older than that. However, we allowed for model ages of up to 20 Gyr that is the oldest age BC03 provides as a check on the fits. The BC03 uses 221 age steps from 0.1 Myr to 20 Gyr, which are not equally-spaced in logarithmic scale. The \textit{SEDfit} resamples this 221 to 51 equally-spaced logarithmic age steps both to speed up the calculations and to avoid having to deal with the unequally-spaced scale of the original 221 models.

We took into account H$\alpha$ emission line in the model spectrum. The H$\alpha$ emission line comes into the 3.6$\mu$m band if the redshifts of the LBGs are $4.0\la z \la5.0$. The procedure of putting the H$\alpha$ line in the model spectrum is not equipped in the \textit{SEDfit}. The H$\alpha$ luminosity is calculated from the model star formation rate (i.e., intrinsic star formation rate) by using the \citet{kennicutt98} relation and is put into the model spectrum. \citet{stark07} estimated that the contribution of the H$\alpha$ line to the flux density in the 3.6$\mu$m band is $10-20\%$ for $z\sim5$ LBGs and does not affect the stellar mass significantly. \citet{eyles07} also reported the H$\alpha$ contribution for $z\sim6$ LBGs is $\la10\%$ and the results in their paper do not change within errors. We examine this effect for our sample LBGs at $z\sim5$. For the $z=4.8$ galaxies, the effects of H$\alpha$ inclusion on the magnitudes in 3.6$\mu$m-band range from $\sim0.1$ (for model age = 1 Gyr) to $\sim0.7$ mag (for model age=10 Myr), depending on the age of the model spectrum. The contribution of the H$\alpha$ for the model age of 25 Myr, which is the median value of the best-fitted age as we mention in \S\ref{sec:result_other_parameters}, is $\sim0.5$ mag. Since this difference is larger than the typical errors in $3.6\mu$m band and is not negligible, we take into account the H$\alpha$ line and run the SED fitting. It should be worth emphasizing that the inclusion of H$\alpha$ emission line in the model spectrum improves the fit very much without increasing the number of free parameters as shown below. Details for the inclusion of H$\alpha$ line and effects on the results are discussed in Appendix \ref{appendix:halpha}.

Resulting model spectrum was attenuated by internal dust with extinction values ranging from $E(B-V) = 0.0$ to 0.8 mag in a step of 0.01 mag using the extinction law by \citet{calzetti00}. We also tested the effects of using the Small Magellanic Cloud (SMC), Large Magellanic Cloud (LMC), and Milky Way (MW) extinction laws (see details in Appendix \ref{appendix:effects}). Finally, the spectrum was attenuated by intergalactic medium (IGM) using the prescription by \citet{madau95}.

\begin{figure*}
\epsscale{0.90}
\plotone{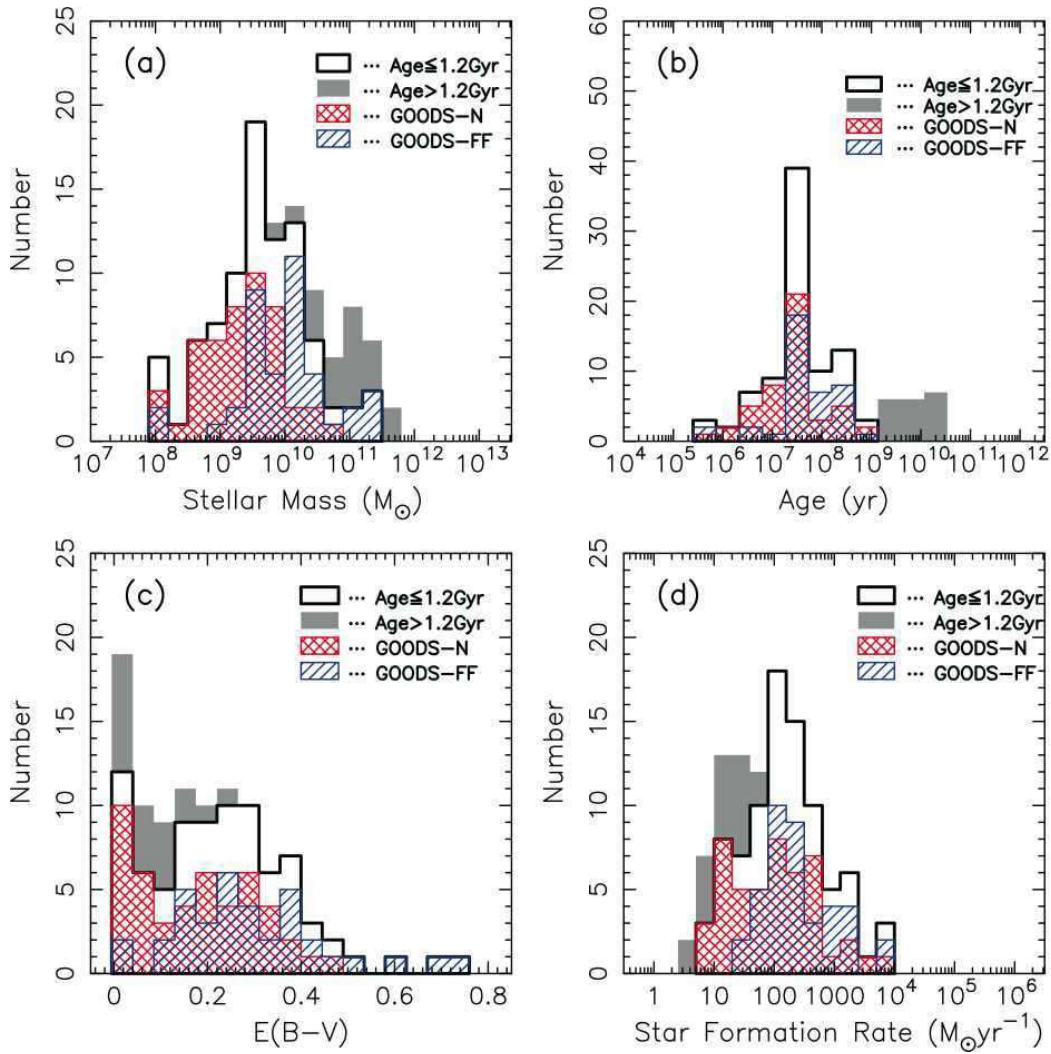}
\caption{Distributions of the best fit parameters. The distributions of the best-fitted stellar masses, ages, color excesses, and SFRs are plotted in panel (a), (b), (c), and (d), respectively. The distributions of sample in the GOODS-N and in the GOODS-FF are indicated by a cross-hatched and a hatched region, respectively, in each panel. Shaded regions indicate the overage LBGs; these overage histograms are plotted cumulatively on top of the $<1.2$ Gyr histograms. See text for details.\label{fig:params_distribution}}
\end{figure*}

We fixed the redshift to $z=4.8$ in order to reduce the number of free parameters. According to the selection function by \citet{iwata07}, the expected redshifts of the LBGs range from $z=4.3$ to $z=5.3$ and the average redshift is $z=4.8$. Also, the spectroscopic study by \citet{ando07} shows that the distribution of the identified redshifts is broadly consistent with the expected distribution with the mean redshift of $z=4.8$. We thus adopted the redshift of the objects in our sample as 4.8 in the SED fitting. Note that two objects in the sample are spectroscopically confirmed ($z = 4.70$ and 4.62), but we take the redshifts of these objects to be 4.8. We examine the effects of this assumption on the stellar mass in Appendix \ref{appendix:redshift}, and find that fixing the redshift induces a systematic offset of only 0.06 dex and the uncertainty is 0.22 dex compared to making the redshift free. We additionally assess the uncertainty by assigning the redshift randomly along the expected distribution by \citet{iwata07}. The uncertainties are $3-5$ times smaller than the errors in the SED fitting (see Appendix \ref{appendix:redshift}).

In this SED fitting analysis, except for a testing case of redshift free fitting, we used the data in $I_C$, $z'$, IRAC 3.6$\mu$m, and 4.5$\mu$m bands. We did not use $V$-band magnitude to avoid the uncertainty due to the fluctuation of IGM absorption. For Category 2 and 3 objects, the upper limit magnitudes were not used in the SED fitting. The free parameters in the SED fitting are age, color excess, and scaling normalization (stellar mass and SFR).

%
%

\section{Results}
\label{sec:results}
%
%
\subsection{Stellar Mass}
\label{sec:results_mass}
We select via eye inspection 170 objects that are isolated and are not contaminated by neighboring objects in $z'$-band and IRAC images out of the entire LBG sample consisting of 617 objects. We made the SED fitting for 105 objects that are detected in IRAC 3.6$\mu$m and/or 4.5$\mu$m among the 170 objects. We did not make the SED fitting for the other 65 objects detected neither in 3.6$\mu$m nor 4.5$\mu$m. Figure \ref{fig:fit_results} shows representative examples of the observed SEDs and the best-fitted models for 12 objects (6 objects from the Category 1, 3 objects from the Category 2, and 3 objects from the Category 3). The best-fitted parameters (stellar mass, stellar age, color excess, and star formation rate) of each object are summarized in Table \ref{table2}. Almost all of the SEDs are well reproduced. The effect of including H$\alpha$ emission can be seen particularly in the second row of Figure \ref{fig:fit_results}; the excess due to the H$\alpha$ emission is significant in the IRAC 3.6$\mu$m band. However as seen in the bottom row of Figure \ref{fig:fit_results}, for the Category 3 the model flux density is larger than the upper limit in the 3.6$\mu$m band. This may be caused by the assumption of $z=4.8$. Category 3 objects may show an excess in IRAC 4.5$\mu$m band rather than in 3.6$\mu$m. Since the H$\alpha$ emission comes into the 4.5$\mu$m band if $z>5.1$, the redshifts of these LBGs may be larger than 5.1. In fact, if we take the redshift as a free parameter in the range from $z=3.8$ to 5.5 with $\Delta z=0.1$, the value of the chi-square is minimum at $z\sim5.2$ or is almost comparable to those at $z\la5.0$, suggesting the presence of H$\alpha$  emission in the 4.5$\mu$m band. We proceed, however, assuming that all of our sample objects are at $z=4.8$.

\begin{figure}
\begin{center}
\epsscale{1.00}
\plotone{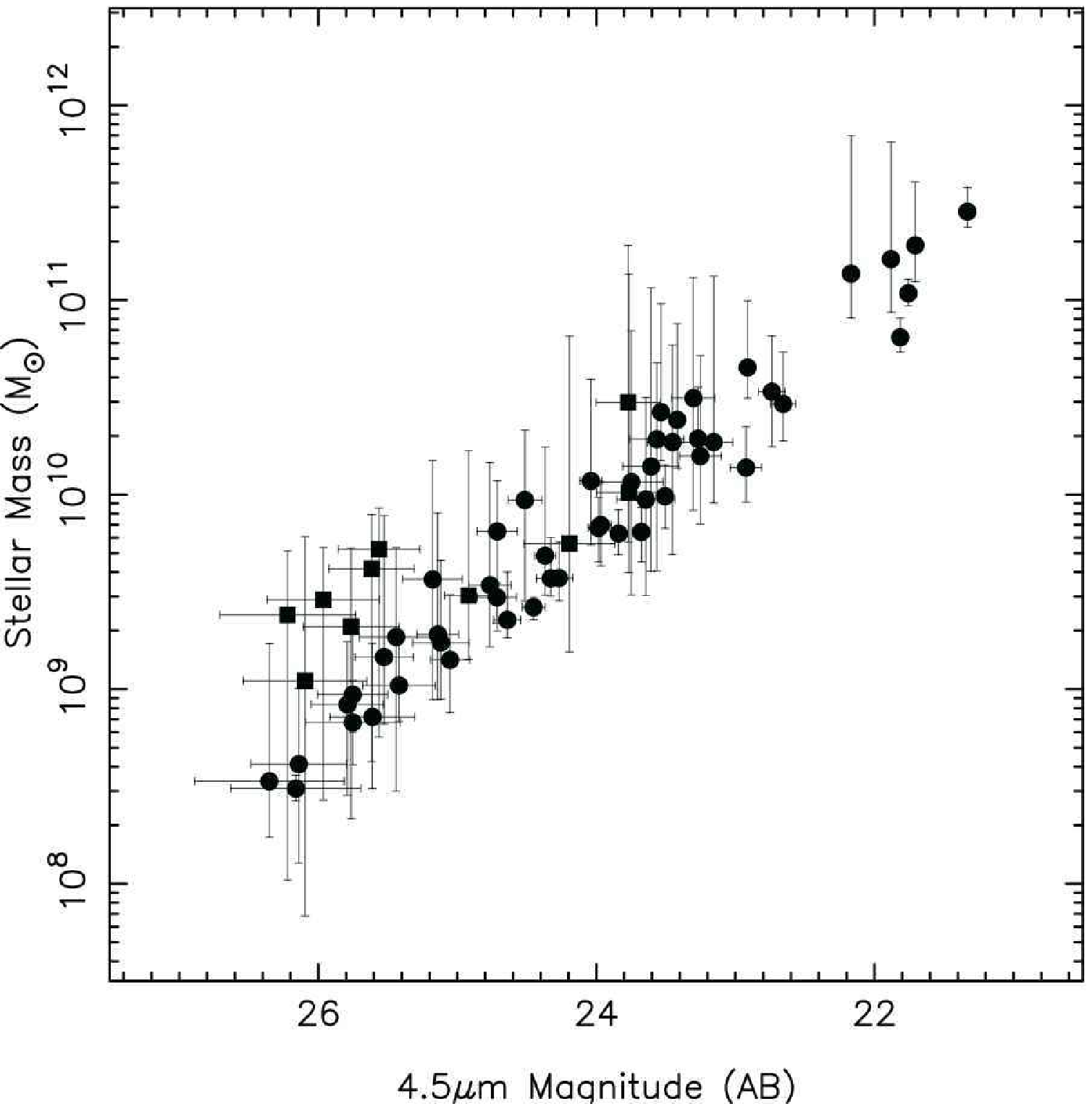}
\caption{$4.5\mu$m magnitudes vs. the derived stellar masses. Category 1 objects and Category 3 objects are indicated by circles and squares, respectively. The vertical error bars are 90\% confidence level and the horizontal bars show 1$\sigma$ errors in magnitudes.\label{fig:ch2_mass}}
\end{center}
\end{figure}

The distribution of the best-fitted stellar masses is shown in Figure \ref{fig:params_distribution}a. The derived stellar masses of the whole sample galaxies range from $10^{8}$ to $10^{11}M_{\sun}$ with a median value of $4.1 \times 10^{9} M_{\sun}$. The typical error at the $90\%$ confidence level in each SED fitting is $\sim$0.4 dex. As we discuss in Appendix \ref{appendix:effects}, the differing star formation histories, metallicities, and dust extinction laws affect the output parameters of the fitting to some degree. Age and color excess are most affected and stellar mass is least affected. The uncertainties of these effects on the stellar mass are $\sim$0.6 dex at most. Figure \ref{fig:params_distribution}a indicates that some massive galaxies ($>10^{11} M_{\sun}$) have already been assembled at $z\sim5$ when the universe was only 1.2 billion years old. The best-fitted stellar ages of 20 objects are, however, older than the cosmic age ($\sim$1.2 Gyr) at $z\sim5$. In the distributions of Figure \ref{fig:params_distribution}, these objects are shown as shaded regions. It is noteworthy that the stellar masses of these overage objects are typically large. A cause for the large ages is considered to be due to the assumption of constant star formation history \citep{sawicki98, papovich01}. In general, the derived ages are older than those by assuming other star formation histories such as an instantaneous burst or an exponentially declining star-formation history. Although we focus on results for objects whose best-fitted ages are younger than 1.2 Gyr, the presence of this population should be kept in mind. We will come back to this problem when we derive the stellar mass function and density.

A relationship between magnitudes in the 4.5$\mu$m band (which corresponds to approximately the rest-frame $i'$ band) and stellar masses is shown in Figure \ref{fig:ch2_mass}. A clear correlation between the 4.5$\mu$m magnitudes and the stellar masses ($\textrm{log}(M_{*}/M_{\odot})=-0.56\times m_{4.5\mu m} + 23.27$) is seen (especially for Category 1), indicating that the rest-frame optical flux is a good indicator of the stellar masses. The correlation is not exactly linear; the mass-to-light ratio is larger in the brighter objects. 

\begin{figure*}
\begin{center}
\epsscale{0.90}
\plotone{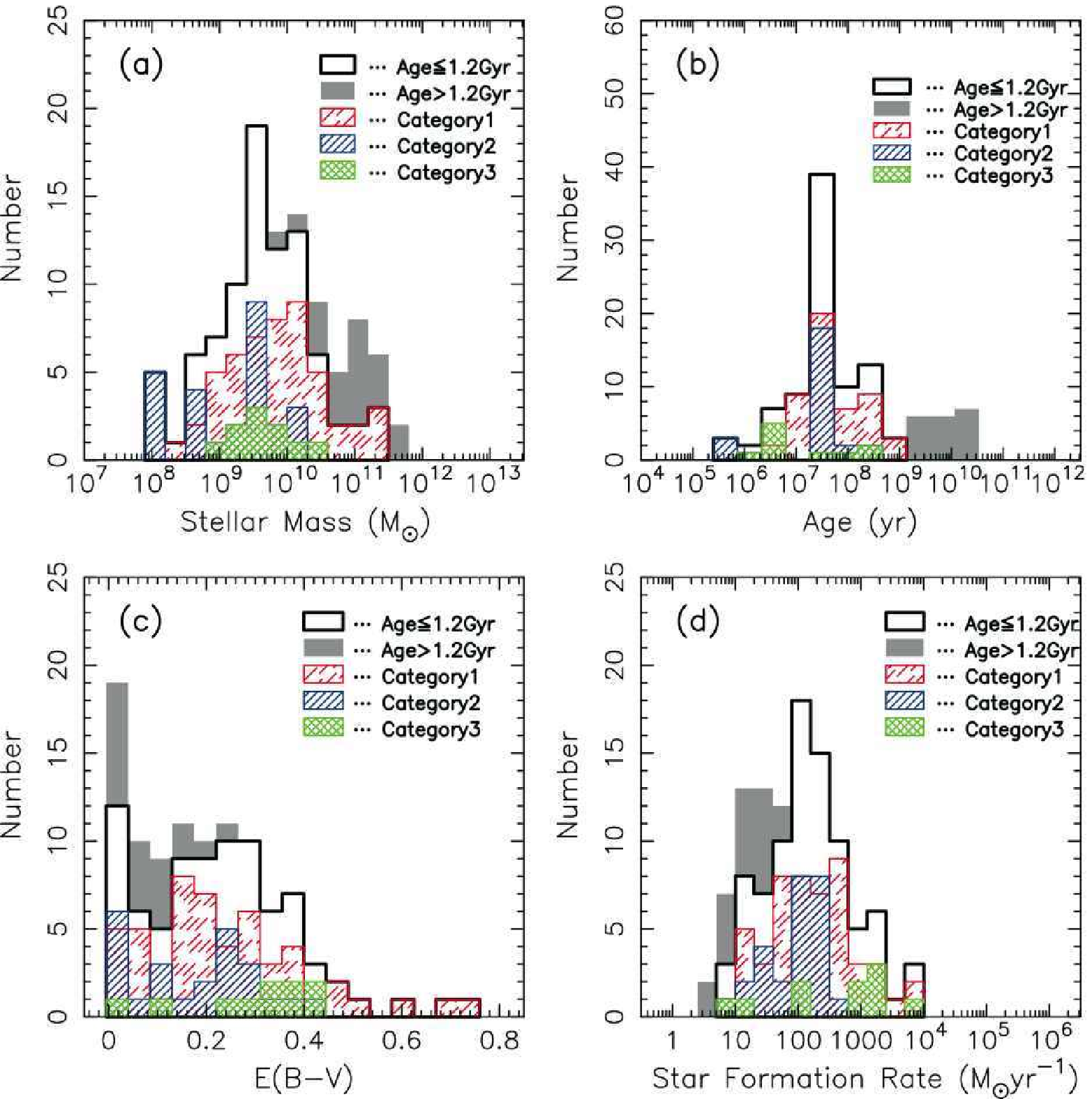}
\caption{Distributions of the best-fitted parameters. The distributions of the best-fitted stellar masses, ages, color excesses, SFRs are plotted in panel (a), (b), (c), and (d), respectively. The LBG sample is divided into four categories (see text). Shaded regions indicate the overage LBGs; these overage histograms are plotted cumulatively on top of the $<1.2$ Gyr histograms. See text for details.\label{fig:params_distribution_category}}
\end{center}
\end{figure*}

The Figure \ref{fig:params_distribution}a shows that the objects in the GOODS-N are relatively less massive than those in the GOODS-FF. The median stellar masses of the objects in the GOODS-N and the GOODS-FF are $2.4 \times 10^{9} M_{\sun}$ and $1.0 \times 10^{10} M_{\sun}$, respectively. It is reasonable because in the GOODS-FF the LBGs faint in IRAC bands are absent, while the bright ones tend to reside in. The number of massive ($>10^{10}M_{\odot}$) objects in the GOODS-N is smaller than that in the GOODS-FF. This deficit of the massive objects in the GOODS-N may be due to cosmic variance. Figure \ref{fig:params_distribution_category} shows distributions of stellar masses for the Category 1, 2, and 3 objects. The median stellar masses of the Category 1, 2, and 3 objects are $6.9 \times 10^{9} M_{\sun}$, $2.9 \times 10^{9} M_{\sun}$, and $3.6 \times 10^{9} M_{\sun}$, respectively. As a whole, the Category 2 or 3 objects are relatively less massive than the Category 1 objects. This is also reasonable if we recall the faintness of the Category 2 and 3 objects in the rest-frame optical wavelength (Figure \ref{fig:ch1_ch2_distribution_category}).

\begin{figure}
\begin{center}
\epsscale{1.00}
\plotone{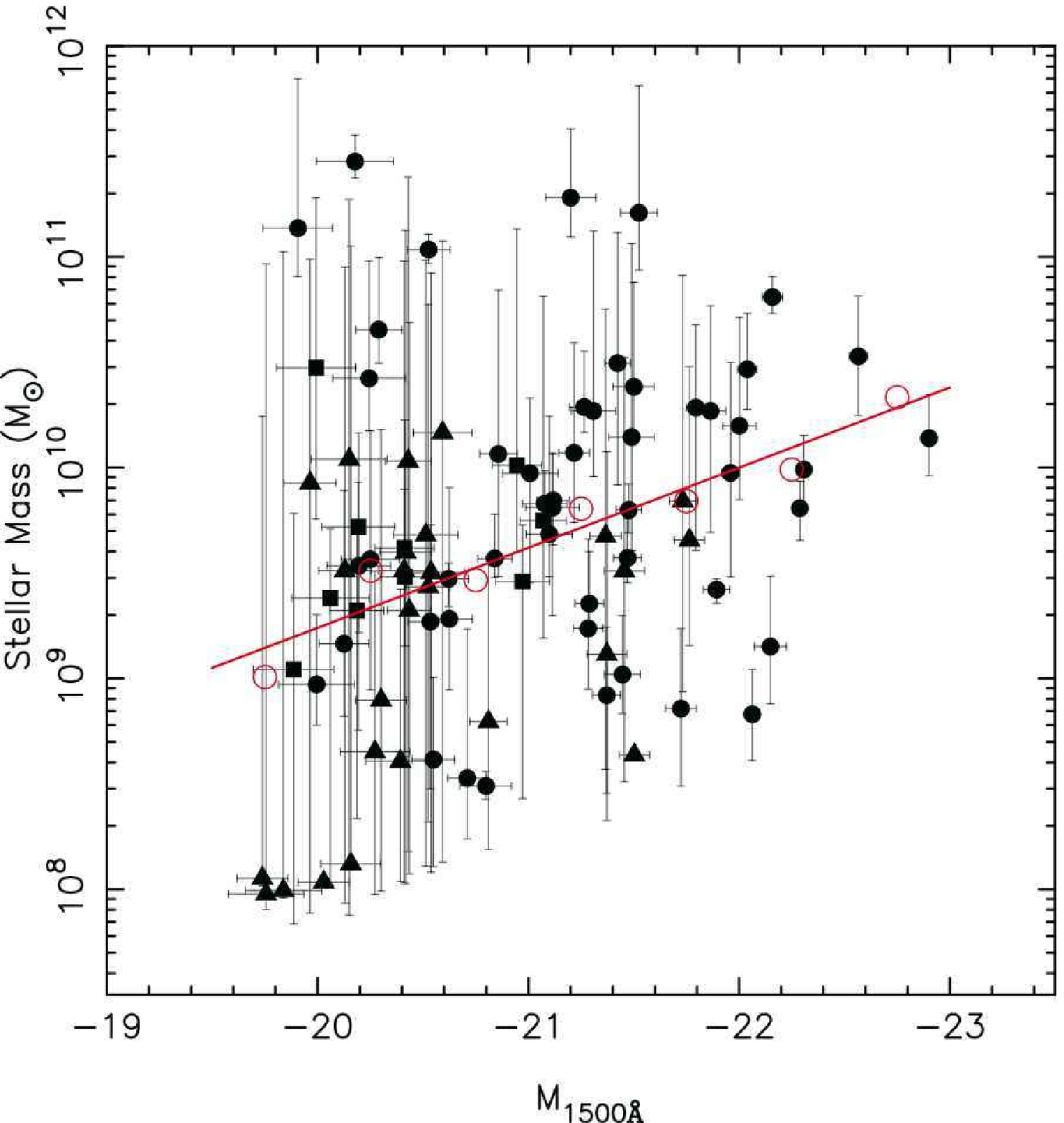}
\caption{Absolute UV magnitudes (uncorrected for dust extinction) vs. stellar masses for Category 1 objects (\textit{circles}), Category 2 objects (\textit{triangles}), and Category 3 objects (\textit{squares}). Large open circles show median stellar masses in 0.5 mag bin. Solid line is the regression line for these points.\label{fig:muv_mass}}
\end{center}
\end{figure}

As shown in Figure \ref{fig:muv_mass}, there seems to be no clear correlation between the rest-frame UV absolute magnitudes (uncorrected for dust extinction) and the stellar masses in LBGs at $z\sim5$. However, we compute a correlation coefficient for the relation, and obtain $r=-0.3$. We can reject the null hypothesis that there is no correlation between the UV absolute magnitude and the stellar mass at the 5\% confidence level. The median stellar masses in 0.5 mag bins show a loose correlation between the rest-frame UV absolute magnitudes and the stellar masses: $\textrm{log}(M_{*}/M_{\odot})=-0.38\times M_{\textrm{1500\AA}}+1.64$. \citet{shapley01,shapley05} found no correlation between the UV absolute magnitude and the stellar mass for LBGs at $z=2-3$. However, \citet{sawicki06b} found that sub-$L^{*}$ LBGs at $z\sim2$ that are much fainter than those studied by \citet{shapley05} show the correlation. \citet{papovich01} also found the correlation between the UV  absolute magnitude and the stellar mass for LBGs at $z\sim3$ by using a deeper sample than that by \citet{shapley01}. The correlation between the rest-frame UV magnitude and the stellar mass for $z\sim5$ sample is much weaker than that found at $z\sim2$ (\citealt{sawicki06b}; Sawicki, in prep.).

%
%
\subsection{Ages, Dust, and Star Formation Rates}
\label{sec:result_other_parameters}
The distributions of the best-fitted ages and color excesses are also presented in Figure \ref{fig:params_distribution}b and \ref{fig:params_distribution}c, respectively. The typical error of the age for each object is $\sim$1.0 dex. The median value of the ages estimated for our sample is 25 Myr. The median values of the ages estimated for the GOODS-N and the GOODS-FF sample are 19 Myr and 31 Myr, respectively. The median values of the ages of the Category 1, 2, and 3 objects are 35 Myr, 25 Myr, and 5 Myr, respectively. The typical error of the color excess for each object is $\sim$0.1 mag. The median value of the color excesses estimated for the total LBG sample is 0.22 mag. The median values of the color excesses estimated for the GOODS-N and the GOODS-FF sample are 0.18 mag and 0.25 mag, respectively. The median values of the color excesses of the Category 1, 2, and 3 objects are 0.20 mag, 0.20 mag, and 0.33 mag, respectively.

It is known, however,  that rest-frame UV-optical colors are degenerate with respect to age and dust extinction. Whether the colors are explained by extinction or age is hard to be specified uniquely. Generally speaking, for LBGs at $z\sim5$, this age-dust degeneracy may be broken by adding NIR data to the SEDs, and we may be able to improve our estimation of ages and dust content. To test this, by using the sample by \citet{stark07} including objects with spec-z and phot-z, for which both $J$ and $K_{s}$ data are available, we compare results derived by using the $J$ and $K_{s}$ data and those without using the data. The details are presented in Appendix \ref{appendix:NIR}. The results show that differences with and without NIR data are not large, although error bars are large. The stellar masses with and without the $J$ and $K_{s}$ data agree with each other within a factor of $\sim3$. However, the errors in the $J$ and $K_{s}$ data used in the test are generally larger than those in other bands and the weights of the NIR data to the SED fitting are relatively small. Thus this may cause the small differences in the test. Therefore, sufficiently deep NIR data are desirable to better constrain age and color excess.

The distribution of the best-fitted SFRs is presented in Figure \ref{fig:params_distribution}d. The typical error of the SFR for each object is $\sim$0.5 dex. The median value of the SFRs estimated for the total LBG sample is $141M_{\sun}\textrm{yr}^{-1}$. The median values of the SFRs estimated for the GOODS-N and the GOODS-FF sample are $104M_{\sun}\textrm{yr}^{-1}$ and $191M_{\sun}\textrm{yr}^{-1}$, respectively. The median SFRs of the Category 1, 2, and 3 objects are $170M_{\sun}\textrm{yr}^{-1}$, $111M_{\sun}\textrm{yr}^{-1}$, and $1023M_{\sun}\textrm{yr}^{-1}$, respectively. Figure \ref{fig:params_distribution}d shows the existence of galaxies which show high SFRs. This is because of the existence of a large amount of dust as presented above. The apparent SFR derived from $L_{1500\textrm{\AA}}^{*}$ at $z\sim 5$ derived by \citet{iwata07} using \citet{madau96} relation is $\sim 20M_{\sun}\textrm{yr}^{-1}$. If we use the median value of color excess of the galaxies of 0.22 mag, the extinction corrected SFR is $\sim 160M_{\sun}\textrm{yr}^{-1}$ using \citet{calzetti00} extinction law, which is consistent with the value derived from SED fitting.

%
%

\section{Discussion}
\label{sec:discussions}

In \S \ref{sec:results}, we showed the stellar properties of LBGs at $z\sim5$ derived from SED fitting. In this section, we compare the results in this work with previous studies for galaxies at other redshifts. We also construct a stellar mass function of LBGs at $z\sim5$ and derive the stellar mass density at $z\sim5$.

\begin{figure*}
\begin{center}
\epsscale{0.90}
\plotone{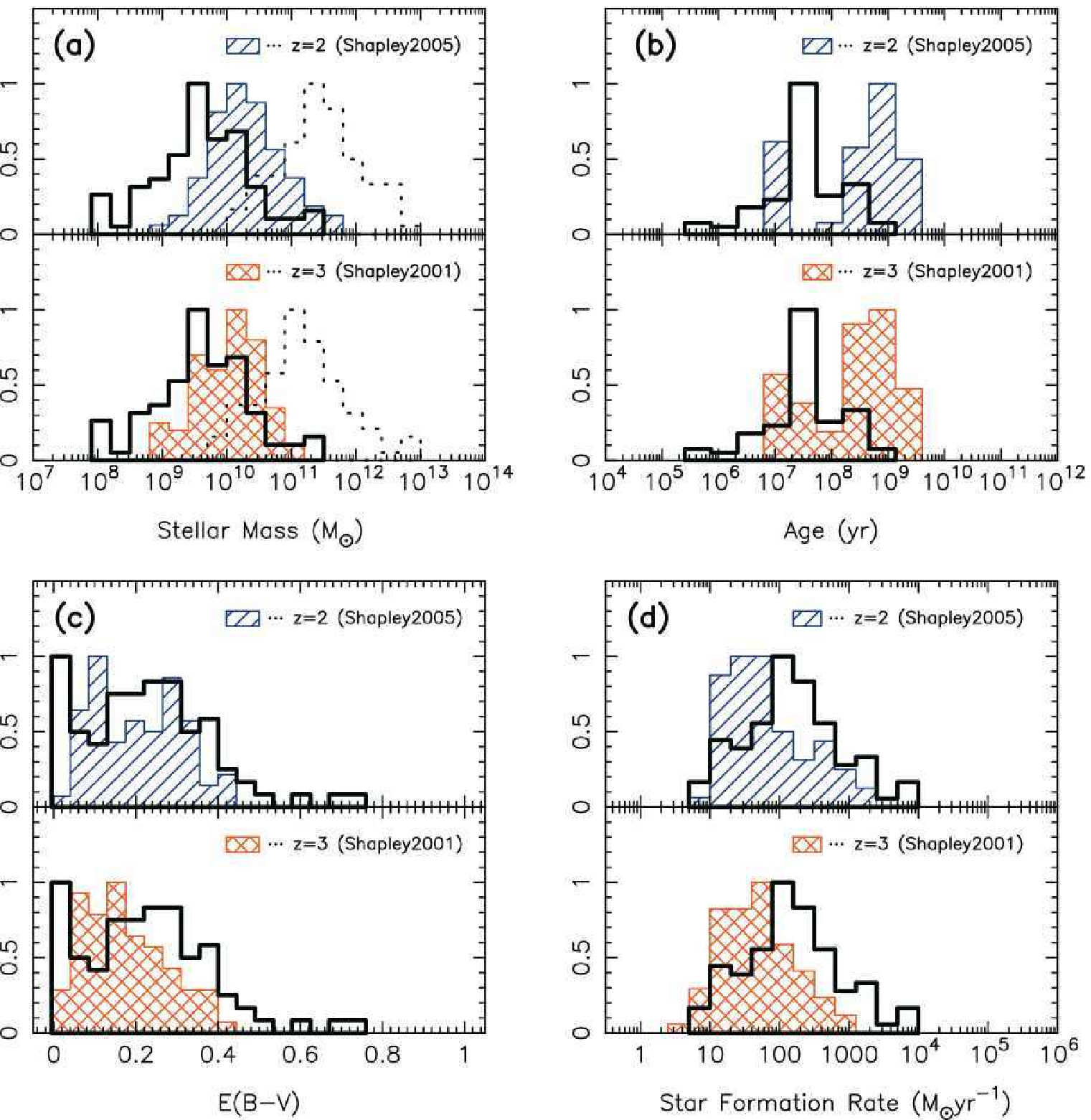}
\caption{Distributions of the best-fitted parameters of our $z\sim5$ sample (\textit{thick line}) and those of $z=2$ sample (\textit{hatched}) from \citet{shapley05} and $z=3$ sample (\textit{cross-hatched}) from \citet{shapley01}. The distributions of the best-fitted stellar mass, age, color excess, and star formation rate are plotted in panel (a), (b), (c), and (d), respectively. In panel (a), dotted lines of top and bottom sub-panels indicate distributions of expected stellar masses at $z=2$ and $z=3$, respectively, if our sample galaxies at $z\sim5$ keep the SFRs derived from the SED fitting. For comparison, the peaks of all distributions are normalized to unity.
\label{fig:params_distribution_z23}}
\end{center}
\end{figure*}

%
%
\subsection{Comparisons with LBGs at $z=2-3$}
\label{sec:discussions_z23}
First we compare the properties of LBGs at $z\sim5$ with those at $z=2-3$ (cosmic age of 3.2 Gyr$-$2.1 Gyr). Here we use the terminology of LBGs including BM/BX \citep{steidel04}. The distributions of the output parameters for our sample are compared with those of $z=2-3$ samples in Figure \ref{fig:params_distribution_z23}, where the histograms are normalized so that its peak value equals unity for comparison. For sample LBGs at $z=2$ and 3, we use \citet{shapley05} and \citet{shapley01}, respectively. All three samples are fitted using models by \citet{bruzual03} with the Salpeter IMF, constant star formation history, and the \citet{calzetti00} extinction law.

\begin{figure*}
\begin{center}
\epsscale{0.90}
\plotone{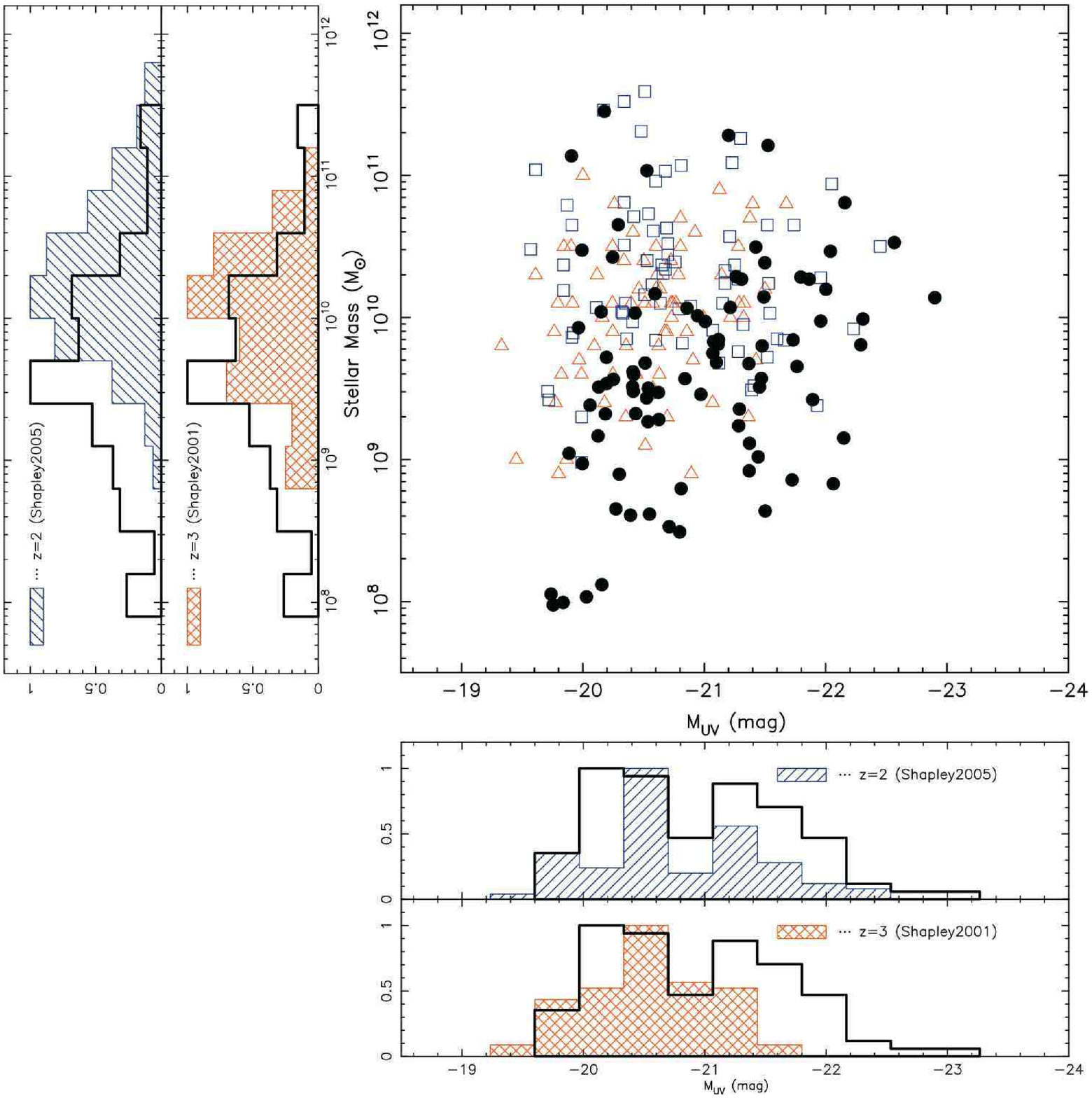}
\caption{\textit{Upper right}: Rest-frame UV (1500$\textrm{\AA}$) absolute magnitudes (uncorrected for dust extinction) vs. stellar masses for $z\sim5$ sample (\textit{filled circles}), $z=3$ sample (\textit{open triangles}) by \citet{shapley01}, and $z=2$ sample (\textit{open squares}) by \citet{shapley05}. \textit{Lower right}: Distributions of the rest-frame UV absolute magnitudes of the $z\sim5$ sample (\textit{thick line}), $z=3$ sample (\textit{cross-hatched}), and $z=2$ sample (\textit{hatched}). \textit{Upper left}: Distributions of the stellar masses of the $z\sim5$ sample (\textit{thick line}), $z=3$ sample (\textit{cross-hatched}), and $z=2$ sample (\textit{hatched}).
\label{fig:muv_distribution_z23}}
\end{center}
\end{figure*}

\begin{figure*}
\begin{center}
\epsscale{0.90}
\plotone{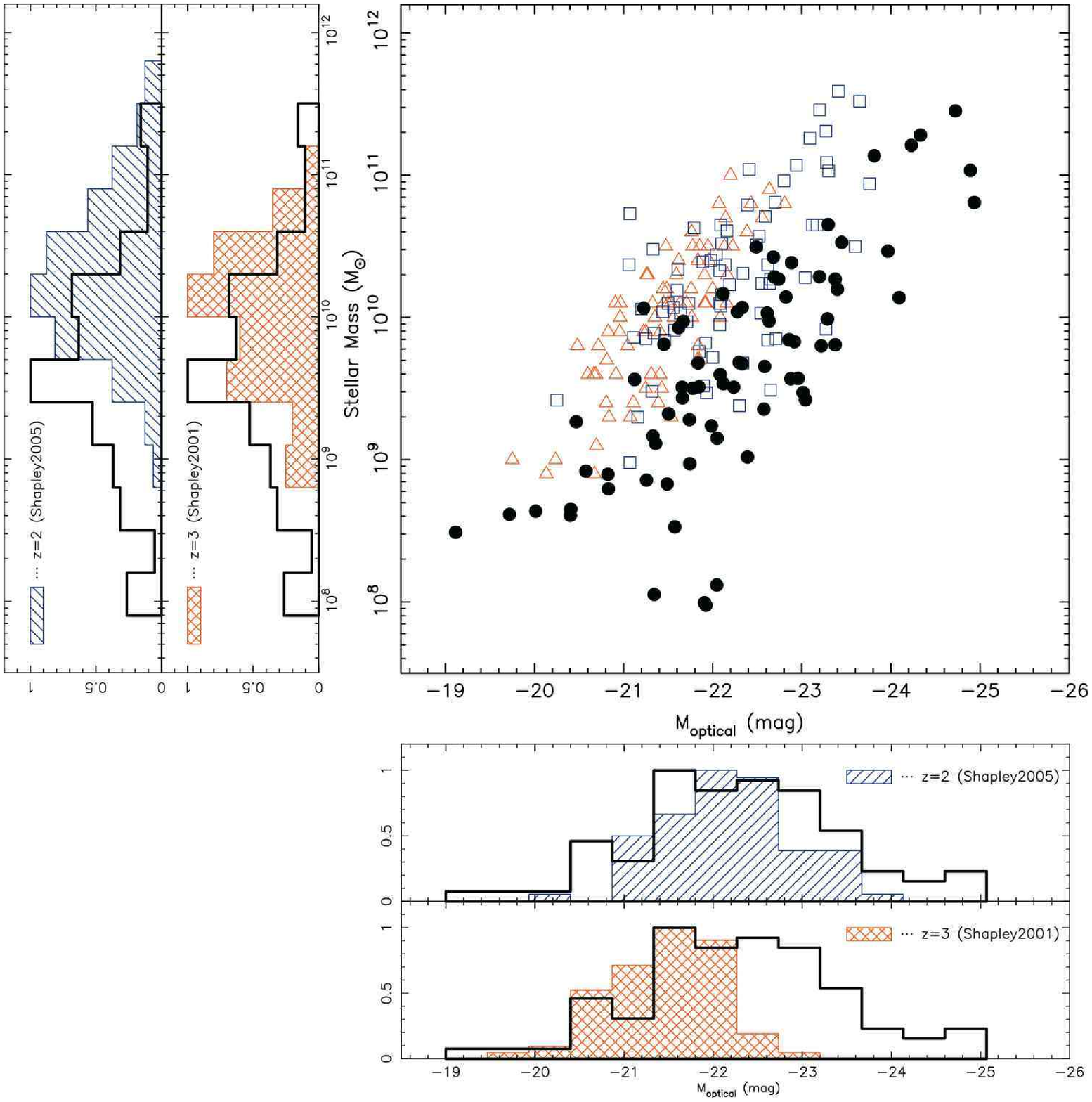}
\caption{\textit{Upper right}: Rest-frame optical (5500$\textrm{\AA}$) absolute magnitudes (uncorrected for dust extinction) vs. stellar masses for $z\sim5$ sample (\textit{filled circles}), $z=3$ sample (\textit{open triangles}) by \citet{shapley01}, and $z=2$ sample (\textit{open squares}) by \citet{shapley05}. \textit{Lower right}: Distributions of the rest-frame optical absolute magnitudes of the $z\sim5$ sample (\textit{thick line}), $z=3$ sample (\textit{cross-hatched}), and $z=2$ sample (\textit{hatched}). \textit{Upper left}: Distributions of the stellar masses of the $z\sim5$ sample (\textit{thick line}), $z=3$ sample (\textit{cross-hatched}), and $z=2$ sample (\textit{hatched}).
\label{fig:mopt_distribution_z23}}
\end{center}
\end{figure*}

In order to compare the samples fairly, their faintest UV luminosities at 1500$\textrm{\AA}$ are on an equal footing with those of ours. The distributions of the rest-frame UV absolute magnitudes of our $z\sim5$ sample, the $z=3$ sample by \citet{shapley01}, and the $z=2$ sample by \citet{shapley05} are presented in Figure \ref{fig:muv_distribution_z23}. While the UV absolute magnitudes of the $z=3$ and the $z=2$ samples lie on the range of $-19.3$ mag to $-21.7$ mag and $-19.6$ mag to $-22.5$ mag, respectively, that of our sample ranges from $-19.7$ mag to $-23.0$ mag. We use the sample galaxies whose UV absolute magnitudes are brighter than $-19.7$ mag. Figure \ref{fig:mopt_distribution_z23} shows the distribution of the rest-frame optical (5500$\textrm{\AA}$) absolute magnitudes of the $z\sim5$ sample, $z=3$ sample by \citet{shapley01}, and $z=2$ sample by \citet{shapley05}. While the faintest magnitudes of these samples are almost the same (from $-19$ to $-20$ mag), the brightest magnitudes are somewhat different. The optical absolute magnitudes of the $z=5$ sample range from $-19.1$ mag to $-24.9$ mag, while those of the $z=3$ and $z=2$ samples range from $-19.7$ mag to $-22.8$ mag and from $-20.3$ mag to $-23.8$ mag, respectively. There are clear deficits in the bright parts of both $M_{UV}$ and $M_{optical}$ distributions of the $z=3$ sample compared to the $z\sim5$ sample. This is probably due to a smaller survey volume for the $z=3$ sample than that for the $z\sim5$ sample. The differences may also be attributed to cosmic variance.

In the upper right panel of Figure \ref{fig:muv_distribution_z23}, the stellar masses against the rest-frame UV absolute magnitudes of the $z=2$, $z=3$ samples and our $z\sim5$ sample are plotted. The relation between the stellar mass and the UV absolute magnitude varies from $z\sim5$ to $z=2-3$ toward large masses at a fixed UV absolute magnitude, although the correlation is not strong (see details in \S \ref{sec:results_mass}). In the upper right panel of Figure \ref{fig:mopt_distribution_z23}, the stellar masses against the rest-frame optical absolute magnitudes of the $z=2$, $z=3$ samples and our $z\sim5$ sample are plotted. The correlation between the stellar mass and the optical absolute magnitude also varies toward large masses at a fixed UV absolute magnitude from $z\sim5$ to $z=2-3$.

In Figure \ref{fig:params_distribution_z23}a, the distribution of stellar masses for LBGs at $z\sim5$ superposed on those at $z=2$ and 3 is presented (The comparison of the distribution of stellar masses is also presented in the upper left panels of Figure \ref{fig:muv_distribution_z23} and Figure \ref{fig:mopt_distribution_z23}). Figure \ref{fig:params_distribution_z23}a shows that the stellar masses of LBGs at $z\sim5$ are smaller than those at $z=2-3$ on average. While the median of the stellar masses of the $z\sim5$ LBGs is $4.1\times10^{9} M_{\sun}$, the medians of the masses of the $z=2$ and 3 LBGs are $1.7 \times10^{10} M_{\sun}$ and $1.3\times10^{10} M_{\sun}$, respectively. Therefore, the median stellar mass of $z\sim5$ LBGs are smaller by a factor of $3-4$ than that of $z=2-3$ LBGs. Note that while \citet{shapley01} and \citet{shapley05} used solar metallicity models ($1.0Z_{\sun}$), we use sub-solar metallicity models ($0.2Z_{\sun}$). If we assume the metallicity of $1.0Z_{\sun}$, the stellar mass decreases by a factor of 1.2, and the difference between the distributions is even more significant. \citet{verma07} also found that the typical stellar mass of $z\sim5$ LBGs is $5-10$ times lower than the $z=3$ LBGs. The stellar masses of these $z\sim5$ LBGs are almost comparable to those of much fainter (sub-$L_{UV}^{*}$) LBGs at $z\sim2$ (\citealt{sawicki06b}; Sawicki, in prep.).

As a whole, star formation ages of $z\sim5$ LBGs are younger than those of $z=2-3$ LBGs (Fig. \ref{fig:params_distribution_z23}b). While the median age of our sample is 25 Myr, the median ages of $z=2$ and $z=3$ LBGs are $\sim600$ Myr and $\sim300$ Myr, respectively. The bimodal distribution in $z=2-3$ samples is not seen in the $z\sim5$ sample. \citet{verma07} also found that $z\sim5$ LBGs are significantly younger than the $z=3$ LBGs. However, typical ages are  $\la10$ Myr, which are about one order of magnitude smaller than our results. The cause of this difference is not clear. The distribution of color excess of the $z\sim5$ LBGs seems to suggest that the amount of dust extinction in $z\sim5$ LBGs may be slightly larger than that of the $z\sim3$ LBGs and similar to that of the $z\sim2$ LBGs (Figure \ref{fig:params_distribution_z23}c). The median color excess of our sample is 0.22 mag, and the median color excesses for $z=2$ and $3$ LBGs are 0.20 mag and 0.16 mag, respectively. \citet{verma07} also found that $z\sim$5 LBGs have a typical color excess of $\sim0.2$ mag. The SFR is higher than those of $z=2-3$ LBGs (Fig. \ref{fig:params_distribution_z23}d). While the median SFR of $z\sim5$ LBGs is 141$M_{\sun}\textrm{yr}^{-1}$, the medians of SFRs are 52$M_{\sun}\textrm{yr}^{-1}$ and 43$M_{\sun}\textrm{yr}^{-1}$ for $z=2$ and $z=3$ samples, respectively. Specific SFR is also larger in $z\sim5$ LBGs than that in $z=2-3$ LBGs. \citet{verma07} also found that the typical SFR ($\sim500$ $M_{\odot}\textrm{yr}^{-1}$) of $z\sim5$ LBGs is $\sim10$ times higher than the $z=3$ sample. From these comparisons, we suggest that galaxies at $z\sim5$ are forming stars very actively, and in consequence, they are dusty and we may see the early phase of these activities; we may witness the evolution of stellar populations of galaxies from $z=5$ to $z=2$.

In the bottom sub-panel of Figure \ref{fig:params_distribution_z23}a, we show distribution of stellar masses of $z=3$ LBGs assuming that each galaxy of our sample keeps the SFR derived from the SED fitting until $z=3$. The distribution shifts toward larger mass and the median value of the distribution is $1.3\times10^{11}M_{\sun}$. Likewise, we plot the expected distribution of stellar masses of $z=2$ LBGs in the top sub-panel of Figure \ref{fig:params_distribution_z23}a. Again, the distribution shifts toward larger mass than observed at $z=2$, and the median value of the distribution is $2.9\times10^{11}M_{\sun}$. Thus, these suggest that the SFR decreases from $z\sim5$ to $3-2$, provided that $z=2-3$ LBGs are direct descendant of the $z\sim5$ LBGs. Alternatively, the descendant of the $z\sim5$ LBGs may be massive objects at $z=2-3$ such as DRGs or sBzKs.

%
%
\subsection{Comparisons with LBGs at $z=5-6$}
\label{sec:discussions_z56}
We compare the properties of our sample galaxies with those at $z=5-6$. For the $z=5$ samples, we use samples by \citet{stark07} and \citet{verma07}. The sample by \citet{stark07} consists of 14 spectroscopically confirmed objects at $z\sim5$ and the sample by \citet{verma07} consists of 21 $V$-dropouts, 6 of which are confirmed to be at $z\sim5$ by spectroscopy. For the $z=6$ samples, we use samples by \citet{yan06} and \citet{eyles07}. The sample by \citet{yan06} and that by \citet{eyles07} consist of 53 and 9 $i'$-dropouts, respectively.

For fair comparison, we checked the limiting magnitudes for these samples both in the rest-frame UV and in the rest-frame optical band. The rest-frame UV and optical magnitude ranges of the other studies are almost the same as ours. The faintest limits of the absolute magnitude in the rest-frame UV and the rest-frame optical are $M_{1500\textrm{\AA}}\sim-20$ mag and $M_{5500\textrm{\AA}}\sim-20$ mag, respectively.

\begin{figure*}
\begin{center}
\epsscale{0.90}
\plotone{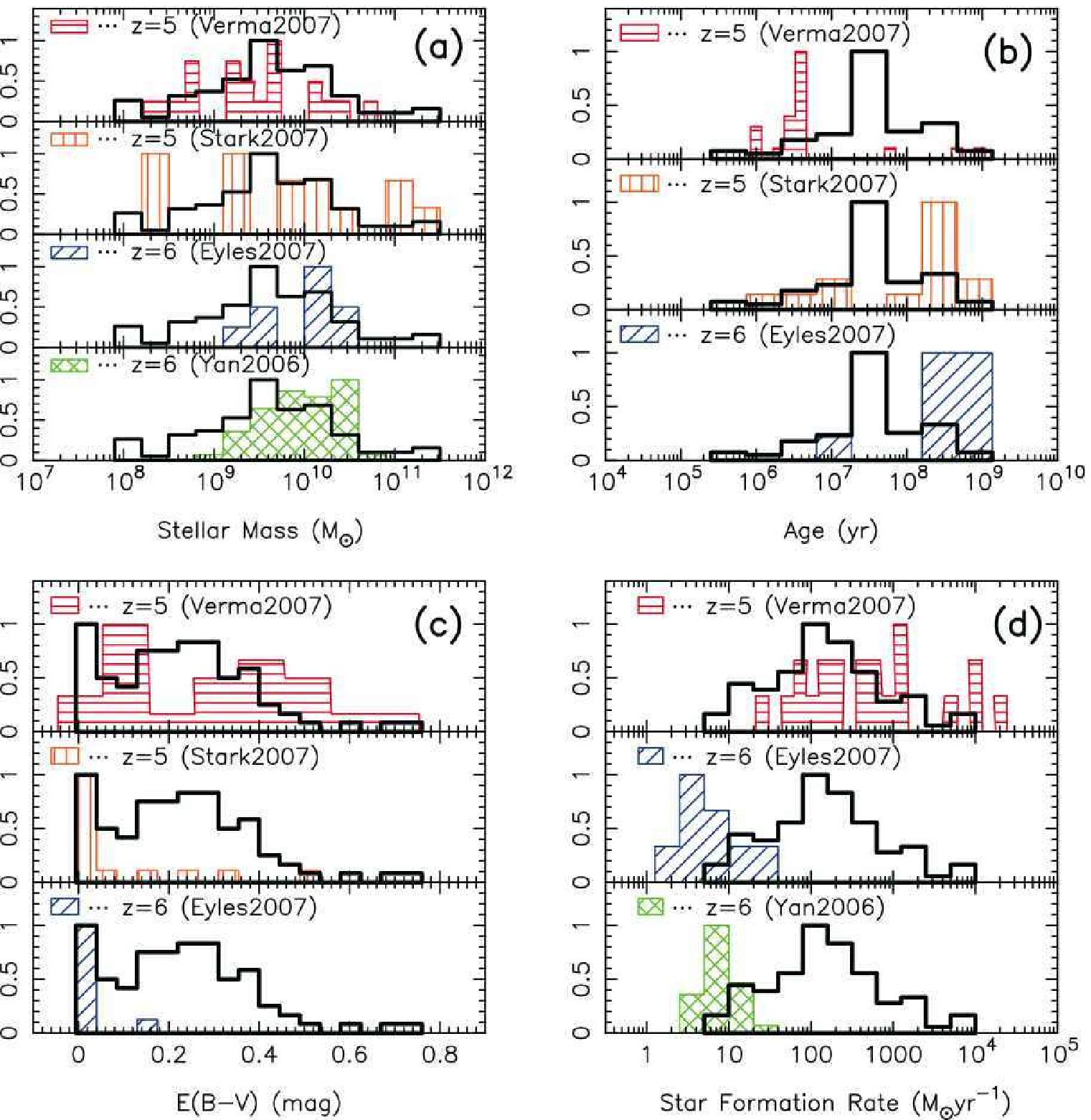}
\caption{Distributions of the best-fitted parameters of our $z\sim5$ sample and those of $z=5-6$ samples from the literature (\citet{stark07} and \citet{verma07} for $z\sim5$, and \citet{yan06} and \citet{eyles07} for $z=6$). The distributions of the stellar mass, age, color excess, and star formation rate are plotted in panel (a), (b), (c), and (d), respectively. For comparison, the peaks of all distributions are normalized to unity.
\label{fig:params_distribution_z56}}
\end{center}
\end{figure*}

In Figure \ref{fig:params_distribution_z56}, we plot the distributions of derived parameters for the various samples, where the peaks of the histograms are normalized to be unity for comparison. Note that because the age and the color excess are not shown explicitly in the paper by \citet{yan06}, we do not plot them in the figure.

As illustrated in Figure \ref{fig:params_distribution_z56}a, the range of the stellar masses for our sample agrees with other studies. The stellar masses are widely distributed from $10^{8}M_{\sun}$ to $10^{11}M_{\sun}$. While the median stellar mass of our sample is $4.1\times10^{9}M_{\odot}$, those of the samples of \citet{stark07}, \citet{verma07}, \citet{eyles07}, and \citet{yan06} are $7.9\times10^{9}M_{\odot}$, $2.0\times10^{9}M_{\odot}$, $1.6\times10^{10}M_{\odot}$, and $9.6\times10^{9}M_{\odot}$, respectively. It seems to be somewhat strange that the representative stellar mass for the $z=5$ LBGs is less massive than that for the $z=6$ LBGs. It is also noteworthy that the models used in the SED fitting for the $z=6$ sample are slightly different from those we used; we will discuss the effects of differing models below.

Figure \ref{fig:params_distribution_z56}b shows the comparisons of the distributions of the star formation ages for our sample and other studies. The resulting age distribution of the other studies tend to be younger or older than our sample: While the median value of the age of our sample is 25 Myr, and the median ages for the $z=5$ sample by \citet{stark07} and for the $z=6$ sample by \citet{eyles07} are 255 Myr and 400 Myr, respectively. On the other hand, the ages for the $z=5$ sample by \citet{verma07} are typically younger than 10 Myr.
In Figure \ref{fig:params_distribution_z56}c, we plot the distribution of color excesses for our sample and the other studies. The derived color excesses for $z=5$ \citep{stark07} and $z=6$ samples \citep{eyles07}, most of which are close to zero (the median values are 0.01 mag and 0.00 mag for \citet{stark07} and \citet{eyles07}, respectively), are much smaller than our result of the median value of 0.22 mag. In contrast, the resulting color excesses for our sample are almost comparable or slightly smaller than those by \citet{verma07}.

In Figure \ref{fig:params_distribution_z56}d, we plot the distribution of SFRs for our sample and the other studies. The SFRs for our sample are much higher than those for $z=6$ LBGs \citep{yan06,eyles07} but slightly lower than those of the $z=5$ sample by \citet{verma07}. The median values of SFRs for our sample and the sample by \citet{verma07} are $141M_{\sun}\textrm{yr}^{-1}$ and $400M_{\sun}\textrm{yr}^{-1}$, respectively. On the other hand, the median SFR of the $z=6$ samples is $\la10M_{\sun}\textrm{yr}^{-1}$. This difference between $z=5$ and $z=6$ samples may be due to the difference of color excesses. Since extinction in $z=6$ sample is negligible, the extinction corrected SFR is low. Meanwhile, for the $z=5$ samples, the moderate amount of dust extinction makes the intrinsic SFR higher.

\begin{figure*}
\begin{center}
\epsscale{0.90}
\plotone{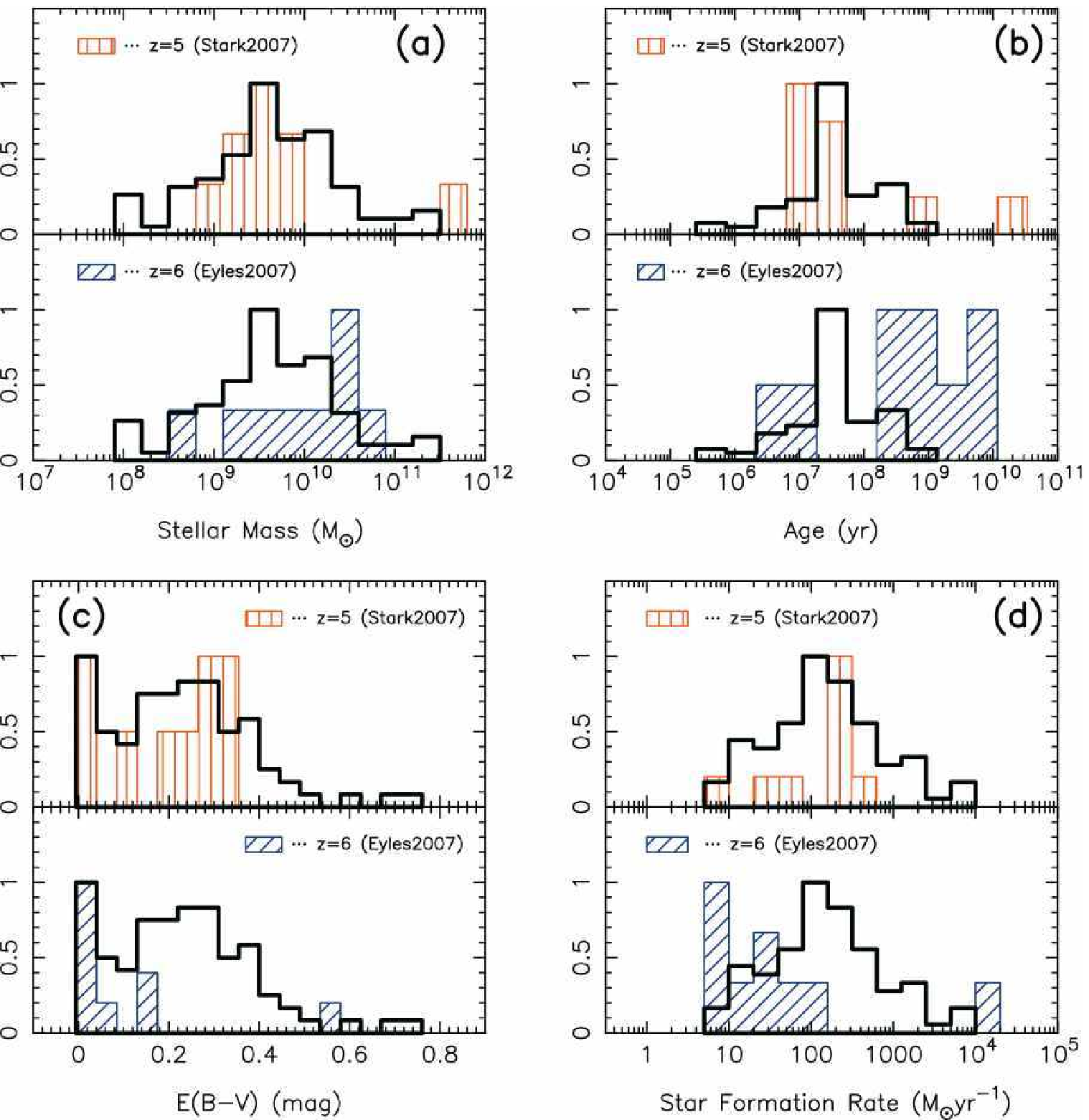}
\caption{Distributions of the best-fitted parameters of our $z\sim5$ sample and those of $z=5-6$ samples from literature (\citet{stark07} for $z=5$ and \citet{eyles07} for $z=6$). The best-fitted parameters of samples by \citet{stark07} and \citet{eyles07} are derived by fitting with the same SED models as we used in this paper. The peaks of all distributions are normalized to unity. 
\label{fig:params_distribution_z56_comparison}}
\end{center}
\end{figure*}

It is worth noting that model ingredients in the SED fitting of these studies for deriving the parameters are different. Thus, it is necessary to take into account these effects when we compare our results with other studies. \citet{stark07} and \citet{eyles07} used the BC03 with the Salpeter IMF and the extinction law by \citet{calzetti00}, but they assumed solar metallicity (1.0$Z_{\odot}$) and the various star formation histories. We refit the observed SEDs of the $z=5$ objects by \citet{stark07}, including 14 objects with spec-z and 59 objects with phot-z, and $z=6$ objects by \citet{eyles07} with the same models as we used for our sample: BC03, Salpeter IMF, $0.2Z_{\sun}$, constant star formation, and the extinction law by \citet{calzetti00}. The resulting distributions are presented in Figure \ref{fig:params_distribution_z56_comparison}. The refitted parameters for the $z=5$ sample by \citet{stark07} are somewhat different from the original results, while the refitted results for the $z=6$ sample by \citet{eyles07} are almost the same as the original results. The median stellar masses of the the $z=5$ sample and the $z=6$ sample are $2.3\times10^{9}M_{\odot}$ and $1.8\times10^{10}M_{\odot}$, respectively. They decrease by a factor of 3.4 and increase by a factor of 1.1 from original results, respectively. The star formation ages also vary from original results. The refitted ages for the $z=5$ sample decreases from the original result by a factor of $\sim$10 (the median value of 25 Myr) and is comparable to our result. On the other hand, the age for the $z=6$ sample increases by a factor of 1.7 from the original result (the median value of 700 Myr). Note that when the constant star formation history is assumed, ages of some objects in the $z=6$ sample exceed the cosmic age at $z=6$ ($\sim0.9$ Gyr). While the color excesses of both the $z=5$ and $z=6$ samples are $\sim0$ mag, the refitted color excess of the $z=5$ sample is 0.17 mag, which is comparable to our results, and that of the $z=6$ sample is 0.0 mag. The median values of the refitted SFR are $56 M_{\sun}\textrm{yr}^{-1}$ and $20 M_{\sun}\textrm{yr}^{-1}$ for the $z=5$ and $z=6$ sample, respectively, and are lower than our result.

Even though the model ingredients in the SED fitting are the same, the resulting stellar mass, age, color excess, and star formation rate of the $z=6$ samples are different from those of our $z=5$ sample. This may imply that there is a significant evolution of stellar population in galaxies from $z=6$ to $z=5$. Although the time interval between $z=6$ and $z=4.8$ is just $\sim0.3$ Gyr, the galaxies may evolve drastically after the end of reionization epoch at $z\sim6$ \citep{fan06, totani06}. However, the small sample size of the previous studies prevents us from concluding that the differences of the resulting parameters in the SED fitting between $z=5$ and $z=6$ samples are significant.

%
%
\subsection{The Stellar Mass Function of LBGs at $z\sim5$}
\label{sec:discussions_mass_function}

\begin{figure}
\begin{center}
\epsscale{1.00}
\plotone{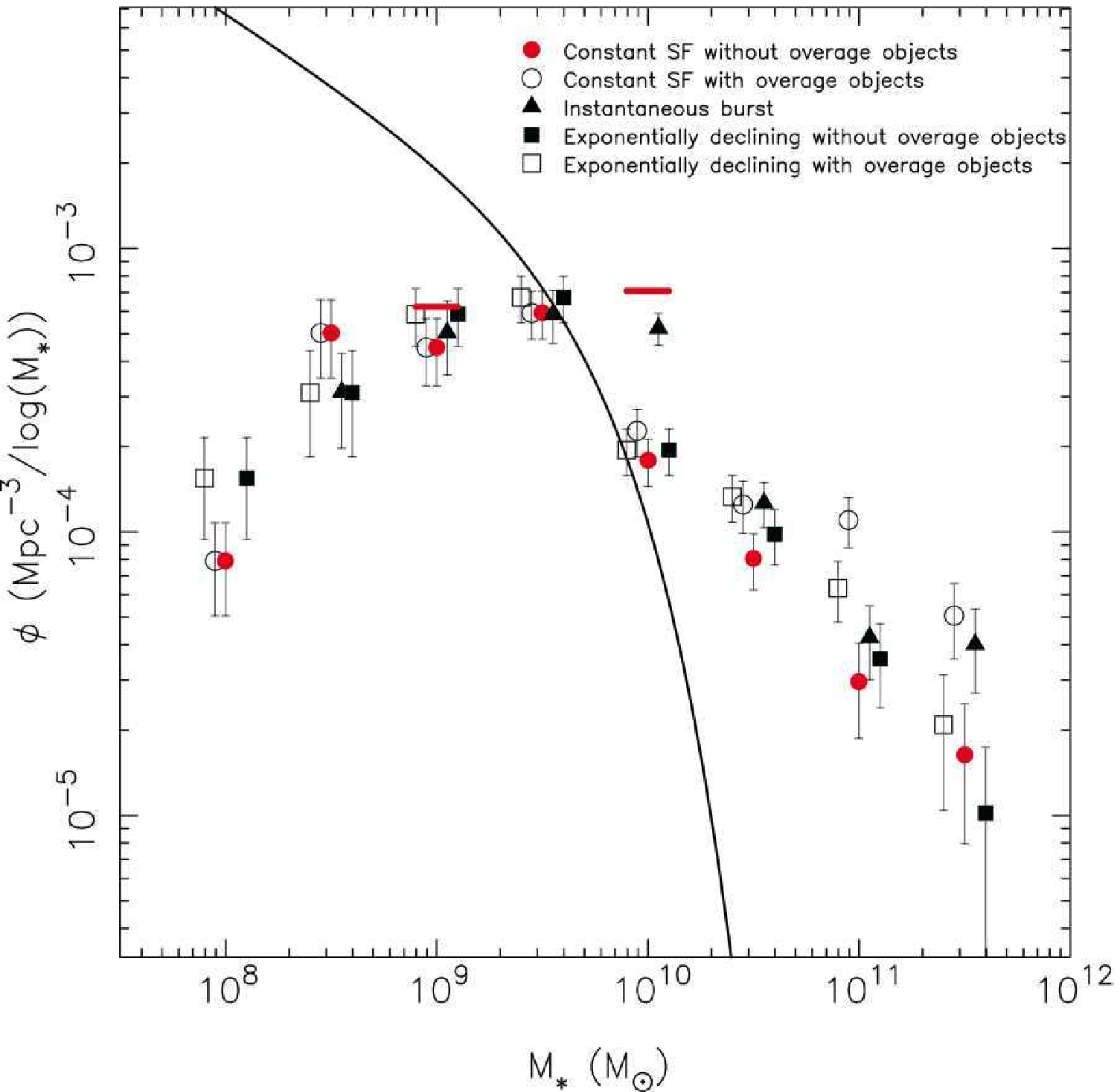}
\caption{Stellar mass functions of LBGs at $z\sim5$ (totaling categories 1-3 objects). The stellar mass functions derived by assuming constant star formation (\textit{circles}), instantaneous burst (\textit{triangles}), and exponentially declining (\textit{squares}) models are plotted. The stellar mass functions excluding objects whose ages are fitted to be older than the cosmic age at $z=5$ are indicated by filled symbols and the functions including these overage objects are indicated by open symbols. For clarity, the data points (\textit{filled-squares}, \textit{filled-triangles}, \textit{open-circles}, and \textit{open-squares}) are shifted horizontally by $+0.1$, $+0.05$, $-0.05$, and $-0.1$ dex, respectively. The expected number densities of the sample including the Category 4 objects are indicated by horizontal bars (see text for the detailed derivation). The solid curve represents the stellar mass function derived by assuming UV luminosity function of $z\sim5$ LBGs and UV luminosity and stellar mass relation (see section \ref{sec:discussions_mass_function}).
\label{fig:mass_function_ours}}
\end{center}
\end{figure}

The large sample of LBGs whose stellar masses are estimated robustly allows us to derive the stellar mass function of LBGs at $z\sim5$. Our sample is originally selected in the optical band, and is selected as uncontaminated objects in IRAC images by eye inspection. The sample is also affected by incompleteness in IRAC images. Therefore, we estimate the number density per log($M_{*}/M_{\odot}$) as follows:
\begin{equation}
\phi(log(M_{*}/M_{\odot}))=\sum_{i,j}\frac{N_{i,j}(1-f_{i}^{int})}{f^{sel}f_{j}^{det}V_{i}^{eff}\Delta log(M_{*}/M_{\odot})}\ ,
\end{equation}
where $i$ and $j$ are bin numbers for $z'$-band and $4.5\mu$m-band magnitudes, respectively, and $N_{i,j}$ is the number of objects entering each $z'$-band bin and $4.5\mu$m-band bin in the $log(M_{*}/M_{\odot})$ bin. $f_{i}^{int}$ refers to a fraction of interlopers estimated in $i$-th $z'$-band bin by \citet{iwata07}. $f^{sel}$ is a fraction of uncontaminated objects in IRAC images. As we discussed in \S \ref{sec:data_sample}, the percentage is independent from the $z'$-band magnitude. $f_{j}^{det}$ is a detection rate in the IRAC $4.5\mu$m-band. For the Category 3 objects, which are undetected in $4.5\mu$m-band, we use a detection rate in $3.6\mu$m-band instead of that in $4.5\mu$m-band. $V_{i}^{eff}$ is an effective volume as a function of $z'$-band magnitude taken from \citet{iwata07}.

We found that 20 out of 105 objects are best-fitted with models which have larger age than the cosmic age at $z\sim5$. It is important to see the contribution from these objects to the stellar mass function, because these objects have generally large stellar masses, and their inclusion would affect the massive end of the stellar mass function. In Figure \ref{fig:mass_function_ours}, we plot the stellar mass function by gathering the Category 1, 2, and 3 objects. The stellar mass function excluding the overage objects is indicated by filled circles, and that including the objects is indicated by open circles. The error bars of the number densities are Poisson errors\footnote{We estimated the error arising from the uncertainty of the stellar mass derived in the SED fitting by the re-sampling method. The stellar mass function is re-calculated by using the stellar masses that perturbed randomly by Gaussian distribution with $1\sigma$ error of the SED fitting. Repeating this process 10000 times, we estimate $1\sigma$ error of the number density in each mass bin. The average error is 0.17 dex, which is 1.5 times larger than the average Poisson error, but is generally smaller than the uncertainties from the choice of the star formation history.}. While the number densities including the overage objects are higher than those without using the overage objects by a factor of $3-4$ in the massive part ($\sim10^{11}M_{\odot}-10^{11.5}M_{\odot}$), the effect of the overage objects is small in less massive part ($<10^{11}M_{\odot}$). If we restrict the maximum age of the models to 1.2 Gyr in the SED fitting, the main results, especially of stellar mass, do not change so much. In this case, the stellar masses are affected only in the massive part. In the most massive bin of the stellar mass function, the number density decrease by a factor of $\sim3$. In the other mass bins, the changes are within a factor of 1.4.

\begin{figure}
\begin{center}
\epsscale{1.00}
\plotone{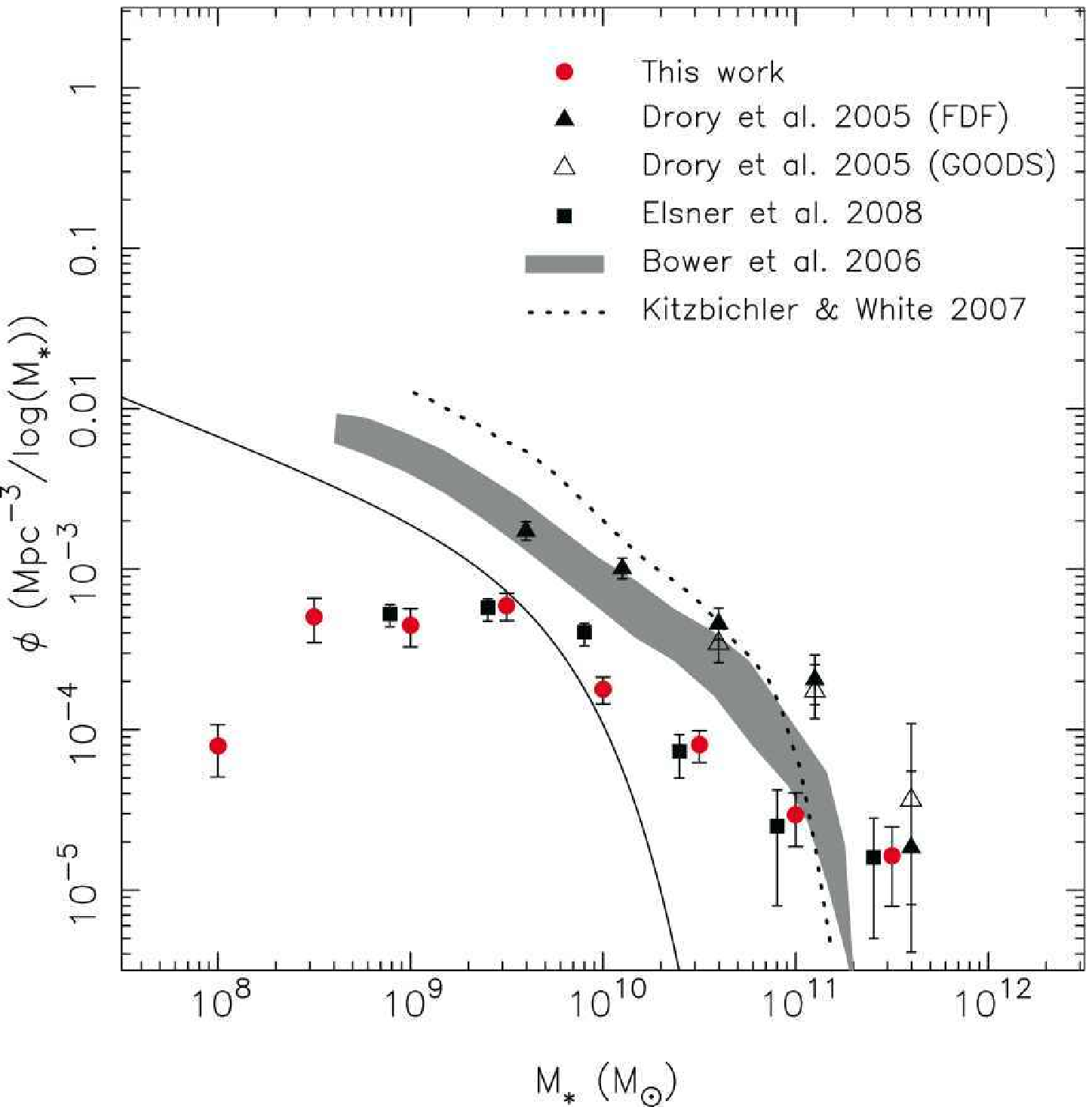}
\caption{Stellar mass function of our LBG sample at $z\sim5$ (\textit{circles}) and that of a sample by \citet{elsner08} (\textit{squares}), that of an FDF sample by \citet{drory05} (\textit{filled triangles}), and that of a GOODS sample by \citet{drory05} (\textit{open triangles}). For clarity, the data points by \citet{drory05} and \citet{elsner08} are shifted horizontally by $+0.1$ and $-0.1$ dex, respectively. The expected stellar mass function from the UVLF is indicated by a solid line (see text for the detailed derivation). The theoretical predictions of semi-analytical models by \citet{bower06} and \citet{kitzbichler07} are indicated by a shaded region and a dotted line, respectively. \label{fig:mass_function}}
\end{center}
\end{figure}

As we discuss in Appendix \ref{appendix:effects}, the most uncertain factor in deriving the stellar mass is the choice of the star formation history.  We examined the instantaneous burst, exponentially declining history, and two-component models as well as constant star formation history. We found that there is no systematic offset between stellar mass derived with different star formation histories, but there are scatters of $\sigma=0.6$ and $0.3$ dex in the case of instantaneous burst and exponentially declining models, respectively, and also we found that, in the case of two-component models, the stellar masses of some objects increase by $\sim1$ dex but those of the majorities are comparable to those in the case of constant star formation history. In Figure \ref{fig:mass_function_ours}, we plot the stellar mass functions by assuming the constant star formation, instantaneous burst, and exponentially declining histories. Although it seems that the choice of the star formation history does not affect the stellar mass function as a whole, in the most massive part, the number density adopting instantaneous burst is larger than that adopting other star formation histories by a factor of $2-3$. The stellar mass function in the least massive part ($M_{*}<10^9M_{\odot}$) decreases regardless of the adopting star formation history. Although we apply the completeness correction, this decrease in the less massive end is probably due to the limitation of our original sample of $z'<26.5$ mag as described below.

We discussed the stellar mass function for the sample galaxies which are detected in $3.6\mu$m and/or $4.5\mu$m. As we mentioned in \S \ref{sec:data_sample}, about 40\% of our sample of 170 objects are detected neither in $3.6\mu$m nor $4.5\mu$m. These Category 4 objects are thought to be less massive than IRAC detected objects (Category 1, 2, and 3). However, since the Category 4 objects are not detected both in $3.6\mu$m and $4.5\mu$m bands, we can not derive the stellar masses for these objects but can only constrain the upper limit of the stellar mass. If we assume the correlation between $4.5\mu$m magnitude and stellar mass in Figure \ref{fig:ch2_mass}, the upper limits ($3\sigma$) on the stellar mass for objects in the GOODS-N and GOODS-FF are $1.6\times10^{9}M_{\sun}$ and $8.9\times10^{9}M_{\sun}$, respectively. In Figure \ref{fig:mass_function_ours}, we show the expected number densities including the Category 4 objects by red horizontal bars. Here we assume that all of the Category 4 objects have these limiting stellar masses. We should emphasize that these Category 4 objects only affect the less massive part of the stellar mass function.

Our sample objects are selected with a criterion of $z'<26.5$ mag, thus objects with $z'>26.5$ mag are missed. We roughly estimate the contribution from the objects that are faint in $z'$-band. As we discussed in \S \ref{sec:results_mass}, the rest-frame UV absolute magnitudes are roughly correlated with the median stellar masses. By using this relationship and the UV luminosity function (UVLF) of LBGs at $z\sim5$ \citep{iwata07}, we estimate the stellar mass function. The resulting stellar mass function is plotted in Figure \ref{fig:mass_function_ours} as a solid curve. It disagrees with the observed stellar mass functions, especially in the massive part, because the scatter of the correlation is large and asymmetric; while there are some objects with faint UV magnitudes and large stellar masses, there is no object with the bright UV magnitude and small stellar mass as seen in Figure \ref{fig:muv_mass}. Nevertheless, this hints that the fainter LBGs ($z'>26.5$ mag) contribute to the less massive part ($M_{*}\la10^{9}M_{\odot}$) of the stellar mass function.

Here we adopt the stellar mass function derived with constant star formation models and with the objects whose age is younger than 1.2 Gyr as a fiducial stellar mass function. The fiducial stellar mass function is forced to be fitted by \citet{schechter76} function. We exclude mass bins of $\textrm{log}(M_{*}/M_{\odot})=8.0$ and $8.5$ from the fit because we probably largely underestimate the number densities at the mass bins as we mentioned above. The best fitted parameters\footnote{The uncertainties are 68\% confidence level.} are $\textrm{log}(M_{*}^{*}/M_{\odot})=13.81^{+0.98}_{-0.70}$, $\phi^{*}=0.60^{+1.49}_{-0.49}\times10^{-7} \textrm{Mpc}^{-3}/\textrm{log}(M_{*})$ , and $\alpha^{*}=-1.83^{+0.17}_{-0.18}$. The stellar mass function from our sample is fitted by only power-law component of the Schechter function in the mass range of $\textrm{log}(M_{*}/M_{\odot})=9.0$ to $11.5$.

The derived stellar mass function for our sample is compared with other observations. In Figure \ref{fig:mass_function}, the fiducial stellar mass function from our sample and the results from photo-$z$ selected samples by \citet{drory05} are plotted. The samples by \citet{drory05} are an $I$-selected sample ($I<26.8$ mag) in the FORS Deep Field (FDF) and a $K_{s}$-selected sample ($K_{s}<25.4$ mag) in the Great Observatories Origins Deep Survey-Sourth (GOODS-S). The stellar masses of the samples by \citet{drory05} are derived without IRAC data. As illustrated in Figure \ref{fig:mass_function}, the stellar mass function of our sample agrees with the results by \citet{drory05} in the most massive end. However, in most of the mass range, the number densities of our sample are significantly smaller than those from the sample by \citet{drory05}, even though we take into account the contribution of the Category 4 objects. Our stellar mass function is also compared with a result by \citet{elsner08}. They used the GOODS-MUSIC catalog and their sample contains objects detected in $z$-band or $K_{s}$-band. The $z$-band and $K_{s}$-band limiting magnitudes are 26.0 and 23.8 mag, respectively. The stellar mass function for our sample is in excellent agreement with that by \citet{elsner08}. As \citet{elsner08} claimed, the stellar masses tend to be overestimated systematically if the IRAC data are not included in the SED fitting, especially at $z\ge4$. The discrepancy between the stellar mass function of our sample and that by \citet{drory05} presumably attributes to this difference. Note that the redshift ranges of the resulting stellar mass functions of the sample by \citet{drory05} and \citet{elsner08} are from $z=4.0$ to 5.0 and the representative redshift ($z=4.5$) is slightly lower than that ($z=4.8$) of our sample. Also note that the discrepancy between the stellar mass function of our sample and that by \citet{drory05} might be due to cosmic variance.

The predictions of the theoretical models are also presented in Figure \ref{fig:mass_function}. We compare our fiducial stellar mass function with the predicted stellar mass functions of the model by \citet{bower06} based on the GALFORM \citep{cole00} and the model by \citet{kitzbichler07} based on \citet{croton06}, both of which are semi-analytic models implemented on the Millennium Simulation and include the feedback effect from AGNs in the galaxy evolution.  In the mass bin of $M_{*}\sim10^{11}M_{\odot}$, our stellar mass function agrees with the theoretical models. In the most massive part ($M_{*}\sim10^{11.5}M_{\odot}$), the number density of our sample is larger than the theoretical predictions. However, if the mass functions of the models are convolved with Gaussian function with a standard deviation of 0.3 dex, which is a typical error in the SED fitting, by considering measurement errors, the models are matched with our result. In the most of the mass range ($\textrm{log}(M_{*}/M_{\odot})\le11.0$), the number density of our sample is significantly lower than the models.

\begin{figure*}
\begin{center}
\epsscale{1.00}
\plotone{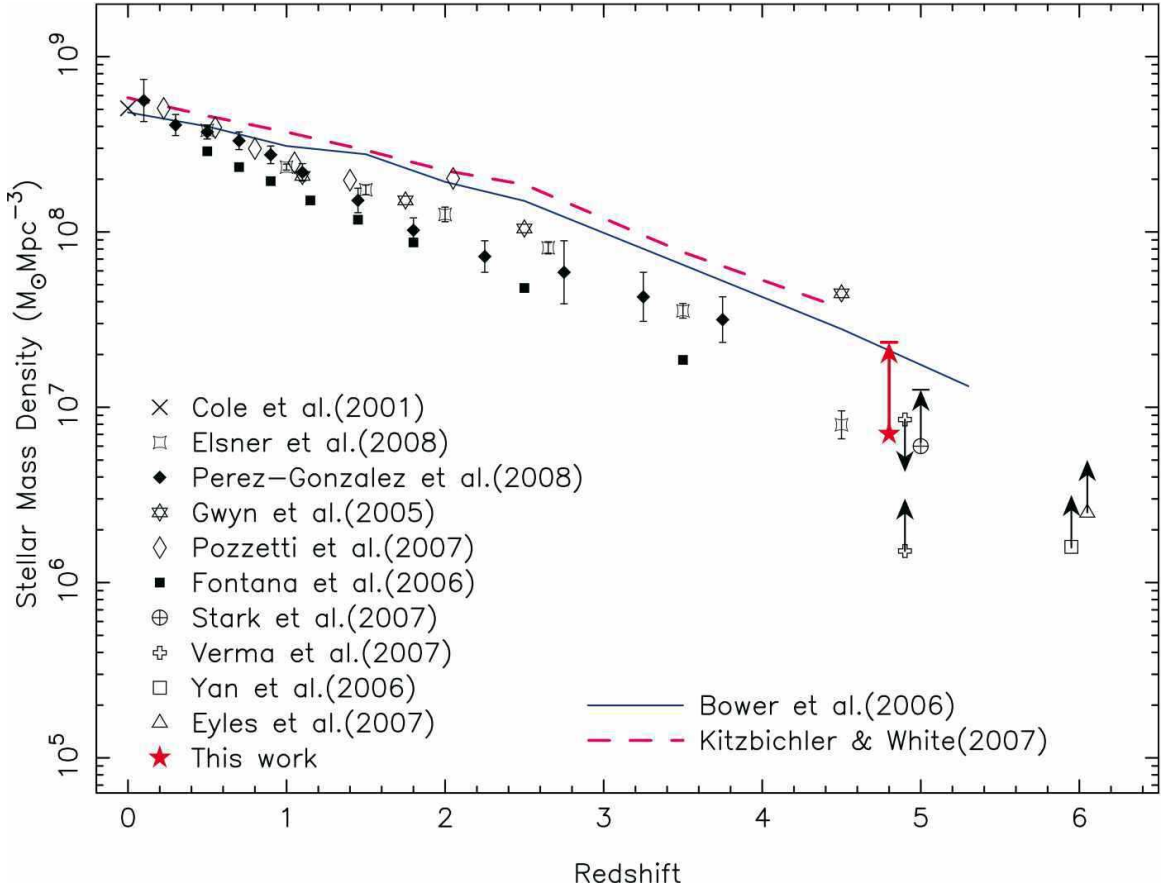}
\caption{Stellar mass density as a function of redshift. The stellar mass density from our sample is indicated by a filled star and the plausible upper limit is presented with a horizontal bar. Theoretical predictions from semi-analytical models are indicated by solid and dashed lines. The stellar mass densities of our sample, other observations at $z\le4.5$, and the theoretical models are calculated by integrating the stellar mass functions down to $10^{8}M_{\odot}$.\label{fig:mass_density}}
\end{center}
\end{figure*}

%
%

\subsection{The Stellar Mass Density at $z\sim5$}
\label{sec:discussions_massdensity}

By integrating the derived stellar mass function, we can calculate the stellar mass density at $z\sim5$. By integrating down to log($M_{*}/M_{\sun}$) = 8.0, we obtained the stellar mass density of 7.0$\times$$10^{6}$$M_{\sun}$$\textrm{Mpc}^{-3}$. As we discussed in \S \ref{sec:discussions_mass_function}, the choice of the assumed star formation history changes the shape of the stellar mass function, and hence the stellar mass density.
In Figure \ref{fig:mass_density}, we plot the stellar mass density calculated by integrating the fiducial stellar mass function (derived from the Category 1, 2, 3 objects with constant star formation history and excluding the overage objects). Assuming other star formation histories makes the mass density larger: For instance, if the instantaneous burst model is assumed, the stellar mass density is $1.4\times10^{7}M_{\sun}\textrm{Mpc}^{-3}$. Also, including the overage objects makes the density larger by a factor of 2.7 and 1.6 for the constant star formation model and the exponentially declining model, respectively. However, the effect of including the Category 4 objects is not considerable: It makes the mass density larger by $\sim20\%$. If we restrict the maximum age of the models to 1.2 Gyr, the integrated stellar mass density is $1.1\times10^{7}M_{\sun}\textrm{Mpc}^{-3}$, which lies between our fiducial value and the plausible upper limit.

The stellar mass density derived above is possibly still underestimated because we miss LBGs with $z'>26.5$ mag and galaxies which are not selected by LBG selection. In \S \ref{sec:discussions_mass_function}, we roughly estimated the contribution from LBGs with $z'>26.5$ mag to the stellar mass function. According to the rough correlation between UV magnitudes and stellar masses, the stellar masses of LBGs with $z'>26.5$ mag are mostly less than$\sim10^{9}M_{\odot}$. By integrating the stellar mass function estimated from the UVLF by \citet{iwata07} between $10^{8}M_{\odot}$ and $10^{9.5}M_{\odot}$, the contribution of the LBGs to the mass density is calculated to be $4.6\times10^{6}M_{\odot}\textrm{Mpc}^{-3}$.

We take the stellar mass density of $2.4\times10^{7}M_{\odot}\textrm{Mpc}^{-3}$ derived with constant star formation models and including the overage objects and the contribution from the $z'>26.5$ LBGs as a plausible upper limit. We show the upper limit indicated by a horizontal bar in Figure \ref{fig:mass_density}. The true stellar mass density at $z\sim5$ probably lies between the value for our sample and the bar in the figure \footnote{We estimated the uncertainty arising from the uncertainty of the stellar mass in the same way described in \S \ref{sec:discussions_mass_function}. The average error in the stellar mass density is $\sim0.1$ dex, which is negligible as compared with the uncertainty from the choice of star formation history.}.

We compare the derived stellar mass density from our sample to other observational studies. The stellar mass densities derived from the samples by \citet{gwyn05}, \citet{fontana06}, \citet{pozzetti07}, \citet{perez-gonzalez08}, and \citet{elsner08} are obtained by integrating their stellar mass functions from $10^{8}M_{\sun}$ to $10^{13}M_{\sun}$. We also plot the stellar mass densities at $z\ga5$ by \citet{yan06}, \citet{stark07}, \citet{eyles07}, and \citet{verma07}. These values are obtained by summing up all stellar masses (ranging from $\sim10^{8}M_{\odot}$ to $\sim10^{11}M_{\odot}$) in their observations. The result by \citet{cole01} is shown as the local value by integrating their stellar mass function down to $10^{8}M_{\odot}$. The estimation of the stellar mass depends heavily on the choice of IMF; the differing IMFs varies the mass systematically. In comparisons with other results, we applied corrections for consistency with our results in which we assumed the Salpeter IMF with lower and upper mass cutoffs of 0.1 and 100$M_{\sun}$, respectively. As illustrated in Figure \ref{fig:mass_density}, our data point, including the plausible upper limit, is on a trend of gradual increase of stellar mass density with time. Our result agrees with other observations for galaxies at $z\sim5$ \citep{stark07, verma07, elsner08} within the uncertainties. It is worth noting that the mass density at $z\sim5$ is dominated by the massive end of stellar mass function.

We also compare these observational results with theoretical predictions of two semi-analytical models by \citet{bower06} and \citet{kitzbichler07}. The mass densities of these models are obtained by integrating their stellar mass function down to $10^{8}M_{\sun}$. We applied corrections for the IMF as in the case of the comparisons with other observations. Figure \ref{fig:mass_density} shows that the number density for our sample is smaller by a factor of $2-3$ than the semi-analytical models. However, if we take into account the plausible upper limit, our result is almost comparable to the models. The theoretical models reproduce the overall trend of increase of mass density with time, though the number densities of the models tend to be somewhat larger than those of the observations in most of the redshift bins.

The stellar mass density at $z\sim5$ derived with a large sample of LBGs roughly agrees with that by the model predictions based on the CDM hierarchical structure formation scenario. However, since our fiducial stellar mass function disagrees with the model predictions as we discussed in \S \ref{sec:discussions_mass_function}. A larger amount of feedback to quench star formation might be needed in the lower-mass part, i.e., mass-dependent feedback process may be needed.

%
%
\section{Conclusions and Summary}
\label{sec:conclusions}

In this paper, we studied the stellar populations of Lyman Break Galaxies at $z\sim5$. We used the LBG sample by \citet{iwata07} obtained by the Suprime-Cam attached to the Subaru Telescope in the area of $\sim500$ $\textrm{arcmin}^2$ around the GOODS-N, which consists of $\sim600$ objects. For the mid-Infrared photometry, we used the publicly available data of the IRAC onboard the Spitzer in the GOODS-N. In addition, we observed the GOODS-N flanking fields (GOODS-FF) with the IRAC to cover the bulk ($\sim80\%$) of the Subaru area. We selected $\sim100$ objects which are isolated and not seriously contaminated by neighboring objects in the IRAC images by eye inspection. For these objects, the rest-frame UV to optical SEDs were constructed. We used \textit{SEDfit} package (\citealt{sawicki98}; Sawicki, in prep.) to derive the properties of these galaxies: stellar mass, star formation age, color excess, and SFR.

We assumed the constant star formation history, the metallicity of 0.2$Z_{\sun}$, the Salpeter IMF ranging from  0.1$M_{\sun}$ to 100$M_{\sun}$, and the extinction law by \citet{calzetti00}, and found that the median values of the stellar mass, age, color excess, and SFR are $4.1\times10^{9}M_{\sun}$, 25 Myr, 0.22 mag, and 141 $M_{\sun}\textrm{yr}^{-1}$, respectively. The comparisons of the distributions of these parameters with those for the $z=2-3$ LBG sample by \citet{shapley01} and \citet{shapley05}, all of which are in the similar rest-frame UV and optical luminosity range, show the increase of the median stellar mass from $z\sim5$ to $z=2-3$ by a factor of $3-4$. The $z\sim5$ LBGs are relatively younger by a factor of $10-20$ than the $z=2-3$ LBGs. The median color excess of our sample might be slightly larger that that at $z\sim3$ and similar to that at $z\sim2$. The median SFR of our sample is higher by a factor of $2-3$ than in the $z=2-3$ LBGs. We suggest that the LBGs at $z\sim5$ are undergoing intense star formation making them dusty and they are dominated by younger stellar populations than in the case of $z=2-3$ LBGs. \citet{verma07} presented similar results with a smaller sample. If each LBG at $z\sim5$ keeps the SFR derived from the SED fitting until $z=2-3$, the expected distribution of stellar mass shifts toward larger than those derived at $z=2-3$. This could imply that the SFR decreases from $z\sim5$ to $z=2-3$. We also compared the results for our sample with other studies for the $z=5-6$ galaxies. Although we found similarities and differences in the distributions of the parameters, we can not conclude their significance due to the small sample sizes of other studies.

The large number of our sample galaxies allows us to derive the stellar mass function of LBGs at $z\sim5$ after applying corrections for both $z'$-band and IRAC-band incompleteness. We compared the resulting stellar mass function with other observational studies. The stellar mass function of our sample agrees with the result by \citet{drory05} in the most massive end. However, in most of the mass range, the number densities of our sample are smaller than those of the sample by \citet{drory05}. Meanwhile, our result agrees well with the result by \citet{elsner08}. The discrepancy between the stellar mass function of our sample and that by \citet{drory05} is considered to be due to the use of IRAC data in our analysis. We also compared our result with the predictions of semi-analytic models involving AGN feedback and found that although the number densities of our sample are comparable to the model predictions in the massive end of the stellar mass function, the observed number densities are smaller than those by the theoretical predictions in the lower-mass part. By integrating the stellar mass function down to $10^{8}M_{\sun}$, the stellar mass density at $z\sim5$ is calculated to be $(0.7-2.4)\times10^{7}M_{\sun}\textrm{Mpc}^{-3}$. The stellar mass density is dominated by the massive part of the stellar mass function. The stellar mass density of our sample is consistent with general trend of the increase of the stellar mass density with time obtained in other observational studies. Our stellar mass density is almost comparable to the models if we take into account the plausible upper limit. The stellar mass density at $z\sim5$ derived with a large sample of LBGs roughly agrees with that by the model predictions based on the CDM hierarchical structure formation scenario. However, since our fiducial stellar mass function disagrees with the model predictions, some alterations may be needed for the theoretical models at high redshift.

\acknowledgments
We thank an anonymous referee for helpful comments to improve the paper. This work is based on observations made with the $Spitzer\ Space\ Telescope$, which is operated by the Jet Propulsion Laboratory, California Institute of Technology under a contract with NASA. Support for this work was provided by NASA through an award issued by JPL/Caltech. Support for this work was also provided by a NASA Spitzer Archival Research grant and by grants from the Natural Sciences and Engineering Research Council of Canada and the Canadian Space Agency. This work is supported by Grant-in-Aid for Scientific Research (17540216) from Japan Society for the Promotion of Science (JSPS) and by Grand-in-Aid for Scientific Research on Priority Areas (19047003) from Ministry of Education, Culture, Sports, Science, and Technology of Japan. MS thanks the JSPS for an Invitational Visitorship which supported his long-term visit to Kyoto University. 

\appendix

\section{Effects of varying star formation history, metallicity, and dust extinction law}
\label{appendix:effects}

\begin{figure}
\begin{center}
\epsscale{0.95}
\plotone{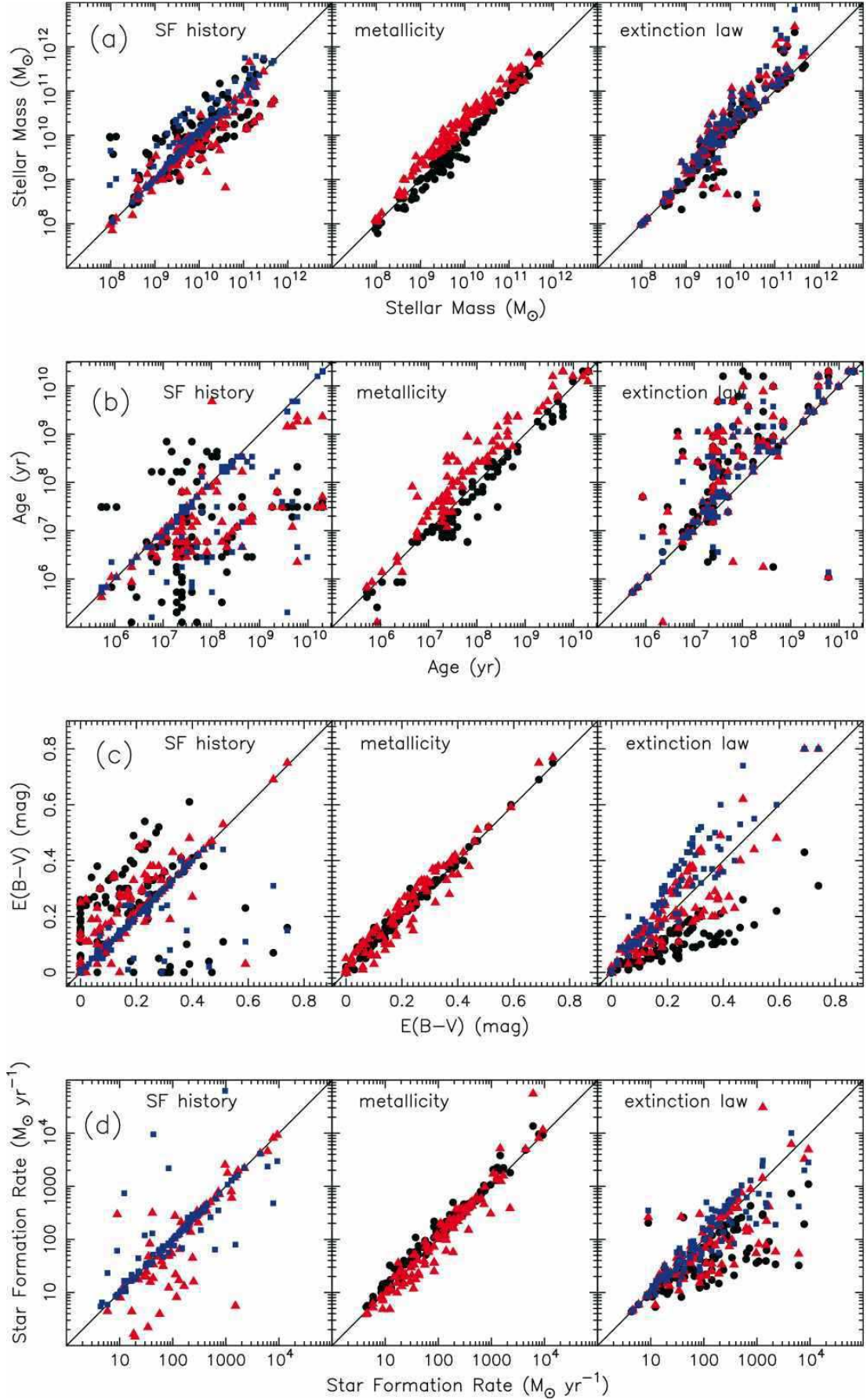}
\caption{Effects of varying star formation history, metallicity, and dust extinction law on the output parameters in the SED fitting. In panels (a), (b), (c), and (d), we show stellar mass, age, color excess, and star formation rate, respectively, derived by our fiducial model used in this work (abscissa) and by changing models (ordinate). In the left sub-panels, the output parameters obtained by assuming instantaneous burst (\textit{circles}), exponentially declining (\textit{triangles}), and two-component (\textit{squares}) models are plotted against those obtained by assuming constant star formation models. In the middle sub-panels, the output parameters obtained by assuming $1.0Z_{\sun}$ (\textit{circles}) and $0.02Z_{\sun}$ (\textit{triangles}) metallicity are plotted against those derived with $0.2Z_{\sun}$ metallicity models. In the right sub-panels, the output parameters obtained with the SMC (\textit{circles}), LMC(\textit{triangles}), and MW(\textit{squares}) dust extinction law are plotted against those derived with the extinction law by \citet{calzetti00}.\label{fig:effects_mass}}
\end{center}
\end{figure}

In this paper, we adopted the models with constant star formation history, $Z=0.2Z_{\sun}$, and extinction law by \citet{calzetti00}. Effects of varying the star formation history, metallicity and dust extinction law have been explored for LBGs at lower redshifts \citep{sawicki98, papovich01} and these effects, especially on stellar mass, are usually not large. Here we examine the effects of these various assumptions on the estimations of the stellar mass, age, color excess, and SFR for LBGs at $z\sim5$.

In the left sub-panel of Figure \ref{fig:effects_mass}a, the stellar masses derived by assuming the instantaneous burst and the exponentially declining star formation history with a time scale of $\tau$ against the stellar mass derived for the case of the constant star formation history are plotted. To derive the stellar mass for the $\tau$-models, we used models with $\tau$=1 Myr, 10 Myr, 100 Myr, 1 Gyr, and 10 Gyr. Although there seems to be no systematic difference ($\la0.1$ dex), the scatters are $\sigma=0.59$ and $0.35$ dex for the instantaneous burst and the $\tau$-models, respectively. We also examine the two-component models, in which we put the additional star formation into the passively evolving component. We assume that the old component is the instantaneous burst model whose age is 890 Myr ($z_{f}\sim13$). The spectrum of this old component multiplied by a flux ratio\footnote{Note that $\tau$ values of the exponentially declining models and the flux ratio in the two-component models are free parameters, and the degree of freedom decreases in the SED fitting.} against the young component, is put into that of the constant star formation model. The stellar masses of some ($\sim10$) objects increase by $\sim1$ dex, but those of most of the objects agree with the stellar masses in the case of constant star formation history with the median difference of 0.02 dex.

We examined systematic effects of metallicity to the estimation of stellar mass. The middle sub-panel of Figure \ref{fig:effects_mass}a shows the output stellar masses with $Z=1.0Z_{\sun}$ and $0.02Z_{\sun}$ against those with $Z=0.2Z_{\sun}$. Using solar metallicity models ($Z=1.0Z_{\sun}$) makes the stellar mass smaller systematically than using sub-solar models. Conversely, using the low metallicity models ($Z=0.02Z_{\sun}$) derives systematically larger stellar masses. Mean offsets are $-0.09$ and $0.28$ dex for $Z=1.0Z_{\sun}$ and $Z=0.02Z_{\sun}$, respectively.

We examined the effects of the choice of extinction laws on the stellar mass estimation. The extinction laws we tested apart from the Calzetti law are SMC \citep{prevot84}, LMC, and MW extinction laws \citep{fitzpatrick86}. These extinction laws are screen-type extinction, and we believe that the choice of the Calzetti extinction curve is appropriate, but we test these effects. In the right sub-panel of Figure \ref{fig:effects_mass}a, the derived stellar masses with using these three extinction laws are plotted against those with the Calzetti extinction law. The choice of these three extinction laws gives larger stellar mass especially in the massive part. On average, the SMC, LMC, and MW extinction laws give larger stellar mass by 0.07, 0.22, and 0.27 dex, respectively.

Age, color excess, and star formation rate are also affected by the choice of star formation history, metallicity, and extinction law. We examined here these effects of varying the assumptions on the output parameters.

As shown in the left sub-panel of Figure \ref{fig:effects_mass}b, varying star formation history affects the age estimation. On average, the age derived by assuming instantaneous burst, exponentially declining, and two-component models decreases by $0.89$, $0.70$, and $0.44$ dex, respectively, with respect to that by assuming constant star formation history. Most of the LBGs showing the old age (larger than 1.2 Gyr) in the SED fitting under constant star formation are younger than 1.2 Gyr under other star formation histories. The middle sub-panel of Figure \ref{fig:effects_mass}b shows the effects of varying metallicity. Metallicity does not change the age estimation drastically as in the case of mass; on average, the age with $1.0Z_{\sun}$ and $0.02Z_{\sun}$ decreases by $0.22$ dex and increases by $0.33$ dex, respectively, as compared with the case of $Z=0.2Z_{\sun}$. The right sub-panel of Figure \ref{fig:effects_mass}b shows the effects of varying the choice of dust extinction law. The choice of other dust extinction laws generally increases the age systematically; on average, the age derived by assuming SMC, LMC, and MW extinction law increases by 0.48 dex, 0.40 dex, and 0.30 dex, respectively.

Varying star formation history also affects the color excess estimation (left sub-panel of Figure \ref{fig:effects_mass}c). For most of the objects, the color excesses derived by assuming instantaneous burst are systematically larger than those by assuming constant star formation. There are two sequences; one shows no offset and the other shows a $\sim0.2$ mag offset, and the average difference is $\sim0.1$ mag. For some ($\sim10$) objects, the color excesses decrease toward E(B-V)$\sim0$ mag. The color excesses by assuming exponentially declining star formation are also larger systematically by, on average, $\sim0.05$ mag than those by assuming constant star formation. There is no systematic difference ($\sim0.03$ dex) for the case of two-component models except for some ($\sim10$) outliers. The color excesses of some objects decrease for all star formation models, especially for the instantaneous burst models. The middle sub-panel of Figure \ref{fig:effects_mass}c shows the effects of varying metallicity. Metallicity does not change the color excess estimation drastically; there seems to be no systematic difference ($\la0.01$ mag) between the estimation of color excess by assuming $1.0Z_{\sun}$ or $0.02Z_{\sun}$ and that by assuming $0.2Z_{\sun}$. The right sub-panel of Figure \ref{fig:effects_mass}c shows the effects of varying the choice of dust extinction law. Assuming the SMC extinction law decreases the color excess systematically; though the difference is larger in the larger color excess, the average difference of the color excess from the Calzetti's law is 0.11 mag. On the contrary, assuming the MW extinction law increases the color excess; again though the difference is larger in the larger color excess, the average difference of the color excess from the Calzetti's law is 0.06 mag. Although assuming the LMC extinction law does not change color excess estimation systematically, there is a scatter of $\pm0.07$ mag.

\begin{figure}
\begin{center}
\epsscale{0.90}
\plotone{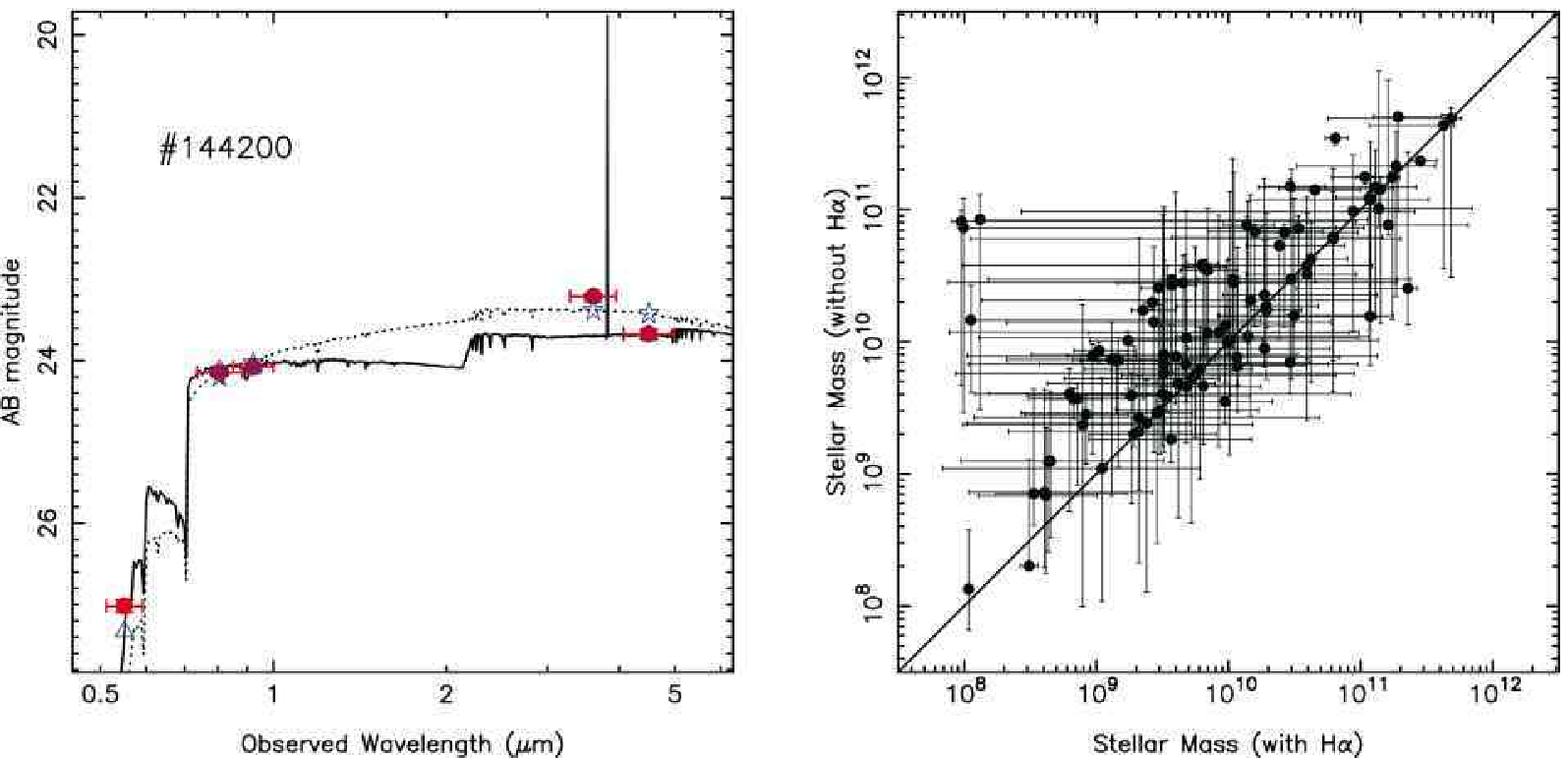}
\caption{(left): Best-fitted spectra of an objects with spectroscopic redshift with (\textit{solid line}) and without (\textit{dotted line}) H$\alpha$ emission line. The observed SED is indicated by filled circles. The SEDs of the best-fitted models are plotted by open triangles and open stars for models with and without H$\alpha$ emission line, respectively. (right): Comparison of the best-fitted stellar mass with and without H$\alpha$ emission line. \label{fig:halpha_effects}}
\end{center}
\end{figure}

Figure \ref{fig:effects_mass}d shows the effects on star formation rate. Varying star formation history (left sub-panel) also affects the star formation rate estimation but not so serious as compared with the age and the color excess, especially for the larger star formation rate. The middle sub-panel of Figure \ref{fig:effects_mass}d shows the effects of varying metallicity. The effects on star formation rate is not large; the average difference between the star formation rate by assuming $Z=1.0Z_{\sun}$ ($0.02Z_{\sun}$) and that by assuming $Z=0.2Z_{\sun}$ is $0.12$ ($-0.03$) dex. The right sub-panel of Figure \ref{fig:effects_mass}d shows the effects of varying the choice of dust extinction law. Assuming the SMC and LMC extinction law decrease the star formation rate systematically ($0.39$ dex and $0.16$ dex, respectively, on average).
Although assuming the MW extinction law does not change star formation rate estimation systematically, there is a scatter of 0.33 dex on average.

\section{Effects of H$\alpha$ emission line}
\label{appendix:halpha}

About 70 percent of the SEDs of our sample LBGs show excesses in $3.6\mu$m-band. This is very likely due to the presence of the H$\alpha$ emission that shifts into the 3.6$\mu$m band. Thus we took into account the effect of H$\alpha$ emission line in the SED fitting procedure.

The H$\alpha$ luminosity $L$(H$\alpha$) is calculated from the SFR of the given model by using the \citet{kennicutt98} relation:
\begin{equation}
L(\textrm{H}\alpha)\ (\textrm{ergs}\ \textrm{s}^{-1})\ =\ 1.26\times 10^{41}\ \textrm{SFR}\ (M_{\sun}\ \textrm{yr}^{-1})
\end{equation}
We examined the metallicity dependence of this relation. By using the BC03 models with metallicities of $Z=0.02$, $0.2$, and $1.0Z_{\sun}$, we derive the relation assuming all the ionizing photons are used for ionization and case B recombination. The systematic differences from the case of $1.0Z_{\odot}$ are factors of $1.4$ and $1.6$ in the case of $0.2Z_{\odot}$ and $0.02Z_{\odot}$, respectively. For $0.2Z_{\sun}$, the deviation from the Kennicutt relation is a factor of $1.3$ and our estimation of $L$(H$\alpha$) is considered to be underestimated with this factor if the metallicity is 0.2$Z_{\odot}$. Although the relation also depends on the assumed star formation history, the difference is less than a factor of $1.3$ among $\tau$ models.

\begin{figure}
\begin{center}
\epsscale{1.00}
\plotone{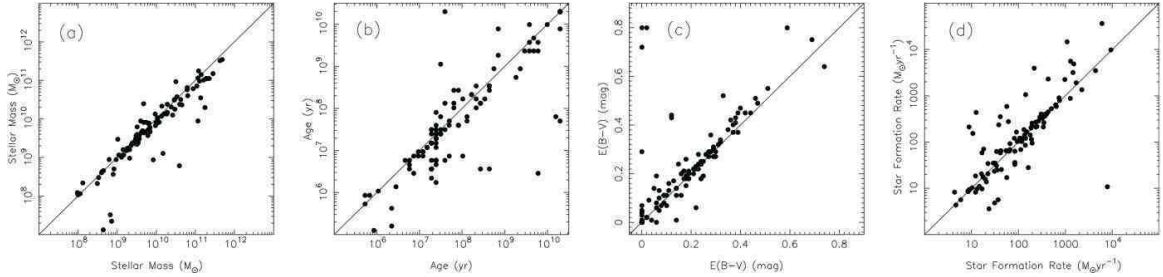}
\caption{Effects of varying redshift on the output parameters in the SED fitting. In panels (a), (b), (c), and (d), we show stellar mass, age, color excess, and star formation rate, respectively, derived with fixed redshift (abscissa) and with free redshift (ordinate).\label{fig:effects_redshift}}
\end{center}
\end{figure}

The H$\alpha$ luminosity calculated from its SFR is put into the model spectrum and the spectrum is attenuated with the \citet{calzetti00} extinction law. \citet{calzetti00} argued the difference between the color excess for the stellar continuum $E_{S}(B-V)$ and that for the gas emission $E(B-V)$, and presented the relation of $E_{S}(B-V)=(0.44\pm0.03)E(B-V)$. Thus, there is a possibility that the $L(\textrm{H}\alpha)$ is overestimated by a factor of $2.3$ in our prescription. While the $L(\textrm{H}\alpha)$ may be underestimated by a factor of $1.3$ for the sub-solar abundance, it may be overestimated by a factor of $2.3$ due to the differing extinctions for the stellar continuum and the emission line. Considering these uncertainties, we adopt the Kennicutt relation.

In the left panel of Figure \ref{fig:halpha_effects}, we show the best fitted spectra with and without H$\alpha$ emission line for an object (\#144200). The redshift of this object was confirmed to be $z=4.69$ by spectroscopy \citep{ando04}. It is clear that the model with H$\alpha$ emission fits the observed SED much better than that without H$\alpha$. We emphasize that we can obtain much better fit without increasing a free parameter. 

In the right panel of Figure \ref{fig:halpha_effects}, the comparison of the best-fitted stellar mass with and without H$\alpha$ emission is presented. By taking into account H$\alpha$ emission into the models, the stellar masses tend to be smaller than those estimated using models without H$\alpha$; there are two sequences, and the average difference is 0.32 dex. We examine the cases if H$\alpha$ line is not included; the stellar mass function derived in \S \ref{sec:discussions_mass_function} shifts toward larger mass systematically by $\sim0.3$ dex, and the stellar mass density discussed in \S \ref{sec:discussions_massdensity} increases by a factor of $\sim2$. If H$\alpha$ emission is not included, the medina age decreases by 0.83 dex, the median color excess increases by 0.14 mag, and the median SFR increases by 1.00 dex, on average.

\section{Effects of varying redshift}
\label{appendix:redshift}

\begin{figure}
\begin{center}
\epsscale{0.90}
\plotone{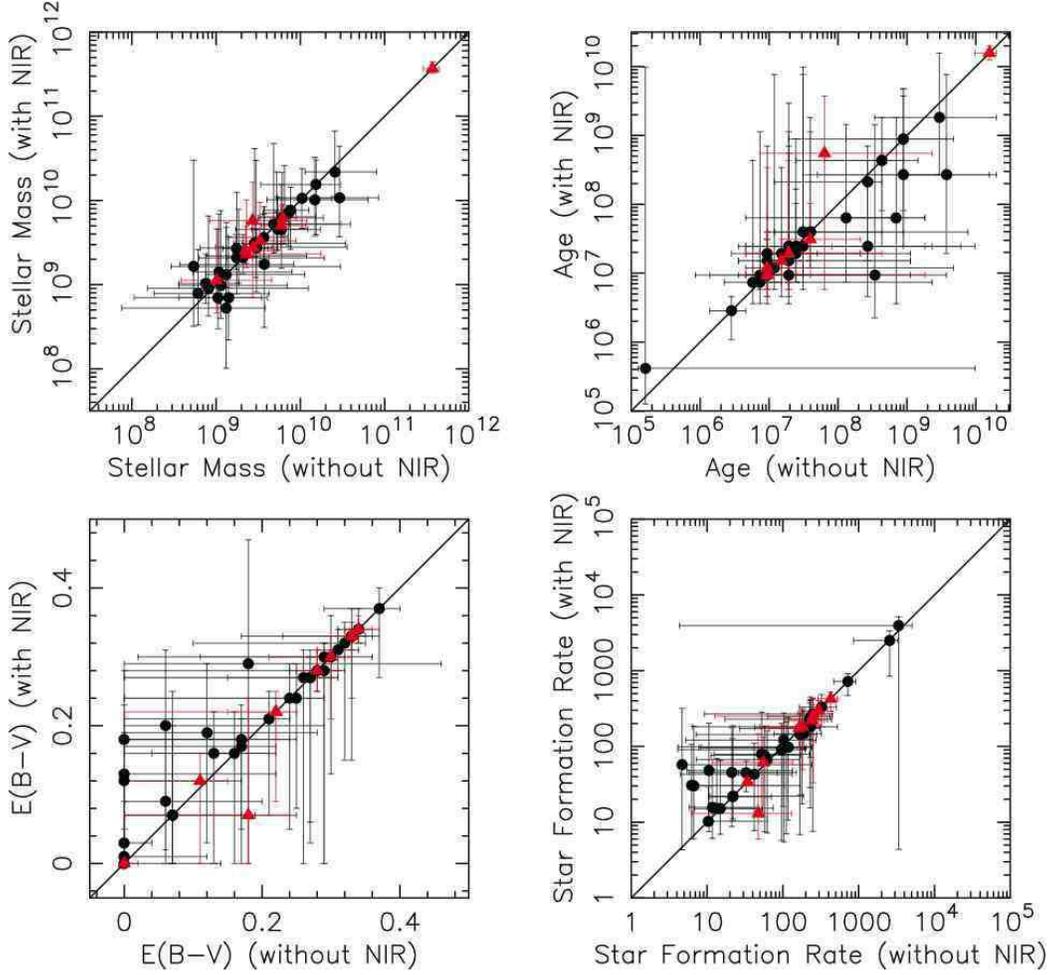}
\caption{Comparisons of the output parameters in the SED fitting for the $z\sim5$ LBG sample by \citet{stark07} with and without including NIR data ($J$ and $K_{s}$). The photometric sample is indicated by \textit{circles} and the spectroscopic sample is indicated by \textit{triangles}. The comparisons of the stellar mass, age, color excess, and star formation rate are plotted in panel (a), (b), (c), and (d), respectively. The error bars are 90\% confidence levels.
\label{fig:NIR_data}}
\end{center}
\end{figure}

As we described in \S \ref{sec:models}, we fixed the redshift of our sample objects to $z=4.8$. Here we examine the effects on the stellar mass as well as other output parameters when we take redshift as a free parameter. We refit the observed SEDs of the Category $1-3$ objects with the model SEDs, which are now parameterized by the redshift ranging from $z=0.0$ to $6.0$ in a step of 0.1. Although we did not use $V$-band photometry data in the fitting in the text, we incorporate $V$-band data here. Most of the objects ($\sim89\%$) have the best-fitted redshifts within $z=4.6\pm1.0$, although some objects ($\sim9\%$) can be fitted as low-redshift ($z\leqq2$) objects like ellipticals as well. Eighty-four percent of the objects show no secondary minimum of $\chi^{2}$ value in lower redshift. These support that contaminants in our sample is small. If we exclude these possible low-redshift objects, the number density of the stellar mass function decreases by $\sim30\%$ on average and the integrated stellar mass density decreases by $\sim30\%$.

As illustrated in Figure \ref{fig:effects_redshift}a, there is no significance difference between the stellar masses with the fixed redshifts and those with the free redshifts except for some outliers. Excluding these outliers, the median of log($M_{*}$(redshift free))$-$log($M_{*}$(redshift fixed)) is $-0.06$ dex and the scatter is $\sigma=0.22$ dex. Thus, even if we treat the redshift as a free parameter in the fitting, the change in estimated stellar masses is small.

In the Figure \ref{fig:effects_redshift}b, \ref{fig:effects_redshift}c, and \ref{fig:effects_redshift}d, the age, color excess, and star formation rate, respectively, derived by SED fitting taking redshift as a free parameter are plotted versus those obtained with fixed redshift of $z=4.8$. Except for the outliers, fixing the redshift to $z=4.8$ does not introduce a significant systematic offset and scatters are $\sigma\sim0.4$ dex for the age and star formation rate and $\sigma\sim0.1$ mag for the color excess.

We additionally assess the uncertainty from the redshift in the following way: For each object, we assign the redshift randomly along the expected redshift distribution by \citet{iwata07} and run a large set of SED fitting. On average, the scatters of the obtained distributions of the parameters are $3-5$ times smaller than the errors in the SED fitting.

\section{Effects of the presence of the near-Infrared data}
\label{appendix:NIR}

In \S \ref{sec:results} we showed the results of the SED fitting analysis; the stellar mass, age, color excess, and star formation rate of $z\sim5$ LBGs were derived. In this work, since we have no NIR data in the GOODS-N and the GOODS-FF, we used observed SEDs without NIR data. Thus we examined how large the discrepancy in the best fit parameters with and without NIR data. We did this test with the \textit{SEDfit} by using a sample of $z\sim5$ LBGs by \citet{stark07}. The sample by \citet{stark07} consists of 9 objects with spec-z and 34 objects with phot-z, for which both $J$ and $K_{s}$ data are available.

Figure \ref{fig:NIR_data} shows comparisons of the stellar mass, age, color excess, and star formation rate obtained with the $J$ and $K_{s}$ data and those without the $J$ and $K_{s}$ data. There seems to be no large difference between the stellar mass derived with and without NIR data; the stellar masses agree with each other within a factor of $\sim3$. Thus, the estimation of the stellar mass is robust regardless of the presence of the NIR data.

Figure \ref{fig:NIR_data}b and \ref{fig:NIR_data}c show the comparisons of the derived age and color excess, respectively, with and without the NIR data. Although the error bars are large, there seems to be no  significant systematic difference between these derived parameters except for some outliers. We also plot the comparison of the star formation rate with and without the NIR data in Figure \ref{fig:NIR_data}d; the star formation rate seems to be rather securely determined except for some objects.

However, the errors in the $J$ and $K_{s}$ data we use in the test are generally larger than those in other bands and the weights of the NIR data to the SED fitting are relatively small. The results of the test might be caused by this effects. Better constraints on the properties such as age and color excess are expected if sufficiently deep NIR data are available.



\clearpage
\LongTables
\begin{deluxetable}{lccccc}
\tabletypesize{\footnotesize}
\tablewidth{0pt}
\tablecaption{Photometry of the LBG sample (Category 1, 2, and 3)\label{table1}}
\tablehead{
\colhead{ID} & \colhead{$V$\tablenotemark{a,b}} & \colhead{$I_{c}$\tablenotemark{b}} & \colhead{$z'$\tablenotemark{b}} & \colhead{3.6$\mu$m\tablenotemark{b,c}} & \colhead{4.5$\mu$m\tablenotemark{b,c}}\\
\colhead{} & \colhead{(mag)} & \colhead{(mag)} & \colhead{(mag)} & \colhead{(mag)} & \colhead{(mag)}
}
\startdata
038819 & $<28.09$ & $24.78\pm0.07$ & $24.66\pm0.06$ & $23.48\pm0.11$ & $<24.14$\\
038859 & $<28.09$ & $25.22\pm0.09$ & $25.05\pm0.09$ & $24.97\pm0.37$ & $<24.14$\\
039340 & $<28.09$ & $26.16\pm0.16$ & $26.17\pm0.12$ & $24.42\pm0.24$ & $<24.14$\\
040064 & $28.06\pm0.19$ & $26.02\pm0.15$ & $26.03\pm0.12$ & $25.51\pm0.56$ & $<24.14$\\
046788 & $27.14\pm0.08$ & $25.12\pm0.07$ & $25.10\pm0.07$ & $23.30\pm0.09$ & $<24.14$\\
048421 & $<28.09$ & $25.11\pm0.09$ & $24.93\pm0.11$ & $23.54\pm0.11$ & $23.61\pm0.20$\\
048806 & $26.50\pm0.04$ & $24.57\pm0.04$ & $24.44\pm0.04$ & $23.68\pm0.13$ & $23.64\pm0.21$\\
050272 & $27.37\pm0.10$ & $25.29\pm0.08$ & $25.11\pm0.10$ & $23.03\pm0.07$ & $23.16\pm0.14$\\
051334 & $27.46\pm0.11$ & $24.21\pm0.04$ & $23.91\pm0.04$ & $22.83\pm0.06$ & $22.74\pm0.10$\\
053312 & $<28.09$ & $25.09\pm0.07$ & $24.97\pm0.06$ & $23.65\pm0.12$ & $23.30\pm0.16$\\
061662 & $<28.09$ & $25.50\pm0.10$ & $25.34\pm0.11$ & $<24.75$ & $24.19\pm0.33$\\
062238 & $<28.09$ & $25.55\pm0.11$ & $25.64\pm0.09$ & $23.79\pm0.14$ & $23.40\pm0.17$\\
063161 & $<28.09$ & $26.06\pm0.18$ & $26.35\pm0.18$ & $24.42\pm0.24$ & $<24.14$\\
063736 & $<28.09$ & $26.42\pm0.19$ & $26.24\pm0.18$ & $21.51\pm0.02$ & $21.33\pm0.03$\\
068263 & $27.80\pm0.15$ & $25.83\pm0.11$ & $25.77\pm0.14$ & $24.22\pm0.20$ & $<24.14$\\
070807 & $27.61\pm0.12$ & $25.65\pm0.11$ & $25.46\pm0.13$ & $23.59\pm0.12$ & $22.92\pm0.11$\\
071773 & $<28.09$ & $25.11\pm0.09$ & $25.27\pm0.08$ & $24.07\pm0.18$ & $23.70\pm0.22$\\
071987 & $28.27\pm0.23$ & $25.54\pm0.11$ & $25.49\pm0.13$ & $<24.75$ & $23.01\pm0.12$\\
072556 & $<28.09$ & $25.48\pm0.12$ & $25.25\pm0.12$ & $21.90\pm0.03$ & $21.71\pm0.04$\\
074486 & $<28.09$ & $26.69\pm0.22$ & $26.49\pm0.19$ & $24.27\pm0.21$ & $23.65\pm0.21$\\
074882 & $<28.09$ & $26.22\pm0.15$ & $26.01\pm0.11$ & $23.72\pm0.13$ & $<24.14$\\
075205 & $27.55\pm0.12$ & $24.81\pm0.10$ & $24.81\pm0.09$ & $22.02\pm0.03$ & $21.88\pm0.04$\\
077666 & $<28.09$ & $25.70\pm0.14$ & $25.49\pm0.12$ & $<24.75$ & $23.77\pm0.23$\\
077684 & $<28.09$ & $25.92\pm0.16$ & $25.85\pm0.15$ & $24.50\pm0.25$ & $<24.14$\\
078553 & $<28.09$ & $26.21\pm0.16$ & $26.02\pm0.14$ & $<25.93$ & $25.62\pm0.31$\\
079161 & $<28.09$ & $26.20\pm0.18$ & $26.01\pm0.12$ & $24.25\pm0.21$ & $<24.14$\\
079524 & $<28.09$ & $26.59\pm0.21$ & $26.42\pm0.19$ & $<24.75$ & $23.77\pm0.23$\\
080726 & $<28.09$ & $24.75\pm0.09$ & $24.47\pm0.08$ & $23.05\pm0.07$ & $23.25\pm0.15$\\
082124 & $<28.09$ & $26.04\pm0.12$ & $25.88\pm0.10$ & $21.55\pm0.02$ & $21.76\pm0.04$\\
082855 & $<28.09$ & $25.45\pm0.14$ & $25.63\pm0.15$ & $<24.75$ & $23.86\pm0.25$\\
083563 & $<28.09$ & $26.05\pm0.19$ & $25.88\pm0.16$ & $24.55\pm0.27$ & $<24.14$\\
083925 & $<28.09$ & $26.44\pm0.17$ & $26.28\pm0.12$ & $24.67\pm0.29$ & $<24.14$\\
084637 & $<28.09$ & $25.82\pm0.10$ & $25.59\pm0.09$ & $24.63\pm0.28$ & $23.75\pm0.23$\\
084850 & $<28.09$ & $25.17\pm0.09$ & $25.03\pm0.07$ & $24.00\pm0.17$ & $<24.14$\\
087802 & $<28.09$ & $25.89\pm0.12$ & $25.99\pm0.11$ & $22.99\pm0.02$ & $22.91\pm0.03$\\
088084 & $<28.09$ & $26.51\pm0.19$ & $26.41\pm0.12$ & $24.71\pm0.30$ & $<24.14$\\
089484 & $<28.09$ & $26.07\pm0.12$ & $25.96\pm0.10$ & $24.83\pm0.33$ & $<24.14$\\
090105 & $<28.09$ & $24.93\pm0.08$ & $24.89\pm0.10$ & $24.10\pm0.18$ & $<24.14$\\
091420 & $28.05\pm0.19$ & $24.98\pm0.12$ & $25.03\pm0.10$ & $22.89\pm0.06$ & $22.72\pm0.09$\\
092240 & $25.84\pm0.02$ & $23.67\pm0.02$ & $23.51\pm0.02$ & $22.48\pm0.04$ & $22.92\pm0.11$\\
092242 & $27.75\pm0.14$ & $26.14\pm0.12$ & $25.99\pm0.12$ & $24.49\pm0.25$ & $<24.14$\\
093559 & $28.01\pm0.18$ & $25.24\pm0.09$ & $25.22\pm0.08$ & $23.65\pm0.04$ & $23.97\pm0.07$\\
096484 & $27.35\pm0.10$ & $24.89\pm0.07$ & $24.67\pm0.07$ & $23.75\pm0.13$ & $<24.14$\\
096510 & $<28.09$ & $26.50\pm0.22$ & $26.43\pm0.20$ & $23.18\pm0.08$ & $22.72\pm0.09$\\
098022 & $<28.09$ & $25.67\pm0.15$ & $25.95\pm0.17$ & $24.03\pm0.06$ & $23.63\pm0.06$\\
100184 & $27.85\pm0.16$ & $25.75\pm0.10$ & $25.60\pm0.09$ & $25.50\pm0.22$ & $<25.64$\\
100509 & $27.07\pm0.08$ & $24.58\pm0.05$ & $24.50\pm0.07$ & $23.55\pm0.11$ & $23.45\pm0.18$\\
102671 & $26.18\pm0.03$ & $24.10\pm0.03$ & $24.05\pm0.03$ & $23.20\pm0.03$ & $23.51\pm0.05$\\
103109 & $27.48\pm0.11$ & $24.83\pm0.07$ & $24.92\pm0.07$ & $23.07\pm0.07$ & $23.06\pm0.13$\\
103742 & $<28.09$ & $26.18\pm0.14$ & $25.93\pm0.10$ & $24.67\pm0.29$ & $<24.14$\\
103985 & $28.01\pm0.18$ & $26.13\pm0.18$ & $26.43\pm0.18$ & $24.41\pm0.09$ & $<25.64$\\
104189 & $<28.09$ & $25.52\pm0.10$ & $25.50\pm0.08$ & $23.76\pm0.05$ & $24.33\pm0.10$\\
104766 & $<28.09$ & $24.97\pm0.08$ & $25.04\pm0.07$ & $23.18\pm0.03$ & $23.27\pm0.04$\\
106944 & $27.27\pm0.09$ & $24.56\pm0.04$ & $24.55\pm0.05$ & $23.61\pm0.12$ & $23.56\pm0.20$\\
107878 & $26.71\pm0.05$ & $24.46\pm0.04$ & $24.27\pm0.05$ & $21.54\pm0.01$ & $21.81\pm0.01$\\
108167 & $<28.09$ & $26.16\pm0.20$ & $26.48\pm0.20$ & $23.32\pm0.09$ & $22.71\pm0.09$\\
108384 & $28.02\pm0.18$ & $26.23\pm0.17$ & $26.36\pm0.17$ & $22.39\pm0.04$ & $22.17\pm0.06$\\
108417 & $<28.09$ & $25.76\pm0.20$ & $26.04\pm0.14$ & $24.29\pm0.21$ & $<24.14$\\
110593 & $27.93\pm0.17$ & $25.48\pm0.11$ & $25.58\pm0.09$ & $25.32\pm0.19$ & $26.35\pm0.54$\\
112044 & $<28.09$ & $26.04\pm0.19$ & $25.99\pm0.14$ & $23.86\pm0.15$ & $<24.14$\\
113060 & $27.81\pm0.15$ & $26.09\pm0.15$ & $26.33\pm0.12$ & $25.29\pm0.18$ & $24.99\pm0.18$\\
113120 & $<28.09$ & $25.89\pm0.16$ & $25.92\pm0.17$ & $25.93\pm0.31$ & $<25.64$\\
113749 & $27.50\pm0.11$ & $25.04\pm0.10$ & $24.90\pm0.07$ & $26.32\pm0.42$ & $<25.64$\\
115354 & $<28.09$ & $26.33\pm0.18$ & $26.17\pm0.17$ & $23.61\pm0.04$ & $23.54\pm0.05$\\
115925 & $<28.09$ & $24.55\pm0.10$ & $24.31\pm0.08$ & $24.55\pm0.07$ & $25.05\pm0.14$\\
116678 & $<28.09$ & $26.55\pm0.19$ & $26.41\pm0.18$ & $25.00\pm0.10$ & $25.75\pm0.25$\\
116910 & $<28.09$ & $25.57\pm0.15$ & $25.41\pm0.13$ & $24.61\pm0.10$ & $24.51\pm0.12$\\
117078 & $<28.09$ & $25.54\pm0.11$ & $25.35\pm0.10$ & $23.61\pm0.04$ & $23.98\pm0.08$\\
120190 & $27.00\pm0.07$ & $24.68\pm0.05$ & $24.52\pm0.06$ & $23.70\pm0.03$ & $24.45\pm0.08$\\
120554 & $27.93\pm0.17$ & $26.02\pm0.14$ & $25.95\pm0.11$ & $<25.93$ & $24.92\pm0.17$\\
120838 & $<28.09$ & $25.50\pm0.08$ & $25.14\pm0.07$ & $25.77\pm0.27$ & $25.79\pm0.26$\\
123533 & $<28.09$ & $25.73\pm0.13$ & $25.48\pm0.13$ & $<25.93$ & $25.96\pm0.40$\\
125665 & $27.28\pm0.09$ & $25.07\pm0.07$ & $24.93\pm0.07$ & $23.69\pm0.04$ & $24.27\pm0.10$\\
126010 & $27.14\pm0.08$ & $25.30\pm0.08$ & $25.25\pm0.08$ & $23.79\pm0.05$ & $23.67\pm0.06$\\
126510 & $<28.09$ & $26.45\pm0.19$ & $26.24\pm0.17$ & $<25.93$ & $25.56\pm0.29$\\
127245 & $27.35\pm0.10$ & $24.96\pm0.06$ & $24.89\pm0.06$ & $23.37\pm0.02$ & $23.84\pm0.05$\\
127900 & $27.10\pm0.08$ & $24.90\pm0.06$ & $24.70\pm0.07$ & $25.26\pm0.18$ & $25.61\pm0.30$\\
129670 & $<28.09$ & $26.27\pm0.18$ & $26.33\pm0.18$ & $25.55\pm0.23$ & $25.20\pm0.22$\\
130018 & $<28.09$ & $25.18\pm0.09$ & $24.98\pm0.09$ & $24.46\pm0.09$ & $25.42\pm0.26$\\
130851 & $27.62\pm0.13$ & $25.47\pm0.11$ & $25.63\pm0.11$ & $24.79\pm0.09$ & $25.14\pm0.15$\\
131482 & $26.69\pm0.05$ & $24.60\pm0.04$ & $24.38\pm0.04$ & $25.16\pm0.17$ & $25.75\pm0.34$\\
133419 & $27.75\pm0.14$ & $25.12\pm0.10$ & $25.19\pm0.11$ & $24.15\pm0.05$ & $24.37\pm0.08$\\
133694 & $27.56\pm0.12$ & $25.21\pm0.08$ & $25.10\pm0.07$ & $24.07\pm0.05$ & $24.64\pm0.10$\\
134614 & $<28.09$ & $25.69\pm0.10$ & $25.42\pm0.09$ & $24.61\pm0.10$ & $23.99\pm0.08$\\
135678 & $<28.09$ & $25.20\pm0.11$ & $25.42\pm0.12$ & $26.85\pm0.62$ & $26.16\pm0.47$\\
138613 & $<28.09$ & $26.55\pm0.21$ & $26.36\pm0.18$ & $<25.93$ & $26.22\pm0.49$\\
138763 & $<28.09$ & $25.64\pm0.10$ & $25.69\pm0.09$ & $23.81\pm0.05$ & $24.72\pm0.14$\\
138810 & $27.66\pm0.13$ & $26.04\pm0.12$ & $25.88\pm0.11$ & $25.71\pm0.26$ & $25.44\pm0.27$\\
139906 & $<28.09$ & $26.15\pm0.20$ & $26.09\pm0.16$ & $25.93\pm0.23$ & $9.99\pm9.61$\\
141088 & $<28.09$ & $25.58\pm0.09$ & $25.22\pm0.07$ & $24.62\pm0.10$ & $25.12\pm0.20$\\
141117 & $<28.09$ & $26.52\pm0.17$ & $26.31\pm0.12$ & $25.19\pm0.12$ & $25.53\pm0.21$\\
141368 & $<28.09$ & $25.53\pm0.14$ & $25.32\pm0.13$ & $24.82\pm0.12$ & $24.71\pm0.14$\\
142195 & $27.89\pm0.16$ & $25.20\pm0.10$ & $24.95\pm0.10$ & $23.44\pm0.04$ & $23.42\pm0.05$\\
144200 & $27.02\pm0.07$ & $24.15\pm0.03$ & $24.08\pm0.03$ & $23.21\pm0.03$ & $23.68\pm0.06$\\
145330 & $<28.09$ & $26.24\pm0.15$ & $26.13\pm0.14$ & $25.20\pm0.17$ & $25.18\pm0.21$\\
147153 & $<28.09$ & $26.45\pm0.23$ & $26.27\pm0.18$ & $24.06\pm0.18$ & $<24.14$\\
147965 & $27.51\pm0.11$ & $25.95\pm0.12$ & $26.08\pm0.13$ & $<25.93$ & $25.76\pm0.34$\\
148198 & $<28.09$ & $24.93\pm0.05$ & $24.50\pm0.05$ & $22.47\pm0.04$ & $22.66\pm0.09$\\
149470 & $28.16\pm0.21$ & $26.50\pm0.17$ & $26.37\pm0.12$ & $29.06\pm2.10$ & $<25.64$\\
149604 & $<28.09$ & $26.41\pm0.19$ & $26.43\pm0.19$ & $<25.93$ & $26.09\pm0.44$\\
149667 & $<28.09$ & $26.36\pm0.17$ & $26.36\pm0.17$ & $23.98\pm0.06$ & $23.48\pm0.05$\\
152993 & $<28.09$ & $25.87\pm0.18$ & $25.81\pm0.10$ & $26.45\pm0.35$ & $26.14\pm0.34$\\
156057 & $27.61\pm0.12$ & $25.31\pm0.10$ & $25.18\pm0.07$ & $24.01\pm0.06$ & $24.04\pm0.08$\\
161503 & $27.31\pm0.09$ & $25.70\pm0.13$ & $26.30\pm0.12$ & $24.99\pm0.14$ & $<25.64$\\
164830 & $<28.09$ & $25.93\pm0.15$ & $26.07\pm0.15$ & $24.41\pm0.09$ & $24.77\pm0.15$\\
\enddata
\tablenotetext{a}{Upper limits are 3$\sigma$ values in $1.6\arcsec$ diameter aperture.}
\tablenotetext{b}{Errors are 1$\sigma$ values.}
\tablenotetext{c}{Upper limits are 3$\sigma$ values in $2.4\arcsec$ diameter aperture.}
\end{deluxetable}

\clearpage
\begin{deluxetable}{lccccc}
\tabletypesize{\footnotesize}
\tablewidth{0pt}
\tablecaption{The best-fitted parameters\label{table2}}
\tablehead{
\colhead{ID} & \colhead{log[$M_{*}$ ($M_{\odot}$)]\tablenotemark{a}} & \colhead{log[Age (yr)]\tablenotemark{a}} & \colhead{$E(B-V)$ (mag)\tablenotemark{a}} & \colhead{log[SFR ($M_{\odot}$yr$^{-1}$)]\tablenotemark{a}} & \colhead{$\chi^{2}$}
}
\startdata
038819 & ~9.84$^{+1.07}_{-0.90}$ & ~7.39$^{+2.39}_{-1.14}$ & 0.24$^{+0.03}_{-0.24}$ & 2.48$^{+0.40}_{-1.21}$ & 0.24\\
038859 & ~9.12$^{+0.96}_{-0.79}$ & ~7.39$^{+1.66}_{-1.04}$ & 0.11$^{+0.07}_{-0.11}$ & 1.75$^{+0.57}_{-0.66}$ & 2.20\\
039340 & 10.59$^{+0.50}_{-2.60}$ & ~9.78$^{+0.52}_{-4.16}$ & 0.06$^{+0.31}_{-0.06}$ & 0.95$^{+1.77}_{-0.25}$ & 0.01\\
040064 & ~8.90$^{+1.28}_{-0.91}$ & ~7.39$^{+2.18}_{-1.35}$ & 0.14$^{+0.10}_{-0.14}$ & 1.54$^{+0.69}_{-0.79}$ & 0.01\\
046788 & 10.94$^{+0.47}_{-2.51}$ & ~9.57$^{+0.73}_{-3.85}$ & 0.09$^{+0.25}_{-0.09}$ & 1.50$^{+1.54}_{-0.37}$ & 0.02\\
048421 & 10.14$^{+0.92}_{-0.54}$ & ~7.91$^{+2.18}_{-0.94}$ & 0.25$^{+0.06}_{-0.25}$ & 2.29$^{+0.42}_{-1.16}$ & 0.35\\
048806 & ~9.98$^{+0.52}_{-0.49}$ & ~7.91$^{+1.35}_{-0.83}$ & 0.16$^{+0.06}_{-0.16}$ & 2.12$^{+0.42}_{-0.76}$ & 2.69\\
050272 & 10.27$^{+0.85}_{-0.31}$ & ~7.49$^{+2.18}_{-0.52}$ & 0.37$^{+0.04}_{-0.25}$ & 2.81$^{+0.22}_{-1.22}$ & 0.11\\
051334 & 10.53$^{+0.29}_{-0.28}$ & ~8.22$^{+0.83}_{-0.62}$ & 0.19$^{+0.06}_{-0.11}$ & 2.38$^{+0.34}_{-0.51}$ & 9.78\\
053312 & 10.50$^{+0.62}_{-0.58}$ & ~8.84$^{+1.25}_{-1.35}$ & 0.14$^{+0.14}_{-0.14}$ & 1.75$^{+0.75}_{-0.61}$ & 2.21\\
061662 & ~9.75$^{+1.06}_{-0.56}$ & ~6.76$^{+3.22}_{-1.66}$ & 0.36$^{+0.18}_{-0.36}$ & 2.99$^{+2.44}_{-2.01}$ & 0.02\\
062238 & 11.15$^{+0.09}_{-0.77}$ & 10.30$^{+0.00}_{-1.87}$ & 0.02$^{+0.24}_{-0.02}$ & 1.01$^{+1.02}_{-0.07}$ & 0.49\\
063161 & ~7.99$^{+3.03}_{-0.03}$ & ~5.72$^{+4.58}_{-0.10}$ & 0.00$^{+0.35}_{-0.00}$ & 2.27$^{+0.38}_{-1.59}$ & 0.27\\
063736 & 11.45$^{+0.12}_{-0.08}$ & ~7.60$^{+0.31}_{-0.21}$ & 0.74$^{+0.06}_{-0.04}$ & 3.89$^{+0.16}_{-0.17}$ & 1.07\\
068263 & 10.17$^{+0.91}_{-2.04}$ & ~8.64$^{+1.66}_{-2.91}$ & 0.18$^{+0.16}_{-0.18}$ & 1.62$^{+1.10}_{-0.79}$ & 0.01\\
070807 & 11.36$^{+0.07}_{-0.85}$ & 10.30$^{+0.00}_{-1.87}$ & 0.07$^{+0.23}_{-0.04}$ & 1.22$^{+1.00}_{-0.12}$ & 5.87\\
071773 & 10.62$^{+0.28}_{-0.63}$ & ~9.68$^{+0.31}_{-1.56}$ & 0.00$^{+0.20}_{-0.00}$ & 1.09$^{+0.89}_{-0.02}$ & 1.21\\
071987 & 11.07$^{+0.44}_{-0.91}$ & ~9.57$^{+0.73}_{-4.47}$ & 0.16$^{+0.50}_{-0.12}$ & 1.64$^{+4.26}_{-0.48}$ & 0.02\\
072556 & 11.28$^{+0.32}_{-0.19}$ & ~8.12$^{+0.94}_{-0.42}$ & 0.51$^{+0.06}_{-0.13}$ & 3.23$^{+0.26}_{-0.58}$ & 0.17\\
074486 & 11.07$^{+0.17}_{-1.14}$ & 10.20$^{+0.10}_{-2.60}$ & 0.12$^{+0.32}_{-0.06}$ & 1.03$^{+1.43}_{-0.20}$ & 1.23\\
074882 & 10.03$^{+1.35}_{-1.85}$ & ~7.49$^{+2.81}_{-1.87}$ & 0.40$^{+0.04}_{-0.37}$ & 2.57$^{+0.44}_{-1.56}$ & 0.02\\
075205 & 11.21$^{+0.60}_{-0.27}$ & ~8.43$^{+1.56}_{-0.73}$ & 0.38$^{+0.08}_{-0.21}$ & 2.86$^{+0.42}_{-0.88}$ & 2.08\\
077666 & 10.01$^{+1.12}_{-0.41}$ & ~6.66$^{+3.64}_{-1.56}$ & 0.44$^{+0.16}_{-0.44}$ & 3.35$^{+2.25}_{-2.43}$ & 0.01\\
077684 & ~9.68$^{+1.30}_{-1.57}$ & ~7.80$^{+2.50}_{-1.98}$ & 0.24$^{+0.09}_{-0.24}$ & 1.92$^{+0.70}_{-1.14}$ & 0.01\\
078553 & ~9.62$^{+0.28}_{-0.99}$ & ~6.45$^{+2.70}_{-1.35}$ & 0.38$^{+0.07}_{-0.38}$ & 3.17$^{+1.55}_{-2.47}$ & 0.05\\
079161 & ~9.60$^{+1.53}_{-1.58}$ & ~7.28$^{+3.02}_{-1.66}$ & 0.33$^{+0.06}_{-0.33}$ & 2.34$^{+0.46}_{-1.59}$ & 0.07\\
079524 & 10.47$^{+0.81}_{-0.72}$ & ~8.64$^{+1.66}_{-3.54}$ & 0.32$^{+0.40}_{-0.28}$ & 1.93$^{+3.78}_{-1.16}$ & 0.01\\
080726 & 10.20$^{+0.52}_{-0.35}$ & ~7.60$^{+1.35}_{-0.62}$ & 0.28$^{+0.03}_{-0.15}$ & 2.64$^{+0.25}_{-0.77}$ & 1.75\\
082124 & 11.03$^{+0.07}_{-0.06}$ & ~7.08$^{+0.10}_{-0.10}$ & 0.69$^{+0.02}_{-0.02}$ & 3.97$^{+0.08}_{-0.07}$ & 2.02\\
082855 & 10.79$^{+0.38}_{-1.22}$ & ~9.99$^{+0.31}_{-4.89}$ & 0.00$^{+0.54}_{-0.00}$ & 0.95$^{+4.46}_{-0.02}$ & 0.30\\
083563 & ~9.50$^{+1.42}_{-1.42}$ & ~7.49$^{+2.81}_{-1.66}$ & 0.26$^{+0.07}_{-0.26}$ & 2.04$^{+0.56}_{-1.29}$ & 0.11\\
083925 & ~9.51$^{+1.44}_{-1.58}$ & ~7.49$^{+2.81}_{-1.77}$ & 0.30$^{+0.07}_{-0.30}$ & 2.05$^{+0.57}_{-1.42}$ & 0.05\\
084637 & 10.07$^{+0.78}_{-0.58}$ & ~8.53$^{+1.56}_{-1.14}$ & 0.17$^{+0.11}_{-0.17}$ & 1.62$^{+0.63}_{-0.75}$ & 5.80\\
084850 & ~9.68$^{+1.08}_{-1.11}$ & ~7.49$^{+2.29}_{-1.35}$ & 0.22$^{+0.05}_{-0.22}$ & 2.21$^{+0.50}_{-1.10}$ & 0.47\\
087802 & 10.65$^{+0.34}_{-0.16}$ & ~7.91$^{+1.04}_{-0.31}$ & 0.46$^{+0.05}_{-0.14}$ & 2.80$^{+0.20}_{-0.65}$ & 3.59\\
088084 & ~9.93$^{+1.06}_{-2.04}$ & ~8.43$^{+1.87}_{-2.81}$ & 0.23$^{+0.16}_{-0.23}$ & 1.58$^{+1.07}_{-0.98}$ & 0.01\\
089484 & ~9.32$^{+1.36}_{-1.25}$ & ~7.39$^{+2.70}_{-1.46}$ & 0.24$^{+0.06}_{-0.24}$ & 1.96$^{+0.52}_{-1.21}$ & 0.12\\
090105 & ~9.51$^{+1.01}_{-1.00}$ & ~7.39$^{+2.08}_{-1.25}$ & 0.18$^{+0.05}_{-0.18}$ & 2.15$^{+0.47}_{-0.97}$ & 0.03\\
091420 & 11.27$^{+0.30}_{-0.75}$ & ~9.78$^{+0.52}_{-1.98}$ & 0.11$^{+0.25}_{-0.08}$ & 1.63$^{+1.13}_{-0.30}$ & 0.31\\
092240 & 10.14$^{+0.21}_{-0.18}$ & ~7.28$^{+0.42}_{-0.31}$ & 0.22$^{+0.02}_{-0.02}$ & 2.88$^{+0.16}_{-0.19}$ & 3.87\\
092242 & ~9.51$^{+1.47}_{-1.48}$ & ~7.39$^{+2.91}_{-1.66}$ & 0.29$^{+0.06}_{-0.29}$ & 2.15$^{+0.51}_{-1.42}$ & 0.06\\
093559 & ~9.84$^{+0.22}_{-0.21}$ & ~7.39$^{+0.52}_{-0.31}$ & 0.29$^{+0.02}_{-0.04}$ & 2.48$^{+0.13}_{-0.28}$ & 0.42\\
096484 & ~9.66$^{+0.82}_{-0.50}$ & ~7.28$^{+1.87}_{-0.73}$ & 0.22$^{+0.02}_{-0.19}$ & 2.39$^{+0.33}_{-1.02}$ & 2.01\\
096510 & 11.68$^{+0.08}_{-0.93}$ & 10.30$^{+0.00}_{-2.29}$ & 0.23$^{+0.30}_{-0.04}$ & 1.54$^{+1.28}_{-0.08}$ & 0.09\\
098022 & 11.07$^{+0.05}_{-0.25}$ & 10.30$^{+0.00}_{-0.52}$ & 0.02$^{+0.06}_{-0.02}$ & 0.93$^{+0.25}_{-0.04}$ & 3.97\\
100184 & ~8.80$^{+0.78}_{-0.61}$ & ~7.28$^{+1.46}_{-0.83}$ & 0.10$^{+0.04}_{-0.10}$ & 1.53$^{+0.41}_{-0.64}$ & 1.88\\
100509 & 10.27$^{+0.50}_{-0.58}$ & ~8.43$^{+1.14}_{-1.25}$ & 0.13$^{+0.10}_{-0.13}$ & 1.92$^{+0.61}_{-0.59}$ & 0.61\\
102671 & ~9.99$^{+0.16}_{-0.16}$ & ~7.60$^{+0.42}_{-0.31}$ & 0.18$^{+0.02}_{-0.04}$ & 2.43$^{+0.16}_{-0.23}$ & 0.42\\
103109 & 11.11$^{+0.31}_{-0.88}$ & ~9.78$^{+0.52}_{-2.18}$ & 0.06$^{+0.27}_{-0.06}$ & 1.47$^{+1.26}_{-0.25}$ & 1.39\\
103742 & ~9.43$^{+1.34}_{-1.11}$ & ~7.39$^{+2.81}_{-1.25}$ & 0.27$^{+0.06}_{-0.27}$ & 2.07$^{+0.48}_{-1.34}$ & 0.77\\
103985 & ~7.98$^{+2.99}_{-0.07}$ & ~5.72$^{+4.58}_{-0.10}$ & 0.00$^{+0.35}_{-0.00}$ & 2.25$^{+0.35}_{-1.56}$ & 0.28\\
104189 & ~9.57$^{+0.21}_{-0.09}$ & ~6.97$^{+0.42}_{-0.10}$ & 0.31$^{+0.02}_{-0.03}$ & 2.61$^{+0.11}_{-0.19}$ & 0.57\\
104766 & 10.29$^{+0.26}_{-0.12}$ & ~7.80$^{+0.73}_{-0.31}$ & 0.30$^{+0.04}_{-0.08}$ & 2.53$^{+0.18}_{-0.43}$ & 2.38\\
106944 & 10.29$^{+0.39}_{-0.68}$ & ~8.64$^{+0.83}_{-1.46}$ & 0.09$^{+0.14}_{-0.09}$ & 1.74$^{+0.79}_{-0.40}$ & 0.19\\
107878 & 10.81$^{+0.10}_{-0.08}$ & ~7.18$^{+0.21}_{-0.10}$ & 0.47$^{+0.01}_{-0.02}$ & 3.65$^{+0.04}_{-0.12}$ & 1.03\\
108167 & 11.63$^{+0.08}_{-0.56}$ & 10.30$^{+0.00}_{-1.25}$ & 0.21$^{+0.16}_{-0.04}$ & 1.49$^{+0.65}_{-0.08}$ & 3.69\\
108384 & 11.14$^{+0.71}_{-0.23}$ & ~8.01$^{+1.87}_{-0.52}$ & 0.59$^{+0.08}_{-0.26}$ & 3.18$^{+0.31}_{-1.11}$ & 3.39\\
108417 & ~8.12$^{+2.93}_{-0.03}$ & ~5.83$^{+4.47}_{-0.10}$ & 0.00$^{+0.34}_{-0.00}$ & 2.29$^{+0.39}_{-1.51}$ & 0.21\\
110593 & ~8.53$^{+0.71}_{-0.29}$ & ~6.76$^{+1.56}_{-0.42}$ & 0.08$^{+0.05}_{-0.08}$ & 1.77$^{+0.23}_{-0.78}$ & 0.05\\
112044 & 10.79$^{+0.51}_{-2.75}$ & ~9.57$^{+0.73}_{-4.06}$ & 0.14$^{+0.28}_{-0.14}$ & 1.35$^{+1.58}_{-0.51}$ & 0.01\\
113060 & 10.01$^{+0.25}_{-0.75}$ & ~9.47$^{+0.31}_{-1.87}$ & 0.00$^{+0.23}_{-0.00}$ & 0.67$^{+1.09}_{-0.05}$ & 1.37\\
113120 & ~8.61$^{+0.82}_{-0.57}$ & ~7.39$^{+1.35}_{-1.04}$ & 0.06$^{+0.07}_{-0.06}$ & 1.25$^{+0.54}_{-0.47}$ & 0.04\\
113749 & ~8.64$^{+0.00}_{-0.00}$ & ~7.28$^{+0.00}_{-0.00}$ & 0.00$^{+0.00}_{-0.00}$ & 1.38$^{+0.00}_{-0.00}$ & 12.43\\
115354 & 10.42$^{+0.56}_{-0.25}$ & ~8.12$^{+1.56}_{-0.62}$ & 0.39$^{+0.09}_{-0.22}$ & 2.37$^{+0.37}_{-0.94}$ & 0.07\\
115925 & ~9.15$^{+0.33}_{-0.27}$ & ~7.39$^{+0.73}_{-0.42}$ & 0.05$^{+0.03}_{-0.05}$ & 1.79$^{+0.21}_{-0.36}$ & 5.79\\
116678 & ~8.97$^{+0.33}_{-0.19}$ & ~6.87$^{+0.62}_{-0.21}$ & 0.27$^{+0.05}_{-0.05}$ & 2.11$^{+0.17}_{-0.34}$ & 0.08\\
116910 & ~9.97$^{+0.36}_{-0.52}$ & ~8.84$^{+0.73}_{-1.35}$ & 0.06$^{+0.16}_{-0.06}$ & 1.23$^{+0.78}_{-0.33}$ & 1.16\\
117078 & ~9.83$^{+0.16}_{-0.18}$ & ~7.28$^{+0.31}_{-0.31}$ & 0.32$^{+0.03}_{-0.03}$ & 2.57$^{+0.16}_{-0.18}$ & 0.24\\
120190 & ~9.42$^{+0.05}_{-0.06}$ & ~6.97$^{+0.10}_{-0.10}$ & 0.19$^{+0.01}_{-0.02}$ & 2.46$^{+0.09}_{-0.08}$ & 2.35\\
120554 & ~9.48$^{+0.74}_{-0.33}$ & ~7.60$^{+1.98}_{-2.50}$ & 0.24$^{+0.25}_{-0.24}$ & 1.92$^{+3.06}_{-1.17}$ & 0.01\\
120838 & ~8.92$^{+0.32}_{-0.47}$ & ~7.70$^{+0.52}_{-0.73}$ & 0.03$^{+0.06}_{-0.03}$ & 1.26$^{+0.39}_{-0.21}$ & 15.24\\
123533 & ~9.46$^{+0.27}_{-1.03}$ & ~6.35$^{+2.18}_{-1.25}$ & 0.29$^{+0.07}_{-0.29}$ & 3.11$^{+1.44}_{-2.20}$ & 0.79\\
125665 & ~9.57$^{+0.19}_{-0.12}$ & ~7.08$^{+0.31}_{-0.21}$ & 0.25$^{+0.02}_{-0.03}$ & 2.51$^{+0.13}_{-0.16}$ & 0.73\\
126010 & 10.60$^{+0.37}_{-0.47}$ & ~9.36$^{+0.73}_{-1.35}$ & 0.07$^{+0.18}_{-0.07}$ & 1.35$^{+0.82}_{-0.33}$ & 0.27\\
126510 & ~9.72$^{+0.21}_{-0.97}$ & ~5.93$^{+3.54}_{-0.83}$ & 0.42$^{+0.06}_{-0.42}$ & 3.79$^{+0.96}_{-3.20}$ & 0.02\\
127245 & ~9.80$^{+0.12}_{-0.11}$ & ~7.18$^{+0.21}_{-0.21}$ & 0.28$^{+0.02}_{-0.01}$ & 2.64$^{+0.11}_{-0.08}$ & 0.29\\
127900 & ~8.86$^{+0.38}_{-0.37}$ & ~7.39$^{+0.62}_{-0.62}$ & 0.02$^{+0.03}_{-0.02}$ & 1.50$^{+0.29}_{-0.22}$ & 7.08\\
129670 & ~9.78$^{+0.39}_{-0.88}$ & ~9.26$^{+0.52}_{-2.08}$ & 0.00$^{+0.24}_{-0.00}$ & 0.64$^{+1.16}_{-0.10}$ & 0.86\\
130018 & ~9.02$^{+0.28}_{-0.19}$ & ~6.87$^{+0.52}_{-0.21}$ & 0.15$^{+0.03}_{-0.04}$ & 2.16$^{+0.18}_{-0.28}$ & 2.40\\
130851 & ~9.28$^{+0.62}_{-0.34}$ & ~7.49$^{+1.56}_{-0.52}$ & 0.17$^{+0.04}_{-0.17}$ & 1.82$^{+0.25}_{-0.89}$ & 0.94\\
131482 & ~8.83$^{+0.21}_{-0.22}$ & ~7.28$^{+0.31}_{-0.42}$ & 0.00$^{+0.02}_{-0.00}$ & 1.57$^{+0.26}_{-0.08}$ & 24.30\\
133419 & ~9.68$^{+0.56}_{-0.20}$ & ~7.70$^{+1.56}_{-0.42}$ & 0.19$^{+0.04}_{-0.19}$ & 2.03$^{+0.24}_{-0.92}$ & 0.44\\
133694 & ~9.36$^{+0.25}_{-0.09}$ & ~7.08$^{+0.42}_{-0.21}$ & 0.21$^{+0.03}_{-0.02}$ & 2.29$^{+0.16}_{-0.18}$ & 0.49\\
134614 & 10.46$^{+0.16}_{-0.42}$ & ~9.68$^{+0.21}_{-1.04}$ & 0.00$^{+0.13}_{-0.00}$ & 0.93$^{+0.57}_{-0.04}$ & 17.93\\
135678 & ~8.49$^{+0.07}_{-0.06}$ & ~7.28$^{+0.10}_{-0.10}$ & 0.00$^{+0.00}_{-0.00}$ & 1.23$^{+0.03}_{-0.03}$ & 7.02\\
138613 & ~9.38$^{+0.33}_{-1.36}$ & ~6.35$^{+2.81}_{-1.25}$ & 0.35$^{+0.09}_{-0.35}$ & 3.03$^{+1.53}_{-2.49}$ & 0.05\\
138763 & ~9.47$^{+0.07}_{-0.13}$ & ~6.76$^{+0.21}_{-0.10}$ & 0.32$^{+0.02}_{-0.04}$ & 2.71$^{+0.06}_{-0.18}$ & 1.14\\
138810 & ~9.27$^{+0.46}_{-0.79}$ & ~8.22$^{+0.83}_{-1.35}$ & 0.07$^{+0.11}_{-0.07}$ & 1.12$^{+0.66}_{-0.36}$ & 2.38\\
139906 & ~8.65$^{+0.86}_{-0.68}$ & ~7.39$^{+1.56}_{-1.04}$ & 0.09$^{+0.06}_{-0.09}$ & 1.29$^{+0.51}_{-0.63}$ & 0.17\\
141088 & ~9.24$^{+0.42}_{-0.29}$ & ~7.28$^{+1.04}_{-0.52}$ & 0.18$^{+0.03}_{-0.09}$ & 1.98$^{+0.25}_{-0.57}$ & 7.19\\
141117 & ~9.17$^{+0.73}_{-0.34}$ & ~7.39$^{+1.98}_{-0.52}$ & 0.24$^{+0.04}_{-0.24}$ & 1.80$^{+0.27}_{-1.18}$ & 0.42\\
141368 & ~9.81$^{+0.26}_{-0.51}$ & ~8.74$^{+0.52}_{-1.25}$ & 0.04$^{+0.14}_{-0.04}$ & 1.16$^{+0.73}_{-0.23}$ & 2.05\\
142195 & 10.38$^{+0.49}_{-0.25}$ & ~8.32$^{+1.35}_{-0.62}$ & 0.23$^{+0.07}_{-0.18}$ & 2.13$^{+0.36}_{-0.79}$ & 0.94\\
144200 & ~9.81$^{+0.13}_{-0.15}$ & ~7.28$^{+0.21}_{-0.21}$ & 0.19$^{+0.01}_{-0.02}$ & 2.55$^{+0.11}_{-0.13}$ & 0.63\\
145330 & ~9.57$^{+0.61}_{-0.62}$ & ~8.32$^{+1.35}_{-1.25}$ & 0.14$^{+0.11}_{-0.14}$ & 1.31$^{+0.66}_{-0.68}$ & 0.22\\
147153 & 10.04$^{+1.23}_{-2.16}$ & ~7.80$^{+2.50}_{-2.39}$ & 0.37$^{+0.08}_{-0.37}$ & 2.28$^{+0.64}_{-1.52}$ & 0.01\\
147965 & ~9.32$^{+0.40}_{-0.99}$ & ~8.64$^{+0.42}_{-3.54}$ & 0.00$^{+0.39}_{-0.00}$ & 0.77$^{+3.79}_{-0.03}$ & 0.08\\
148198 & 10.47$^{+0.27}_{-0.19}$ & ~7.39$^{+0.62}_{-0.31}$ & 0.39$^{+0.01}_{-0.05}$ & 3.10$^{+0.11}_{-0.33}$ & 7.60\\
149470 & ~8.03$^{+0.00}_{-0.00}$ & ~7.28$^{+0.00}_{-0.00}$ & 0.00$^{+0.00}_{-0.00}$ & 0.77$^{+0.00}_{-0.00}$ & 9.72\\
149604 & ~9.04$^{+0.74}_{-1.21}$ & ~8.01$^{+1.35}_{-2.91}$ & 0.10$^{+0.35}_{-0.10}$ & 1.09$^{+3.53}_{-0.55}$ & 0.01\\
149667 & 11.24$^{+0.06}_{-0.46}$ & 10.30$^{+0.00}_{-0.94}$ & 0.12$^{+0.11}_{-0.03}$ & 1.10$^{+0.46}_{-0.05}$ & 4.01\\
152993 & ~8.62$^{+0.39}_{-0.51}$ & ~7.70$^{+0.52}_{-0.83}$ & 0.00$^{+0.06}_{-0.00}$ & 0.96$^{+0.34}_{-0.11}$ & 2.24\\
156057 & 10.07$^{+0.52}_{-0.33}$ & ~8.32$^{+1.35}_{-0.83}$ & 0.17$^{+0.08}_{-0.17}$ & 1.82$^{+0.46}_{-0.77}$ & 0.40\\
161503 & ~8.05$^{+2.19}_{-0.04}$ & ~6.04$^{+3.64}_{-0.10}$ & 0.00$^{+0.16}_{-0.00}$ & 2.02$^{+0.20}_{-1.31}$ & 4.29\\
164830 & ~9.53$^{+0.63}_{-0.32}$ & ~7.39$^{+1.77}_{-0.52}$ & 0.29$^{+0.05}_{-0.21}$ & 2.17$^{+0.28}_{-1.06}$ & 1.11\\
\enddata
\tablenotetext{a}{Errors are 90\% confidence and determined as follows: Monte Carlo realizations are computed for each object, and from the $\chi^{2}$ distribution of these realizations we determined the $\Delta\chi^{2}$ which encloses 90\% of the realizations. With this $\Delta\chi^{2}$, we define the 90\% error contour for each object. Projections of the error contours onto the relevant parameter axes give the 90\% errors on the individual parameters.}
\end{deluxetable}

\end{document}